\documentclass[]{interact}

\usepackage{epstopdf}%
\usepackage{subfigure}%

\usepackage{natbib}%
\bibpunct[, ]{(}{)}{;}{a}{}{,}%
\renewcommand\bibfont{\fontsize{10}{12}\selectfont}%

\theoremstyle{plain}%

\theoremstyle{definition}

\theoremstyle{remark}

\usepackage{url}
\usepackage{amsmath,amssymb,amsfonts}
\usepackage{algorithmic}
\usepackage[linesnumbered,ruled,vlined]{algorithm2e}
\usepackage{graphicx}
\usepackage{textcomp}
\usepackage{colortbl, hhline}
\usepackage{xcolor}
\usepackage{multirow,tabularx}
\usepackage{makecell}
\usepackage[export]{adjustbox}

\usepackage{color}
\def\credd{\textcolor{black}}
\def\cred{\textcolor{black}}
\def\cblue{\textcolor{black}}

\definecolor{darkgreen}{rgb}{0,0, 0.}
\definecolor{amber}{rgb}{0, 0, 0.0}
\definecolor{orange}{rgb}{0, 0, 0.0}

\usepackage[normalem]{ulem}

\newcommand\blfootnote[1]{%
  \begingroup
  \renewcommand\thefootnote{}\footnote{#1}%
  \addtocounter{footnote}{-1}%
  \endgroup
}

\begin{document}

\title{Hierarchical Homogeneity-Based Superpixel Segmentation: Application to Hyperspectral Image Analysis}

\author{
\name{L.~C. Ayres\textsuperscript{a}\thanks{CONTACT: Luciano Carvalho Ayres. Email: lucayress@gmail.com}, S.~J.~M. de Almeida\textsuperscript{b}, J.~C.~M. Bermudez\textsuperscript{a,b} and R.~A.~Borsoi\textsuperscript{c}
\thanks{Published in International Journal of Remote Sensing, DOI: 10.1080/01431161.2024.2384098. Copyright belongs to Taylor and Francis.}}
\affil{\textsuperscript{a}Federal University of Santa Catarina, SC Florian\'opolis, Brazil; \\\textsuperscript{b}Catholic University of Pelotas, RS Pelotas, Brazil; \\ \textsuperscript{c}Centre de Recherche en Automatique de Nancy (CRAN), Universit\'e de Lorraine, CNRS, Vandoeuvre-l\`es-Nancy, France}
}

\maketitle

\begin{abstract}
Hyperspectral image (HI) analysis approaches have recently become increasingly complex and sophisticated. Recently, the combination of spectral-spatial information and superpixel techniques have addressed some hyperspectral data issues, such as the higher spatial variability of spectral signatures and dimensionality of the data. However, most existing superpixel approaches do not account for specific HI characteristics resulting from its high spectral dimension. In this work, we propose a multiscale superpixel method that is computationally efficient for processing hyperspectral data. The Simple Linear Iterative Clustering (SLIC) oversegmentation algorithm, on which the technique is based, has been extended hierarchically. Using a novel robust homogeneity testing, the proposed hierarchical approach leads to superpixels of variable sizes but with higher spectral homogeneity when compared to the classical SLIC segmentation. For validation, the proposed homogeneity-based hierarchical method was applied as a preprocessing step in the spectral unmixing and classification tasks carried out using, respectively, the Multiscale sparse Unmixing Algorithm (MUA) and the CNN-Enhanced Graph Convolutional Network (CEGCN) methods. Simulation results with both synthetic and real data show that the technique is competitive with state-of-the-art solutions.
\blfootnote{-- A preliminary version of this work was presented by~\cite{ayres2021}.}
\end{abstract}


\section{Introduction}

Several remote sensing applications have successfully used hyperspectral imaging \citep{schowengerdt2006remote}. This technology has recently been expanded to a number of fields, including military surveillance and reconnaissance \citep{Shimoni2019}, food quality control \citep{liu2017hyperspectral,pu2023recent} and medicine \citep{halicek2019cancer,cihan2022spectral}. 
Various aspects, such as spectral mixing, atmospheric interference and acquisition noise, often make hyperspectral data analysis a complex task that requires sophisticated algorithms. The typically high spectral resolution of hyperspectral sensors, and the increasing temporal resolution required from some applications, lead to hyperspectral data of very high dimensionality and size. This requires computationally efficient solutions that can facilitate the interpretation and exploitation of hyperspectral datasets in multiple applications, such as data fusion, unmixing, classification, target detection, and physical parameter retrieval \citep{Bioucas-dias2013}.

Recently, superpixel segmentation has been frequently employed to address some of these problems in HI analysis. Superpixel structures can be adapted in size and shape based on the spatial and spectral information of adjacent pixels in the observed scene. Although originally designed to work with conventional images, the superpixel concept has been applied to hyperspectral data, and has gained much popularity in HI processing research. There is a wide variety of superpixel generation algorithms, and some works that compare their performances\cred{~\citep{achanta2012slic, machairas2015waterpixels, Wang20177, stutz2018}}. A survey on the use of superpixels as a preprocessing step in HI analysis describes how these segmentation algorithms have been successfully applied to tasks such as classification, unmixing, dimensionality reduction, and band selection, among others \citep{Subudhi2021}. Approximately 80\% of the proposed techniques focused on classification and unmixing problems, which are briefly reviewed in the following.

\subsection{Hyperspectral unmixing}

Large sensor-to-target distances and the low spatial resolution of hyperspectral sensors cause the observed reflectance spectrum of a given HI pixel to frequently represent a combination of different materials ~\citep{Bioucas-Dias2012}.
The technique of \emph{Spectral unmixing} involves breaking down the spectrum of a mixed pixel from a HI into a set of \emph{endmembers}, or spectral signatures of pure materials, and a set of fractional \emph{abundances} that show what proportion of the pixel each endmember represents~\citep{Keshava2002}.
The majority of approaches to the unmixing problem are based on a linear mixing model (LMM) \citep{Keshava2002, Dobigeon2014}, which assumes that each HI pixel is constituted by the linear combination of a few endmembers, weighted by their fractional abundances.

When the endmembers are correctly estimated, the LMM produces fast and accurate unmixing solutions~\citep{Bioucas-Dias2012}.
However, most algorithms that extract the endmembers directly from a given HI rely on the presence of pure pixels (i.e., pixels containing only a single material), or on the data not being heavily mixed~\cite{ma2013signal}. \emph{Sparse regression-based} linear unmixing, which gets over this limitation, makes the assumption that the reflectance of the observed pixels in an HI can be described as a linear combination of a few endmember signatures from a large, \textit{a priori} known spectral library \citep{Iordache2011}.
The unmixing problem consists of finding the subset of signatures in the library and their abundances to best represent each observed pixel. Employing spectral libraries avoids the need for estimating the number of endmembers and their spectral signatures. However, using large libraries makes the unmixing problem ill-posed, and the solution highly sensitive to noise \citep{Iordache2010}.

Traditional unmixing algorithms process the spectrum of each observed image pixel independently, therefore ignoring the spatial organisation and the relationship between adjacent pixels \citep{Iordache2010, Iordache2014, Wang2016, Zheng2016}.
Integrating spatial-contextual information through regularisers can significantly enhance the performance of sparse unmixing \citep{Iordache2012, Zhang2017a, Wang2017a, Zhang2018, QI201997}. Moreover, the use of spatial regularisations also brings significant performance improvements to unmixing algorithms that consider nonidealities such as nonlinear interactions between light and the materials in the scene~\citep{Dobigeon2014,Borsoi2019a}, and the variability of the endmembers spectra across different pixels due to acquisition or physico-chemical variations~\citep{borsoi2021spectralVariabilityReview}.
In particular, recent works have proposed regularisation approaches that leverage the superpixels decomposition of the HI as a representation of its spatial information.
A data-dependent multiscale regularisation model was proposed taking into account the spectral variability of the endmembers, where both the abundances and the endmember signatures were constrained to vary smoothly inside each superpixel \citep{Borsoi2020Data}. Superpixels were used in a multiscale regularisation strategy for nonlinear spectral unmixing based on kernels \citep{Borsoi2019a}. As for the sparsity-based regularisation methods using superpixels, a spatial group-sparsity-regularised non-negative matrix factorisation (NMF) was presented \citep{Wang2017spatial}, which comprises a spatial group sparsity regularisation constraint into the NMF-based unmixing problem. A robust generalised bilinear model (GBM) unmixing solution based on superpixels and low-rank representation was proposed by \cite{Mei2018}. Also, a spatial regularisation technique using superpixels and graph Laplacian regularisation was presented for the sparse unmixing problem \citep{Ince2020,Ince2021}.

A concern about the use of spatial information in the unmixing problem is that it usually increases the computational complexity of the algorithms.
To address this limitation, a fast sparse unmixing algorithm called Multiscale sparse Unmixing Algorithm (MUA) was proposed to efficiently introduce spatial context in the unmixing problem \citep{Borsoi2019}. Applying the Simple Linear Iterative Clustering (SLIC) \citep{achanta2012slic} oversegmentation algorithm, MUA demonstrated its capacity to estimate abundances with a quality that is comparable to the state-of-the-art S$^2$WSU \citep{Zhang2018} (or even superior in noisy circumstances), with significantly lower execution time.
A variation of MUA based on an $\ell_{1/2}$ norm regularisation has also been proposed by \cite{Zou2019}.
A superpixel-based collaborative sparse and low-rank regularisation algorithm was proposed to solve the sparse unmixing \citep{chen2022superpixel}. Also, a reweighted low-rank and joint-sparse unmixing method \citep{zhang2022reweighted} divided the problem into two stages: a superpixel-based rough unmixing stage, and a fine-tuning unmixing stage. More recently, \cite{ye2022combining} combined low-rank constraints for similar superpixels and
total variation sparse unmixing for HIs. A spectral-spatial-sparse unmixing with superpixel oriented graph Laplacian was also presented by \cite{li2023spectral}. \cred{A superpixel-based multiscale approach was also proposed in~\citep{ayres2024generalizedMUA} for unmixing with endmember bundles and structured sparsity penalties.}

Superpixel-based algorithms such as MUA have become a well-established approach to perform unmixing while taking spatial information into account. To obtain significant performance improvements, these methods, however, heavily rely on two aspects of the oversegmentation results: superpixels must 1) aggregate a large number of pixels, and 2) have spectrally uniform pixels inside of them, with the exception of noise. However, conventional superpixel (or image segmentation) methods~\citep{MacQueenJamesandothers1967,Veganzones2014,achanta2012slic,Beucher1993,Shi2000,Felzenszwalb2004,Levinshtein2009} aren't made to maximise these requirements. Using superpixel-based unmixing algorithms to HIs with irregular spatial arrangements thus indicated a higher sensitivity to the image content.
Nevertheless, these algorithms rely strongly on two characteristics of the oversegmentation results to achieve meaningful performance improvements: 1) the superpixels must group a large number of pixels, and 2) the pixels within a superpixel must be spectrally homogeneous, except for the influence of noise. However, traditional superpixel (or image segmentation) algorithms~\citep{MacQueenJamesandothers1967,Veganzones2014,achanta2012slic,Beucher1993,Shi2000,Felzenszwalb2004,Levinshtein2009} are not designed to optimise these criteria. Thus, superpixel-based unmixing algorithms revealed a greater sensitivity to the image content when applied to images with irregular spatial compositions.

\subsection{Hyperspectral image classification}

The extensive spectral information available from the many narrow bands recorded by hyperspectral instruments makes it feasible to accurately discriminate between different materials. This information embedded in hyperspectral data allows the characterisation and identification of elements on the surface of the observed scene with greater accuracy and robustness.
In HI analysis, the classification task tries to allocate a unique class label to each pixel of the image, generating a classification map of the scene as a result \citep{Landgrebe2002}. Nevertheless, classification is not a simple task due to several factors, such as the existence of redundant features, the quality and availability of training samples, and the high dimensionality of the data.

Classification approaches for HIs can be organised according to several criteria \citep{P.Ghamisi2017}, such as whether or not labelled training data is required. Supervised methods classify the input image using training data, while unsupervised approaches do not consider the labels of the training samples. Semi-supervised methods, on the other hand, combine a small amount of labelled data with a large amount of unlabelled data to generate a classification map.
Various supervised and unsupervised techniques have been proposed to handle the problem of classifying HIs \citep{P.Ghamisi2017}, such as artificial neural networks \citep{civco1993, bischof1998}, support vector machines (SVMs) and other kernel-based approaches \citep{melgani2004}. The major limitation of supervised strategies is that the training process relies strongly on the quality of the labelled dataset. Moreover, the training set is often limited or not available due to the high cost of accurate sample labelling. Unsupervised classification techniques, on the other hand, are based on clustering methods, such as spectral clustering~\citep{zhao2019}.
In addition, prior knowledge of the scene is needed, as a preliminary preprocessing step is typically performed to reduce the high dimensionality of the input, such as, e.g., band selection \citep{yang2018superpixel} or feature extraction using sparse autoencoders \citep{tao2015unsupervised}. Still, unsupervised methods have shown satisfactory results in HI classification tasks \citep{zhao2019}. Semi-supervised classification strategies offer the possibility of exploiting effectively both labelled and unlabelled data, improving performance in small sample problems. \cred{Several semi-supervised approaches have been used for HSI classification, ranging from graph-based~\citep{Kotzagiannidis2021a} to neural-network-based approaches~\citep{cao2023transformerContrastiveLossClassification,cao2023semiSupervisedDisjointClassification}.}

To relieve the shortage of labelled samples and increase the classification accuracy of HIs, \cite{Sellars2020} \cred{use SLIC superpixels} to generate local regions that have a high probability of sharing the same classification label. 
Then, spectral and spatial features are extracted from these regions to be used to produce a weighted graph representation, where each vertex represents a region (superpixel) rather than a pixel. The graph is then fed into a graph-based semisupervised classifier that provides the final classification map. In a similar more recent work \citep{Kotzagiannidis2021a}, graphs are also used to model the semi-supervised classification problem by propagating labels among nearest neighbours. A superpixel-based augmentation technique was further presented to address the issue of limited labelled samples, with the purpose of expanding the number of training instances \citep{li2023hyperspectral}.

Like in unmixing, the use of spatial information in addition to spectral properties can lead to improved HI classification performance \citep{He2018, Mookambiga2021}.
Several deep learning algorithms have achieved promising results in HI classification \citep{audebert2019, Kumar2022}. For instance, convolutional neural networks (CNNs) have been widely used to extract HI spectral-spatial features. Recently, it has been verified through semi-supervised approaches which represent hyperspectral data in graphs that the spatial contextual structure of the HI can be better modelled by graph convolutional networks (GCNs) \citep{qin2018}.
However, GCNs are complex structures which require a significant training time. The segmentation of an HI using superpixels has been proposed to significantly reduce the graph size and thus make GCNs more feasible for HI classification \citep{wan2019multiscale, DING2022246}. Despite the fact that such approaches can achieve state-of-the-art performance~\citep{Liu2021}, the classification quality is, as in unmixing, directly linked to the quality of the oversegmentation maps.

\cred{Moreover, we note that superpixels algorithms have also been used to improve other aspects of hyperspectral or multispectral image classification. For instance, in \citep{dieste2023resbagan} and \citep{arguello2021watershed}, the waterpixels algorithm~\citep{machairas2015waterpixels} was used to reduce the computational load of a classification process for multispectral images. In the case of \citep{accion2020dual}, SLIC was used for data augmentation before the classification of HIs.}
\cred{Note that when superpixel segmentation algorithms are used as preprocessing to pixel-wise unmixing or classification tasks (so that only one spectral vector related to each superpixel has to be unmixed/classified), the computational complexity of these downstream tasks can be reduced by the ratio between the number of superpixels in the decomposition and the total number of pixels in the image.}

\subsection{Challenges, contributions and organisation}

\cred{The trend towards incorporating spatial information into HI analysis by means of superpixels is well-established in the literature. Besides HI classification and unmixing, this approach has also been applied to other problems in hyperspectral and multispectral imaging such as change detection~\citep{naik2023spatio}. Moreover, different superpixel segmentation strategies have been used for HI processing in the literature, including SLIC, which is clustering-based, the waterpixels~\citep{machairas2015waterpixels}, which is a gradient based algorithm and presents high adherence to image edges~\citep{dieste2023resbagan,arguello2021watershed}, and other algorithms that are based on the optimisation of carefully constructed energy functions~\citep{yao2015real,van2012seeds}. However, most of these} segmentation techniques were not originally intended for the hyperspectral case. This shows the need and opportunity to develop multiscale representations that meet the specific requirements of problems such as spectral unmixing and classification.

In the HI segmentation process, measures of pixel similarity are usually based on Euclidean distance measurements and spectral angles \citep{Keshava2002} between the pixels and the representative vector of its segment. Although such metrics can be used to identify uniform regions, their effectiveness depends on factors such as parameter initialisation, target application and the nature of the algorithm, e.g., graph-based or gradient-ascent-based \citep{achanta2012slic}. Most image segmentation algorithms that have been applied to HIs do not provide guarantees of spectral uniformity within the regions. This points to the need for an oversegmented image, i.e., to reduce the area size of the regions attempting a higher level of homogeneity. On the other hand, oversegmentation algorithms produce superpixels of relatively similar sizes. This can lead to the grouping of different materials within the same region when the size of the superpixels is large; or to a large number of superpixels of very small average area. An unnecessary number of superpixels can compromise the benefits of dimensionality reduction provided by image segmentation. Therefore, an important additional step is to assess the level of homogeneity of the resulting superpixel regions to determine whether the oversegmentation step is satisfactory or if improvements should be made.
Note that although hierarchical superpixel decomposition algorithms have been proposed, these were aimed at conventional images and are not equipped to deal with the challenges in segmenting HIs.
As proposed by \cite{Vasquez2018}, the representation of digital images is done not only at two scales (original and over-segmented), but at different levels of over-segmentation, in a hierarchical structure based on graphs. Lower resolution scales are generated by contractions of the graphs until a desired number of superpixels is obtained. However, for a higher number of bands, as in HIs, this technique can become excessively complex computationally, or may not generate good results.
Likewise, a hierarchical tree algorithm was proposed that is capable of generating superpixels of multiple scales but with one to two orders of magnitude of speed-up \citep{wei2018}.
An image superpixel segmentation based on hierarchical multi-level local information was also proposed by \cite{liao2021}. However, these methods were aimed at segmenting RGB images.

The homogeneity level of the regions formed in HI segmentation remains relatively unexplored in the literature. A similarity threshold based on the maximum spectral angular distance (SAD) between pixel spectra was used to determine whether they belong to the same class of material (endmember) \citep{Saranathan2016}. The homogeneous regions were then represented by the average spectral signature of the superpixel for spectral unmixing. The technique can improve the performance of the original graph-based Felzenszwalb and Huttenlocher (FH) superpixel generation algorithm \citep{Felzenszwalb2004} in the preprocessing of tasks such as unmixing. However, the clusters formed by this method are not necessarily contiguous, and classes that do not correspond to a real endmember can be produced if the similarity threshold is too restrictive. Moreover, the modified FH algorithm is computationally costly and does not offer explicit parameters for controlling the size and compactness of the superpixels.

The homogeneity of superpixels generated using the SLIC algorithm for spectral unmixing was analysed by \cite{Yi2018}. After the HI segmentation, each superpixel was expressed as a matrix composed of the spectra of their inner pixels. Superpixels of nearly rank-1 were then considered homogeneous. The average spectral signature only represents homogeneous superpixels, in contrast to \cite{Saranathan2016}. The heterogeneous ones are characterised by the most representative pixels according to algorithms that solve a column-subset-selection (CSS) problem. Despite providing an enhanced lower-dimensional representation of an HI, this procedure does not increase the level of homogeneity of the superpixels defined using SLIC.
A hierarchical segmentation algorithm called Binary Partition Tree (BPT) was modified to handle the high spectral information contained in an HI aiming at spectral unmixing quality in terms of spectral reconstruction error \citep{Veganzones2014}. Although adapted for unmixing, experiments reported by \cite{Borsoi2019} showed that BPT often creates clusters of pixels of considerably different sizes, corresponding to both small and large objects in the scene. While this makes it possible to represent large regions with homogeneous abundance characteristics without compromising smaller objects, it can lead to the clustering of pixels that share different abundance characteristics. When applied to the unmixing process, this segmentation map generated results that were very sensitive to the image content. Despite generating regions with similar sizes, the SLIC algorithm was able to mitigate these limitations by generating a larger number of regions, what improved the performance of unmixing.
Furthermore, the lack of homogeneity guarantees in existing superpixels algorithms also constitutes a challenge for superpixel-based HI classification methods~\citep{tu2018knn_superpixels_classification}.
Though a few works have proposed superpixel methods that are trainable~\citep{jampani2018superpixelNetworks} or adapted to ignore textured image content~\citep{yuan2021contentAdaptiveSuperpixels}, learning-based approaches require training data and existing methods are not adequate to process HIs.

This paper proposes a novel hierarchical homogeneity-based oversegmentation method adapted for hyperspectral image analysis problems such as unmixing and classification.
The presented approach is based on a reliable superpixel homogeneity test and the SLIC oversegmentation algorithm. The Euclidean distance between a superpixel's median vector and each of its pixels is used in this novel superpixel homogeneity assessment to mitigate the effect of potential outliers on the test result. Those superpixels classified as non-homogeneous are subdivided  into smaller regions by successive rounds of oversegmentation. This leads to superpixels that are less likely to group pixels from different material compositions or classes, while mitigating the influence of noise and outliers. The proposed oversegmentation method is then applied to HI unmixing and to HI classification problems. The proposed method is referred to as Hierarchical Homogeneity-Based Oversegmentation~(H$^2$BO).

In unmixing, the proposed method is used in the first abundance estimation step (coarse domain) of MUA~\citep{Borsoi2019}. This results in the so-called MUA$_{\text{H$^2$BO}}$ algorithm, which leads to a multiscale representation of an HI that improves the efficiency of sparse unmixing. The technique enhances the sparse unmixing performance of the previous MUA$_{\text{SLIC}}$ for HIs with irregular spatial compositions by more accurately describing the spatial organisation of the abundances in an HI with content spread in regions of different sizes and shapes. 
For the HI classification problem, we propose to replace the application of the linear discriminant analysis (LDA) technique in the dimensionality reduction step of the CEGCN algorithm \citep{Liu2021}. Our hierarchical oversegmentation method is then applied in place of SLIC in the superpixel-based graph constructor (SGC) that defines the graph encoder and decoder of the CEGCN algorithm. The resulting method is termed CEGCN$_{\text{H$^2$BO}}$.

It is important to notice that the proposed hierarchical representation based on superpixel homogeneity assessment can be used as a preprocessing step in other superpixel-based algorithms besides MUA and CEGCN.

The main contributions of the paper are:
\begin{itemize}
    \item We propose a new, computationally efficient hierarchical superpixel method adapted to hyperspectral imaging tasks such as classification and unmixing.
    \item To account for peculiarities of hyperspectral data, such as noise and outliers, a new robust homogeneity test is devised.
    \item The performance of the method is demonstrated with two hyperspectral image analysis tasks (unmixing and classification). 
\end{itemize}

The paper is organised as follows. Section \ref{sec:proposed_method} introduces the design of the proposed multiscale representation of the hyperspectral data. Section \ref{sec:unmixing} presents a brief explanation for the sparse unmixing problem and its use as a MUA, as well as experimental results and discussion with synthetic and real data. Section~\ref{sec:classification} brings the application of the proposed technique to the classification task by adapting the CEGCN method. The conclusions are presented in Section~\ref{sec:conclusions}.

\section{Proposed Hierarchical Homogeneity-Based Oversegmentation} \label{sec:proposed_method}

The results of using image segmentation to introduce spatial information in hyperspectral image analysis strongly depend on the image pixels being adequately grouped into spectrally homogeneous regions. Existing oversegmentation algorithms usually do not guarantee a decomposition with an appropriate homogeneity level, particularly in low signal-to-noise ratio (SNR) scenarios. To address this limitation, we propose a new multiscale image segmentation algorithm specifically designed for hyperspectral image analysis problems, such as sparse unmixing and classification. 

Note that the content in an HI can be distributed in regions of irregular sizes and shapes.
Hence, characterising the spatial organisation of the endmembers in a scene may not be suitable when adopting an oversegmentation approach with a single average superpixel size for the entire region. Large (small) superpixels should be used to segment large (small) patterns. For this, we suggest gradually oversegmenting the HI in a number of scales. Initially, use large superpixels and an appropriate metric to evaluate the homogeneity of these regions following each oversegmentation stage. Thereafter, more cycles of oversegmentation with gradually smaller average superpixel sizes are applied to the regions designated as non-homogeneous. In this manner, a sufficient amount of homogeneous superpixels that can adapt to the various pattern shapes in the HI can be produced.
Beginning from the observed HI $\mathbf{Y} \in \mathbb{R}^{L \times N}$ with $L$ bands and $N$ pixels, to reach a predetermined level of homogeneity, we consider gradually decomposing its non-homogeneous superpixels (up to $R$ representation scales). Moreover, to adequately account for the high dimensionality of hyperspectral data, we propose a new robust homogeneity test.

\begin{figure}[!hbt]
\centerline{\includegraphics[width=\linewidth]{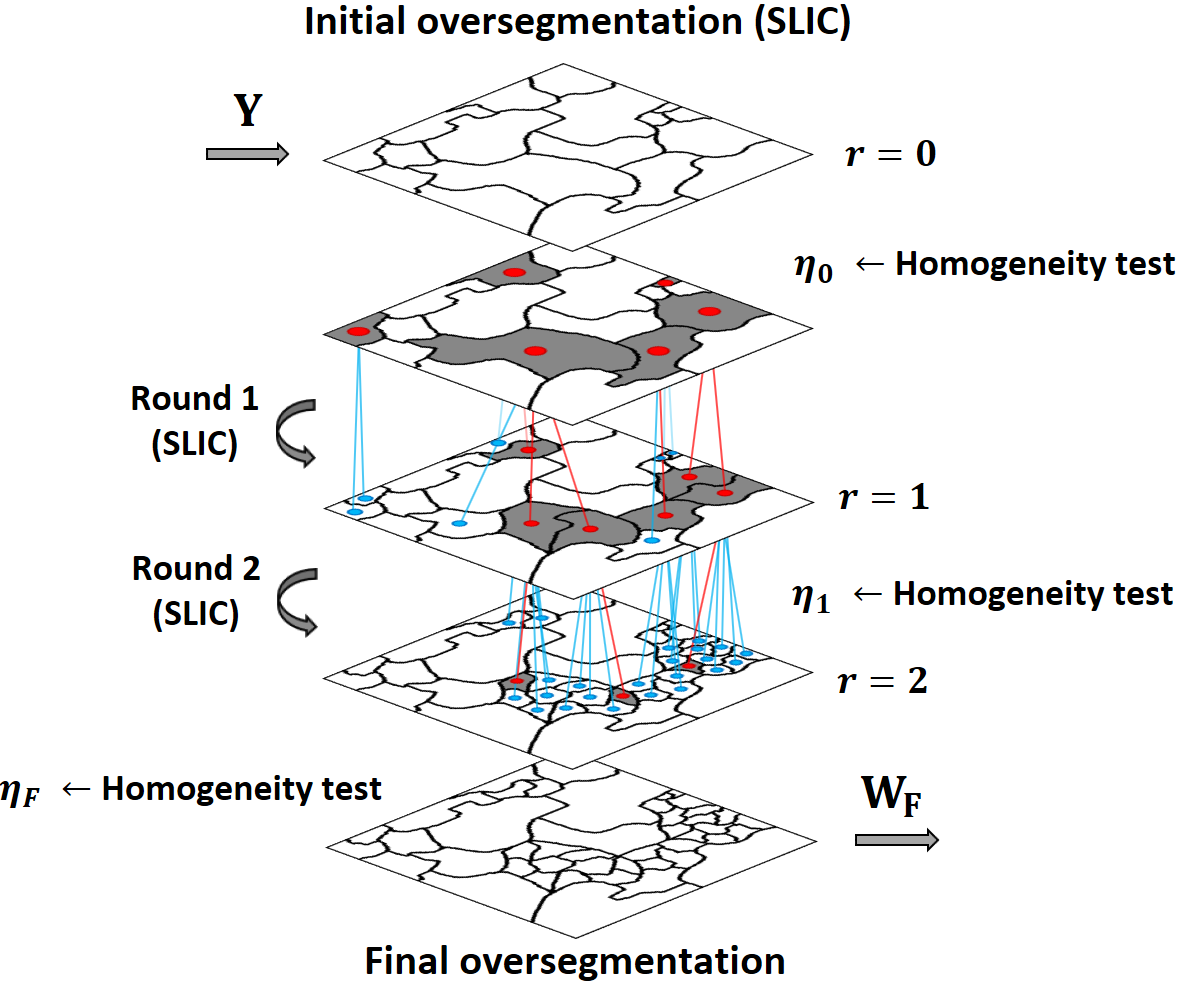}}
\caption{Scheme of proposed method procedure with SLIC for $R=2$. The superpixels identified as non-homogeneous (homogeneous) are represented by the grey regions with red dots (white regions with blue dots). Non-homogeneous superpixels in scale $r$ are divided into homogeneous (non-homogeneous) superpixels in scale $r+1$ indicated by blue (red) lines.
}
\label{fig:metodo_esquema}
\end{figure}

Figure~\ref{fig:metodo_esquema} shows the SLIC method to decompose an HI on multiple scales for $R=2$. The superpixels identified as non-homogeneous (homogeneous) are represented by the grey regions with red dots (white regions with blue dots). Non-homogeneous superpixels in scale $r$ are divided into homogeneous (non-homogeneous) superpixels in scale $r+1$ indicated by blue (red) lines. The steps of the proposed hierarchical oversegmentation algorithm are detailed in the following.

\subsection*{\textbf{Step 1} -- Initial image oversegmentation}

The SLIC algorithm forms a segmentation map that subdivides the image into $K_0$ superpixels with average region size determined by parameter $\sigma_0 = \sqrt{N/K_0}$ and compactness controlled by parameter $\gamma$ \citep{achanta2012slic}\footnote{$\sigma_0$ and $\gamma$ correspond to parameters $S$ and $m$, in the original reference, respectively.}.
Consider the matrix $\mathbf{S}_{r,k} \in \mathbb{R} ^{L\times|\mathcal{B}_{r,k}|}$ whose columns constitute the pixels contained in superpixel $k\in\{1, 2,\ldots,K_r\}$ in a scale of representation $r\in\{0, 1,\ldots,R\}$. The set $\mathcal{B}_{r,k}$ contains the indexes $I_n$, $n=1, 2,\ldots, |\mathcal{B}_{r,k}|$ ($|\cdot|$ denoting set cardinality) of each pixel in the $k^{\rm th}$ superpixel at the $r^{\rm th}$ representation scale. Then,
\begin{equation} \label{eq:sppx_S}
    \mathbf{S}_{r,k} = \big[ \mathbf{y}_{I_1}, \mathbf{y}_{I_2},\ldots, \mathbf{y}_{I_{|\mathcal{B}_{r,k}|}} \big] \,,
\end{equation}
where $\mathbf{y}_i$ is the $i^{\rm th}$ column of the original image $\mathbf{Y}$ and $\{I_1, I_2,\ldots, I_{|\mathcal{B}_{r,k}|}\} \ = \mathcal{B}_{r,k}$. We also define by $\mathbf{W}_r\in\mathbb{R}^{N\times K_r}$ an image \emph{coarsening} transformation promoted by the segmentation map $\mathbf{S}_{r,k}$ at the $r^{\rm th}$ representation level, for $r\in\{0,\ldots,R\}$.

The transformation $\mathbf{W}_r$ computes the average of all HI pixels inside each of the superpixels at the scale $r$. At the final decomposition scale (i.e., the smallest $r\leq R$ such that all superpixels are homogeneous) this transformation is denoted by $\mathbf{W}_F\in\mathbb{R}^{N\times K_F}$ and contains $K_F<N$ superpixels. It defines a coarse-scale spatial decomposition which can be applied to the HI as:
\begin{equation} \label{eq:sparse_Yc}
    \mathbf{Y}_{\!\mathcal{C}} = \mathbf{YW}_F
\end{equation}
where $\mathbf{Y}_{\!\mathcal{C}} \in \mathbb{R}^{L \times K_F}$, and the $\mathcal{C}$ subscript refers to the new approximate (coarse) domain of the image. The relation $K_F<N$ means that there are lesser superpixels in $\mathbf{Y}_{\!\mathcal{C}}$ than pixels in $\mathbf{Y}$. Columns of $\mathbf{Y}_{\!\mathcal{C}}$ are the average of all pixels in each superpixel.

The coarse image $\mathbf{Y}_{\!\mathcal{C}}$ can also be represented in the original image domain (i.e., a regular spatial grid with $N$ pixels), denoted by $\mathcal{D}$. We consider a representation in which all pixels inside each superpixel share the same spectrum, which can be useful to hyperspectral image analysis tasks.
To this end, consider a conjugate transformation by a matrix $\mathbf{W}^*_F\in\mathbb{R}^{K_F \times N}$, represented by $\mathbf{{Y}}_{\!\mathcal{D}} = \mathbf{Y}_{\!\mathcal{C}}\mathbf{W}^*_F \in \mathbb{R}^{L \times N}$.
Operation $\mathbf{{Y}}_{\!\mathcal{C}}\mathbf{W}^*_F$ applies the average reflectance value of each superpixel in $\mathbf{Y}_{\!\mathcal{C}}$ to every pixel in the corresponding superpixel region in the original domain. An example of the computation of $\mathbf{W}^*_F$ for an image with 5
pixels and 3 superpixels ($N = 5$ and $K = 3$), where pixels 1 and 2 belong to superpixel 1, pixel 3 belongs to superpixel 3 and pixels 4 and 5 belong to superpixel 2 is shown in \eqref{eq:ex_superpixel_w_ast}.
\begin{equation} \label{eq:ex_superpixel_w_ast}
    	\underbrace{
    	\left[\begin{array}{ccccc}
        | & | & | & | & | \\
    	\mathbf{y}_{\mathcal{C}_1} & \mathbf{y}_{\mathcal{C}_1} & \mathbf{y}_{\mathcal{C}_3} & \mathbf{y}_{\mathcal{C}_2} & \mathbf{y}_{\mathcal{C}_2} \\
        | & | & | & | & |
    	\end{array}\right]
        }_{\mathbf{{Y}}_\mathcal{D}}
        =
        \underbrace{
        \left[\begin{array}{ccc}
        | & | & | \\
    	\mathbf{y}_{\mathcal{C}_1} & \mathbf{y}_{\mathcal{C}_2} & \mathbf{y}_{\mathcal{C}_3} \\
        | & | & | 
    	\end{array}\right]
        }_{\mathbf{{Y}}_\mathcal{C}}
        \cdot
        \underbrace{
        \left[\begin{array}{ccccc}
        1 & 1 & 0 & 0 & 0 \\
    	0 & 0 & 0 & 1 & 1 \\
        0 & 0 & 1 & 0 & 0
    	\end{array}\right]
        }_{\mathbf{W}^\ast_F}
\end{equation}

\subsection*{\textbf{Step 2} -- Superpixels homogeneity test}

A crucial step of the proposed strategy is deciding whether or not a superpixel is homogeneous. This is not straightforward, since HIs can be contaminated by noise and outliers. To address this challenge, we propose a new robust homogeneity test.
Let $\mathbf{m}_k$ be the median of the pixels in $\mathbf{S}_{r,k}$, which belong to the $k^{\rm th}$ superpixel at a given representation scale~$r$:
\begin{equation} \label{eq:sppx_m}
    \mathbf{m}_k = 
    \begin{bmatrix}
    m_{1} \\
    \vdots \\
    m_{L}
    \end{bmatrix} =
    \begin{bmatrix}
    \text{med}\big([y_{1,I_1},\ldots,y_{1,I_{|\mathcal{B}_{r,k}|}}]\big) \\
    \vdots \\
    \text{med}\big([y_{L,I_1},\ldots,y_{L,I_{|\mathcal{B}_{r,k}|}}]\big)
    \end{bmatrix},
\end{equation}
where $y_{\ell,i}$ is the $\ell^{\rm th}$ band of $\mathbf{y}_i$.
Define $\mathbf{d}_k$ as the vector of Euclidean distances between each pixel $\mathbf{y}_{I_n}$ and $\mathbf{m}_k$:
\begin{equation}\label{eq:sppx_dist_px}
    \mathbf{d}_k=\big[d_1,d_2,\ldots,d_{|\mathcal{B}_{r,k}|} \big]^{\top}, \quad
    d_n=\|\mathbf{m}_k - \mathbf{y}_{I_n}\|_2 \,.
\end{equation}
While evaluating homogeneity, it is important to keep in mind that even for fully homogeneous superpixels, the values in $\mathbf{d}_k$ are typically constrained away from zero due to the influence of measurement perturbations in $\mathbf{Y}$. Also, a small number of outliers may distort the estimate significantly. To solve these issues, we first clear out potential outliers by eliminating a part of the highest values from $\mathbf{d}_k$ as $\tau_{\text{outliers}}$. This results in a reduced distance vector $\mathbf{d}_{k}' \in \mathbb{R}^{\lfloor D \rfloor}$ with $D=(1-\tau_{\text{outliers}})|\mathcal{B}_{r,k}|$, $\lfloor\cdot\rfloor$ being the floor function. The homogeneity measure $\delta_k$ is then established as the deviation between the maximum distance $\text{max}(\mathbf{d}_{k}')$ found in the $k^{\rm th}$ superpixel after removing the outliers, with respect to the average $\overline{\mathbf{d}_{k}'}$ of its distances:

\begin{equation}\label{eq:sppx_delta}
    \delta_k=\frac{\text{max}(\mathbf{d}_{k}')-\overline{\mathbf{d}_{k}'}}{\overline{\mathbf{d}_{k}'}} \,, \qquad \text{Homogeneous:  }\delta_k \leq \tau_{\text{homog}} \,.
\end{equation}
If $\delta_k$ is less than a reasonable threshold, superpixels are considered homogeneous ($\tau_{\text{homog}}$).
The percentage of homogeneous superpixels in the $r^{\rm th}$ scale with $K_r$ superpixels is defined by $\eta_r = (K_{\text{homog}}/K_r)\times 100\%$, where $K_{\text{homog}}$ is the amount of superpixels classified as homogeneous.

\subsection*{\textbf{Step 3} -- Subdivision of non-homogeneous superpixels}

Areas identified in Step~2 as non-homogeneous are subjected to a second oversegmentation phase with a reduced average region size parameter $\sigma_r<\sigma_{r-1}$, $\forall r > 0$. This process is repeated for $r=1,\ldots,R$, or until $\eta_r = 100\%$ is attained. The segmentation scale for which the algorithm executes the subdivision of non-homogeneous superpixels for the final time will be denoted by $r'\leq R$. This produces a series of segmentation maps with increasing spatial definition and superpixel homogeneity: $\mathbf{S}_{1,k},\mathbf{S}_{2,k},\ldots,\mathbf{S}_{r',k}$. The hierarchical oversegmentation related to $\mathbf{S}_{r',k}$ yields the final ($F$ subscript) spatial transformation operator $\mathbf{W}_F$. A pseudocode for the hierarchical homogeneity-based oversegmentation method is presented in Algorithm \ref{alg:metodo}.

\begin{algorithm}[!htb] 
\SetAlgoLined
\KwIn{hyperspectral image $\mathbf{Y}$, parameters $\gamma$, $\sigma_0$, $\sigma_1$, $\ldots$, $\sigma_R$, $\tau_{\text{outliers}}$, $\tau_{\text{homog}}$.}
 $\mathbf{W}_0, \{\mathbf{S}_{0,k}\}_k \leftarrow$ initial oversegmentation of $\mathbf{Y}$\;
 $\mathbf{W}_F \leftarrow \mathbf{W}_0$\;
 $\eta_0 \leftarrow$ homogeneity test of $\{\mathbf{S}_{0,k}\}_k$\;
  \For{$r=1$ \KwTo $R$}{
   \If{$\eta_{r-1} < 100\%$}{
    $\mathbf{W}_r, \{\mathbf{S}_{r,k}\}_k \leftarrow$ oversegmentation of non-homogeneous superpixels of \{$\mathbf{S}_{r-1,k}\}_k$ with $\sigma_r < \sigma_{r-1}$\;
    $\eta_r \leftarrow$ homogeneity test of $\{\mathbf{S}_{r,k}\}_k$ \;
    $\mathbf{W}_F \leftarrow \mathbf{W}_r$\;
   }
  }
 \Return{\rm{the final transformation matrix} $\mathbf{W}_F$};
 \caption{\textit{Hierarchical Homogeneity-Based Oversegmentation~(H$^2$BO)}} \label{alg:metodo}
\end{algorithm}

\begin{color}{black}

\subsection{Parameter selection} \label{sec:parameter_selection}

The proposed H$^2$BO algorithm requires several parameters to be \cblue{selected. However,} choosing most of them is straightforward.  %
The number of scales ($R$) determines the number of hierarchical levels used in the segmentation. While a higher $R$ value leads to more refined segmentations, it also increases the computation time. In practice, due to the rapid \cblue{homogenization} of superpixels within a few iterations, small values such as $R=3$ are often sufficient.
The $\sigma_0,\ldots,\sigma_R$ values represent the size of superpixels at each decomposition level, with $\sigma_i>\sigma_{i+1}$. These values are not overly critical to the overall performance. However, it's important to avoid excessively small initial superpixel sizes ($\sigma_0$). The rationale behind this is that the algorithm relies on the re-segmentation of large, non-homogeneous superpixels.
The outlier threshold ($\tau_{\rm outliers}$) is used to remove outliers from the computation of the metrics employed during segmentation. This parameter is typically set between 0 and 100\%, with values around 10\% being sufficient in most cases. $\tau_{\rm outliers}$ exhibits low sensitivity to specific settings.
The homogeneity threshold ($\tau_{\rm homog}$) plays a crucial role in determining whether a superpixel is deemed homogeneous. It is based on the relative one-sided dispersion of the values in $\mathbf{d}_{k}'$ relative to the median. Due to its dependence on image content, the optimal value for $\tau_{\rm homog}$ can vary. For homogeneous HSIs, values between 20 and 50\% are often adequate. Conversely, highly textured HSIs, such as those containing vegetation or significant noise (common in many real-world HSIs), may require higher thresholds ranging from 100 to 200\%.

\end{color}

\section{Application to spectral unmixing} \label{sec:unmixing}

In this section, we illustrate the performance of the proposed homogeneity-based oversegmentation method with the spectral unmixing problem. To this end, we consider the MUA unmixing algorithm~\citep{Borsoi2019} due to its good trade-off between low computational complexity and good unmixing performance. Moreover, MUA only takes spatial contextual information into account through the HI oversegmentation results, which makes evaluating the quality of H$^2$BO's results easier.

The MUA algorithm solves the spectral unmixing problem by sparse linear regression.
Consider the LMM \citep{Bioucas-Dias2012} of an observed HI $\mathbf{Y} \in \mathbb{R}^{L \times N}$ with $L$ bands and $N$ pixels as $\mathbf{Y} = \mathbf{AX}+\mathbf{N}$, where abundances $\mathbf{X} \in \mathbb{R}^{P \times N}$ are subject to nonnegativity constraints. Matrix $\mathbf{A} \in \mathbb{R}^{L \times P}$ denotes a spectral library containing $P$ endmember signatures and $\mathbf{N} \in \mathbb{R}^{L \times N}$ represents the modelling errors and additive noise. 
In this case, the problem is to estimate the abundance matrix $\widehat{\mathbf{X}}$ that best represents each pixel in $\mathbf{Y}$ as a linear combination of a small subset of the signatures in the library $\mathbf{A}$.
Using image segmentation and superpixel techniques \citep{MacQueenJamesandothers1967, Veganzones2014, achanta2012slic}, MUA divides the sparse unmixing optimisation problem into two spatial domains or scales: one with the actual image ($\mathcal{D}$) and another with its coarse ($\mathcal{C}$) representation composed by the average of the pixels in each superpixel. Prior to solving the optimisation issue at the original scale, spectral unmixing is carried out in the coarse domain to produce initial abundance estimates $\widehat{\mathbf{X}}_\mathcal{C}$, which are then utilised to regularise the optimisation problem to be solved in the original scale, resulting in the final estimated abundance matrix $\widehat{\mathbf{X}}$. MUA uses a spatial transformation represented by $\mathbf{W}$, which is defined analogously to $\mathbf{W}_r$ in Section~\ref{sec:proposed_method}. In~\cite{Borsoi2019}, this transformation was constructed from a modification of the BPT \citep{Veganzones2014}, $k$-means \citep{MacQueenJamesandothers1967} or SLIC \citep{achanta2012slic} algorithms.

\subsection{Experimental setup}

We substituted the spatial transformation operator $\mathbf{W}$ of the original MUA with the $\mathbf{W}_F$ operator resulting from the homogeneity-based hierarchical method proposed in Section~\ref{sec:proposed_method}, obtaining the $\text{MUA}_{\text{H$^2$BO}}$ algorithm.
We compared the proposed $\text{MUA}_{\text{H$^2$BO}}$ to $\text{MUA}_\text{SLIC}$, since the segmentation generated by the SLIC algorithm led to the best experimental results of the original MUA \citep{Borsoi2019}. \cred{As an alternative to SLIC, the waterpixels \citep{machairas2015waterpixels} algorithm (referred to in this paper with the subscript ``WPX'') for generating superpixels was also used with MUA in the experiments, denoted by $\text{MUA}_\text{WPX}$.} We used also the results from the S$^2$WSU algorithm~\citep{Zhang2018} as a baseline. We evaluated the algorithms in terms of abundance estimation quality and computational complexity using both synthetic and real hyperspectral datasets. Note that the choice of the $\text{MUA}_\text{SLIC}$ and S$^2$WSU algorithms is also
justified by the fact that they have already shown better performances compared to others of the same class, e.g., SUnSAL \citep{Iordache2010}, SUnSAL-TV \citep{Iordache2012}, DRSU \citep{Wang2016} and DRSU-TV \citep{Wang2017a}. 

In the simulations, the SLIC oversegmentation was implemented in the \emph{VLFeat} toolbox \citep{Vedaldi2008}, which permits its use for multichannel data. The SLIC method uses two parameters: $\sigma$, which defines the average size of the superpixels ($\sqrt{N/K}$), and $\gamma$, which establishes the importance of the spatial component in the pixel measure of similarity \citep{achanta2012slic}. 
\cred{For the waterpixels technique, the two relevant parameters are $\sigma$, which denotes an initial grid step that in turn determines the number of superpixels; and the spatial regulariser $\kappa$, which operates in a similar way to SLIC's $\gamma$ parameter. The watepixels algorithm also requires the computation of the morphological gradient of the image, which we computed by averaging the morphological gradient at each band of the HI as described in~\citep{noyel2020morphological}.}

The algorithms were conducted in \emph{MATLAB$^{\text{\tiny TM}}$}, on a workstation equipped with a \emph{Intel Core i7 3537U @ 2.00GHz} processor and 8GB RAM. \credd{The codes developed for the proposed \text{H$^2$BO} technique are  available at \url{https://github.com/lucayress/H2BO}.}

\subsection{Simulation results with synthetic data}
\label{sec:simuls_unmix_synth}

\begin{figure}[!htb]
\centering
\subfigure[DC1]{\label{fig:DC1_RGB}\includegraphics[width=11em]{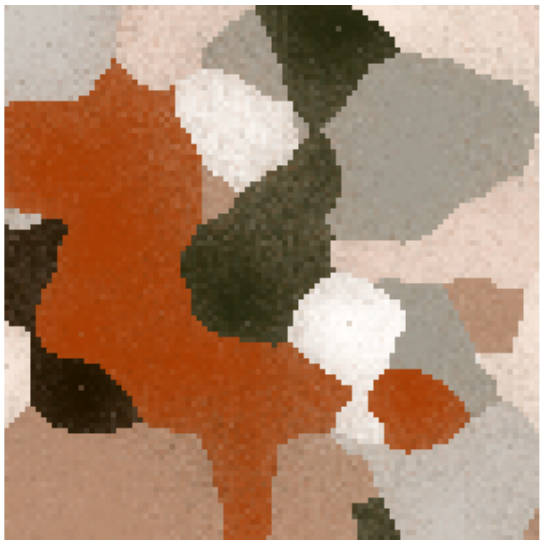}}
\subfigure[DC2]{\label{fig:DC2_RGB}\includegraphics[width=11em]{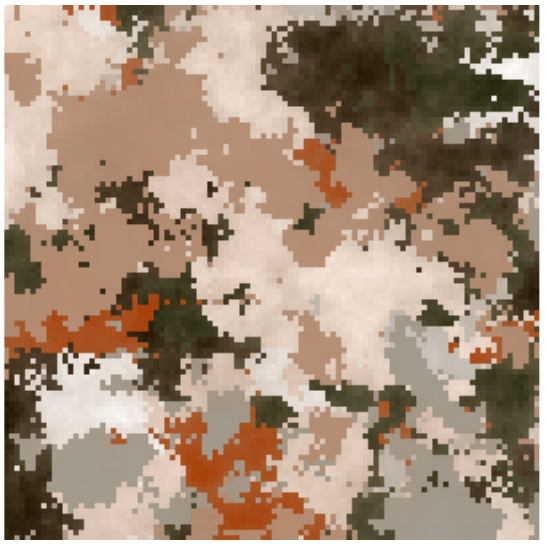}}
\subfigure[DC3]{\label{fig:DC3_RGB}\includegraphics[width=11em]{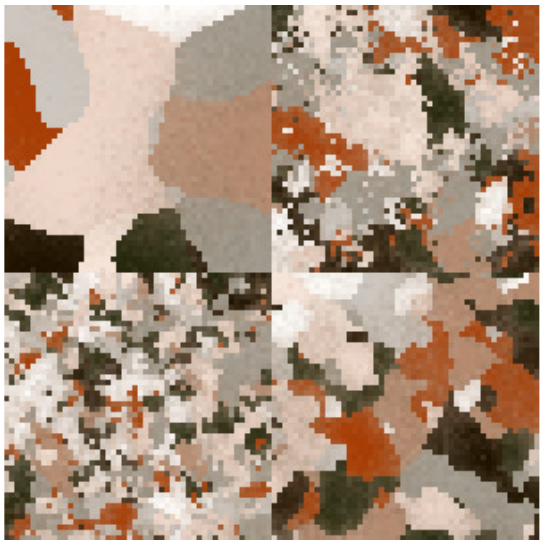}}
\caption{RGB representation of the synthetic HIs.}
\label{fig:DCs_synth}
\end{figure}

Three synthetic HIs (DC1, DC2 and DC3), represented in Figure \ref{fig:DCs_synth} with $100 \times 100$ pixels, were created using nine endmembers chosen from a library $\mathbf{A} \in \mathbb{R}^{224 \times 240}$ formed from a subset of 240 endmember signatures from the USGS \emph{splib06}\footnote{Available online at https://www.usgs.gov/labs/spec-lab/capabilities/spectral-library.} library. This allowed the results to be compared to a known reference (ground-truth). The \emph{Hyperspectral Imagery Synthesis}\footnote{Available online at \url{http://www.ehu.eus/ccwintco.}} tool was used to produce several spatially correlated abundance patterns for the data cubes in order to evaluate the methodologies in various scenarios. DC1 consists of medium and large, uniform areas; DC2 consists of regions with irregular shapes and sizes; and DC3 is a combination of four images, each $50 \times 50$ pixels, with various arrangements. 
In order to get SNRs of 20 and 30 dB, white Gaussian noise was incorporated to the generated images ($\text{SNR} = 10\log_{10}(\mathbb{E}\{\|\mathbf{AX}\|^2_F\}/\mathbb{E}\{\|\mathbf{N}\|^2_F\}$)).  
As a quantitative criterion of the unmixing performance, we used the signal-to-reconstruction error \citep{Iordache2012}, which assesses the estimation of the abundances, 
\begin{equation}\label{eq:SRE}
    \text{SRE (dB)} = 10\log_{10}\left( \frac{\|\mathbf{X}\|^2_F}{\|\mathbf{X}-\widehat{\mathbf{X}}\|^2_F} \right).
\end{equation}

All results are based on optimal parameter values obtained for each HI and shown in Table~\ref{tab:parameters_HMUA}. A grid search was carried out in the following ranges to determine them: Multiscale representation -- rounds $R=3$, regulariser $\gamma \in \{0.00025,0.00125,\ldots,0.1\}$, superpixels size $\sigma_0, \sigma_1, \sigma_2, \sigma_3, \in \{5,6,\ldots,14\}$, considering  $\sigma_0 > \sigma_1 > \sigma_2 > \sigma_3$ and thresholds $\tau_{\text{outliers}} \in \{10\%,20\%,30\%\}$, $\tau_{\text{homog}} \in \{10\%,20\%,\ldots,60\%\}$; \cred{Waterpixels representation -- $\sigma \in \{6,\ldots,20\}$ and $\kappa \in \{0.0001,0.001\ldots,1\}$;}
Sparse unmixing -- regularisation parameters of MUA and S$^2$WSU, $\lambda_{\mathcal{C}}$, $\lambda$ and $\lambda_{\text{S$^2$WSU}}$, were varied according to the values $1,3,5,7,9 \times 10^i$, for $i \in \{-3,-2,-1,0\}$ and $\beta$ in $1,3,5 \times 10^j$, for $j \in \{-1,0,1,2\}$. In all cases removing only $10\%$ of the highest values from $\mathbf{d}_k$ in \eqref{eq:sppx_dist_px} ($\tau_{\text{outliers}}=0.1$) enough to minimise outliers in homogeneity tests. Moreover, Table~\ref{tab:comparacao_algoritmos_superpixels} shows that each scene's homogeneous superpixel percentage increased from the first ($\eta_0$) to the last ($\eta_F$) oversegmentation round, especially for DC2 and DC3.

\begin{table}[!htb] 
	\tbl{Parameters of the proposed algorithm for each \cred{synthetic data}. }
	{\begin{tabular}{cccc|cc|cc}
		\toprule
		\multirow{2}{*}{Algorithm} & \multirow{2}{*}{Parameters} & \multicolumn{2}{c}{DC1} & \multicolumn{2}{c}{DC2} &       \multicolumn{2}{c}{DC3} \\
        & & 30 dB & 20 dB & 30 dB & 20 dB & 30 dB & 20 dB \\
		\midrule
		\multirow{7}{*}{\text{H$^2$BO}} & $\gamma$ & 0.00425 & 0.00425 & 0.00025 & 0.00025 & 0.00225 & 0.00225 \\
		  & $\sigma_0$ & 8 & 12 & 6 & 7 & 7 & 8 \\
		& $\sigma_1$ & 7 & 6 & 5 & 6 & 6 & 7 \\
		& $\sigma_2$ & 3 & 3 & 4 & 4 & 4 & 4 \\
		& $\sigma_3$ & 2 & 2 & 2 & 2 & 2 & 3 \\
		& $\tau_{\text{ outliers}}$ & 10\% & 10\% & 10\% & 10\% & 10\% & 10\%\\
		& $\tau_{\text{ homog}}$ & 50\% & 20\% & 20\% & 20\% & 30\% & 20\%\\
		\midrule
		\multirow{3}{*}{MUA} & $\lambda_{\mathcal{C}}$ & 0.003 & 0.007 & 0.003 & 0.007 & 0.005 & 0.01\\
		& $\lambda$ & 0.03 & 0.1 & 0.03 & 0.1 & 0.05 & 0.1 \\
            & $\beta$ & 3 & 10 & 3 & 3 & 1 & 1 \\
            \midrule
            \multirow{5}{*}{\cred{$\text{MUA}_{\text{WPX}}$}} & \cred{$k$} & \cred{0.01} & \cred{0.01} & \cred{0.1} & \cred{0.1} & \cred{0.1} & \cred{0.1} \\
            & \cred{$\sigma$} & \cred{10} & \cred{10} & \cred{8} & \cred{8} & \cred{8} & \cred{8} \\
            & \cred{$\lambda_{\mathcal{C}}$} & \cred{0.003} & \cred{0.001} & \cred{0.001} & \cred{0.001} & \cred{0.003} & \cred{0.001}\\
		& \cred{$\lambda$} & \cred{0.1} & \cred{0.07} & \cred{0.1} & \cred{0.1} & \cred{0.1} & \cred{0.1} \\
            & \cred{$\beta$} & \cred{1} & \cred{1} & \cred{1} & \cred{1} & \cred{1} & \cred{1} \\
		\bottomrule
	\end{tabular}}
\label{tab:parameters_HMUA}
\end{table}

\begin{figure}[!htb]
\centering
\subfigure{\label{fig:DC3_30dB_SLIC_inicial}\includegraphics[width=10em]{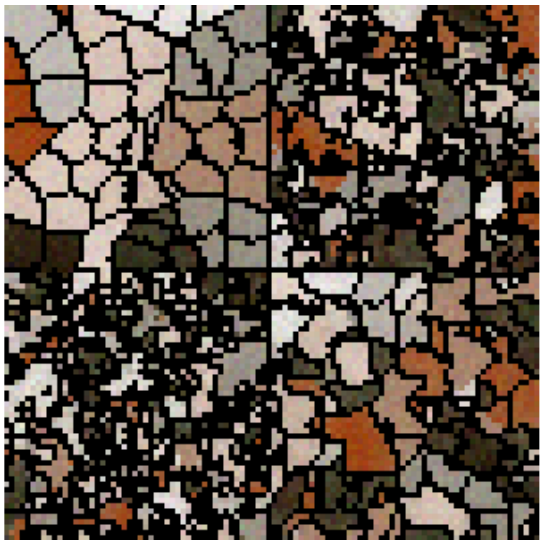}}
\subfigure{\label{fig:DC3_30dB_map_inicial}\includegraphics[width=10em]{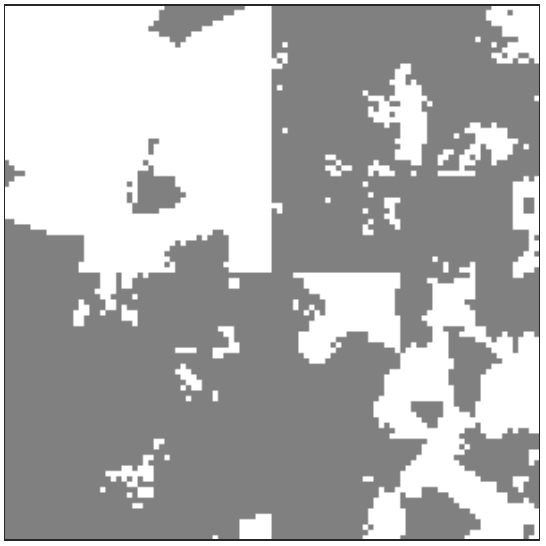}}
\subfigure{\label{fig:DC3_30dB_ratio_inicial}\includegraphics[width=10em]{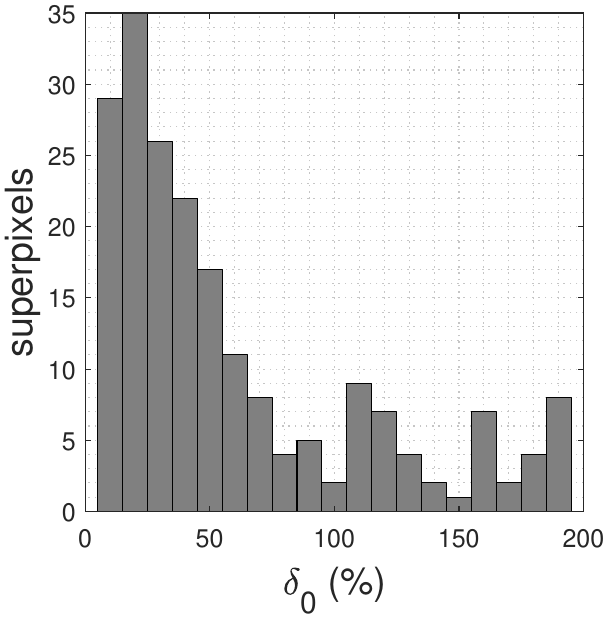}}

\subfigure[Superpixels.]{\label{fig:DC3_30dB_SLIC_final}\includegraphics[width=10em]{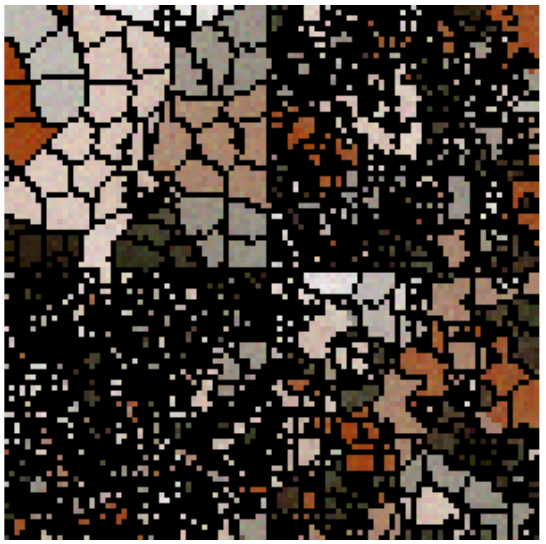}}
\subfigure[Homogeneity map.]{\label{fig:DC3_30dB_map_final}\includegraphics[width=10em]{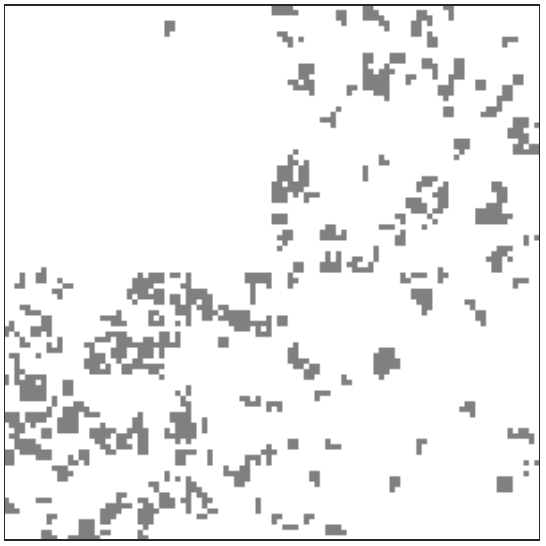}}
\subfigure[Superpixels $\delta_k$ deviation.]{\label{fig:DC3_30dB_ratio_final}\includegraphics[width=10em]{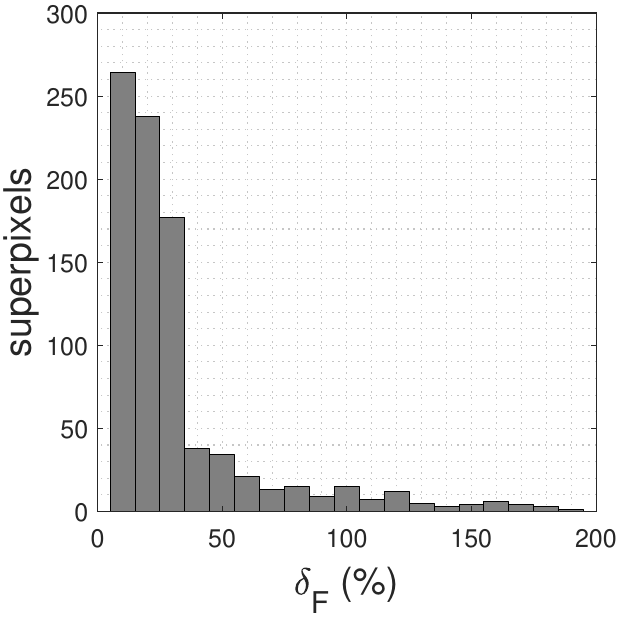}}
\caption{Comparison between initial (top row) and final (bottom row) oversegmentation and superpixels homogeneity results for H$^2$BO in DC3, SNR 30 dB.}
\label{fig:DC3_homog_30dB_results}
\end{figure}

Figure~\ref{fig:DC3_homog_30dB_results} illustrates the comparison between the H$^2$BO results for the initial and final oversegmentation scales, as well as their respective homogeneity maps and the histogram related to the superpixels $\delta_k$ deviation in DC3, 30~dB. The decrease in grey regions (non-homogeneous superpixels) on the final homogeneity map clearly shows the improvement in the overall spectral homogeneity of the HI regions produced by hierarchical oversegmentation with different superpixel sizes. The histograms explain this evolution in terms of $\delta_k$ values.
In order to get a high $\eta_F$ index and a suitable unmixing outcome without compromising the method's reduced computational cost, just three rounds of oversegmentation ($R=3$) were required. Low values of $\eta_0$ result in longer execution durations since more superpixel homogeneity evaluations, subdivisions, and unmixing are necessary.

\begin{figure}[t!]
\centerline{\includegraphics[width=34em]{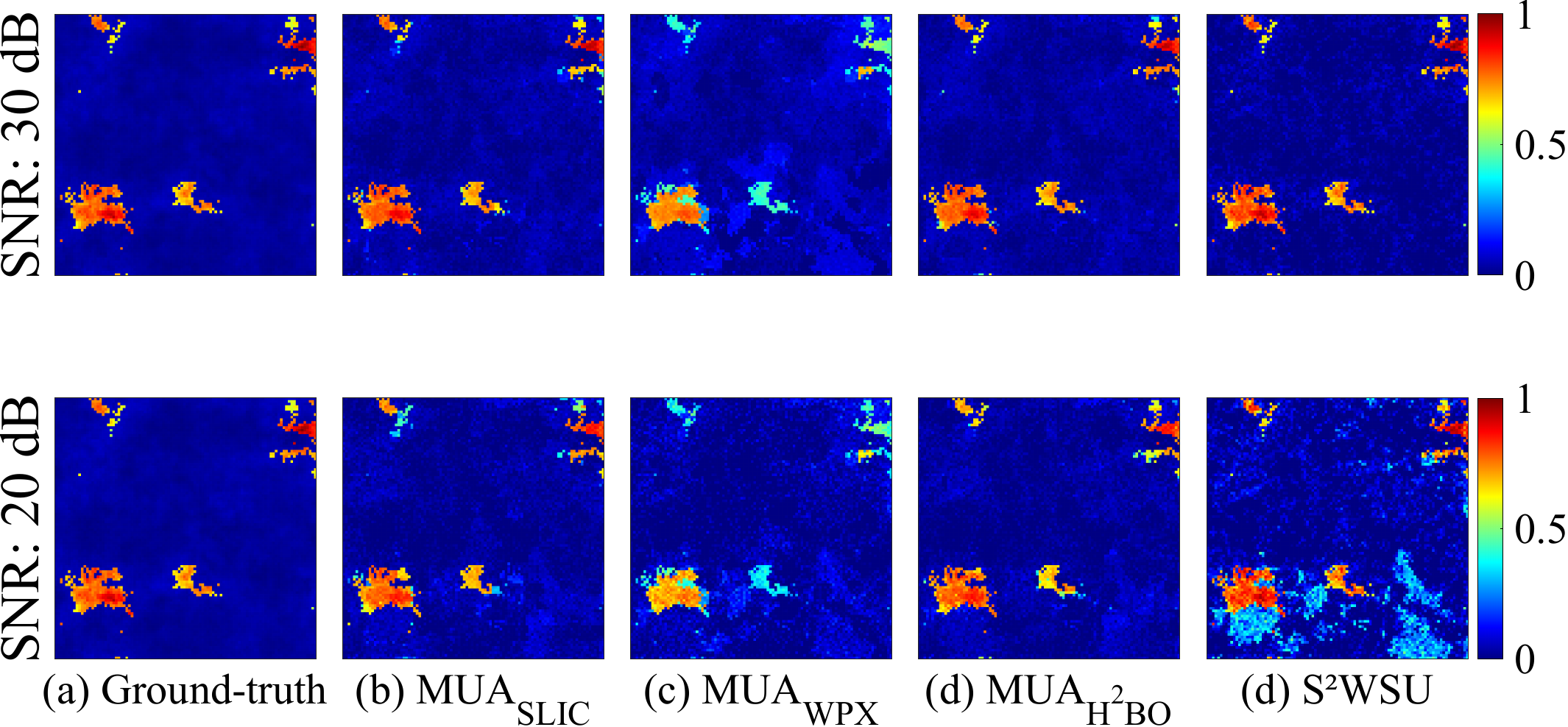}}
\caption{\cred{Abundance estimation results for endmember 3 of DC2.}}
\label{fig:DC2_comaparacao}
\end{figure}

The amount of superpixels used with the original $\text{MUA}_\text{SLIC}$, \cred{$\text{MUA}_\text{WPX}$} and the $\text{MUA}_\text{H$^2$BO}$ are displayed in Table~\ref{tab:comparacao_algoritmos_superpixels}. In more uniform circumstances, like DC1, the proposed technique was effective at lowering the required number of superpixels. Also, the presented approach's final oversegmentation resulted in images that contained a significantly higher percentage of homogeneous superpixels.
\cred{Although $\text{MUA}_\text{WPX}$ generated the smallest number of superpixels, it was the method that showed the lowest homogeneity percentages, according to the criteria proposed in this work. However, we emphasize that the parameters of the waterpixels algorithm were adjusted maximize the unmixing performance, and not the homogeneity of its superpixels.}
The SRE performance of the algorithms is displayed in Table~\ref{tab:comparacao_algoritmos_SRE} in comparison to the optimal value. In a noisy environment, $\text{MUA}_\text{H$^2$BO}$ produced the best quantitative results (SNR 20 dB). When compared to $\text{MUA}_\text{SLIC}$, DC3 showed a considerable improvement, with an approximately 5\% boost in SRE for conditions with 20 and 30 dB SNR.
These results suggest that $\text{MUA}_\text{H$^2$BO}$ tends to work better when the spatial content of abundances varies throughout the scene.
\cred{The $\text{MUA}_\text{WPX}$ showed the lowest SRE results, particularly for DC3, which illustrates its limitation in segmenting regions with heterogeneous spatial sizes.} %
Notwithstanding the additional stages of homogeneity assessment and further oversegmentations, the execution time of the suggested $\text{MUA}_\text{H$^2$BO}$, as shown in Table~\ref{tab:comparacao_algoritmos_tempo}, was comparable to that of $\text{MUA}_\text{SLIC}$ and much slower than that of S$^2$WSU. \cred{The execution \cblue{times of $\text{MUA}_\text{SLIC}$ and $\text{MUA}_\text{WPX}$ were} virtually the same.}
The maps of endmember 3 of DC2's true and reconstructed abundances are shown in Figure~\ref{fig:DC2_comaparacao}.
With a high SNR (30~dB), S$^2$WSU's abundance estimates are visually the most accurate, but $\text{MUA}_\text{H$^2$BO}$ and $\text{MUA}_\text{SLIC}$'s results were more comparable. The outcomes by $\text{MUA}_\text{H$^2$BO}$ indicate a clear improvement over $\text{MUA}_\text{SLIC}$, which illustrates its effectiveness for noisier images because of its robust homogeneity assessment. Nevertheless, for an SNR of 20~dB, the performance of S$^2$WSU dramatically declines.

\begin{table}[t!]
      \tbl{Number of generated superpixels.}{
        \begin{tabular}{cccc|cc|cc} 
            \toprule
            \multirow{2}{*}{Data} & \multirow{2}{*}{SNR} & \multicolumn{2}{c|}{$\text{MUA}_\text{SLIC}$} & \multicolumn{2}{c|}{\cred{$\text{MUA}_{\text{WPX}}$}} & \multicolumn{2}{c}{$\text{MUA}_\text{H$^2$BO}$}    \\
            & & \emph{superpixels} & $\eta\equiv\eta_0$ & \cred{\emph{superpixels}} & \cred{$\eta\equiv\eta_F$} & \emph{superpixels} & $\eta\equiv\eta_F$ \\
            \midrule
            \multirow{2}{*}{DC1} & 30 dB & 225 & 89\% & \cred{42} & \cred{14\%} & \textbf{218} & \textbf{99\%} \\
                                  & 20 dB & 169 & 97\% & \cred{42} & \cred{76\%} & \textbf{218} & \textbf{99\%}  \\
            \midrule
            \multirow{2}{*}{DC2} & 30 dB & \textbf{625} & 69\% & \cred{60} & \cred{18\%} & 1018 & \textbf{90\%}  \\
                                  & 20 dB & \textbf{289} & 81\% & \cred{60}  & \cred{8\%} & 444  & \textbf{94\%} \\
            \midrule
            \multirow{2}{*}{DC3} & 30 dB & \textbf{624} & 55\% & \cred{60}  & \cred{13\%} & 1566  & \textbf{84\%}\\
                                  & 20 dB & 625 & 80\% & \cred{60}  & \cred{36\%} & \textbf{610}  & \textbf{84\%} \\
            \bottomrule
        \end{tabular}}
\label{tab:comparacao_algoritmos_superpixels}
\end{table}

\begin{table}[t!]
      \tbl{SRE results.}{
        \begin{tabular}{cccccc} 
            \toprule
            Data & SNR & S$^2$WSU & $\text{MUA}_\text{SLIC}$ & \cred{$\text{MUA}_{\text{WPX}}$} & $\text{MUA}_\text{H$^2$BO}$   \\
            \midrule
            \multirow{2}{*}{DC1} & 30 dB &  \textbf{21.668 dB} & 18.117 dB & \cred{14.41 dB} & 18.339 dB \\ %
                                  & 20 dB &   9.332 dB & 14.854 dB & \cred{11.93 dB} & \textbf{15.104 dB} \\  %
            \midrule
            \multirow{2}{*}{DC2} & 30 dB & \textbf{18.741 dB} & 11.737 dB & \cred{6.18 dB} & 11.780 dB \\
                                  & 20 dB &  5.689 dB &  8.416 dB &  \cred{5.30 dB} &  \textbf{8.561 dB} \\
            \midrule
            \multirow{2}{*}{DC3} & 30 dB & \textbf{19.798 dB} & 10.841 dB & \cred{5.15 dB} & 11.398 dB \\
                                  & 20 dB &  6.899 dB &  7.776 dB & \cred{4.36 dB} &  \textbf{8.185 dB}\\
            \bottomrule
        \end{tabular}}
        \label{tab:comparacao_algoritmos_SRE}
\end{table}

\begin{table}[t!]
\tbl{Average execution times.}
        {\begin{tabular}{cccccc} 
        \toprule
            Data & SNR & S$^2$WSU & $\text{MUA}_\text{SLIC}$ & \cred{$\text{MUA}_{\text{WPX}}$} & $\text{MUA}_\text{H$^2$BO}$     \\
            \midrule
            \multirow{2}{*}{DC1} & 30 dB & 239 s & \textbf{9  s} & \cred{9 s} & 11 s \\
                                  & 20 dB & 235 s & \textbf{15 s} & \cred{15 s} & 16 s \\
            \midrule
            \multirow{2}{*}{DC2} & 30 dB & 233 s & \textbf{10 s} & \cred{10 s} & 16 s \\
                                  & 20 dB & 232 s & \textbf{10 s} & \cred{10 s} & 16 s \\
            \midrule
            \multirow{2}{*}{DC3} & 30 dB & 233 s & \textbf{12 s} & \cred{12 s} & 21 s \\
                                  & 20 dB & 232 s & \textbf{12 s} & \cred{12 s} & 15 s \\
            \bottomrule
        \end{tabular}}
        \label{tab:comparacao_algoritmos_tempo}
\end{table}

\begin{figure}[!htb]
\centering
\subfigure{\label{fig:DCs_sensitivity_gamma}\includegraphics[width=10em]{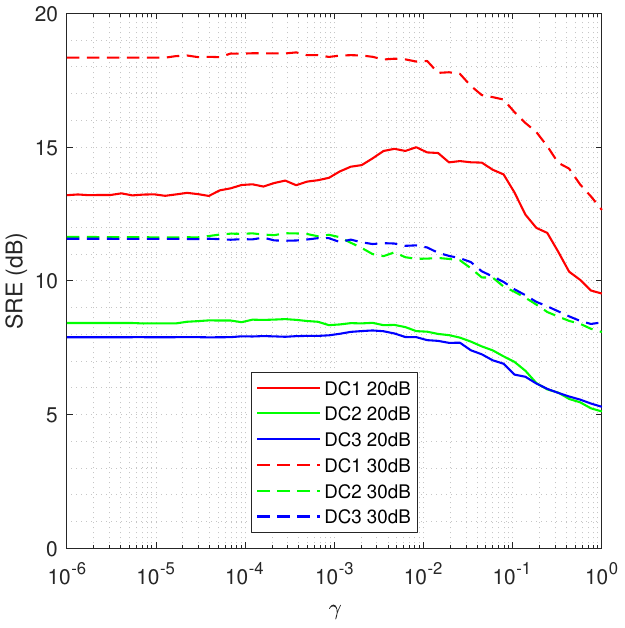}}
\subfigure{\label{fig:DCs_sensitivity_sigma_SRE}\includegraphics[width=10em]{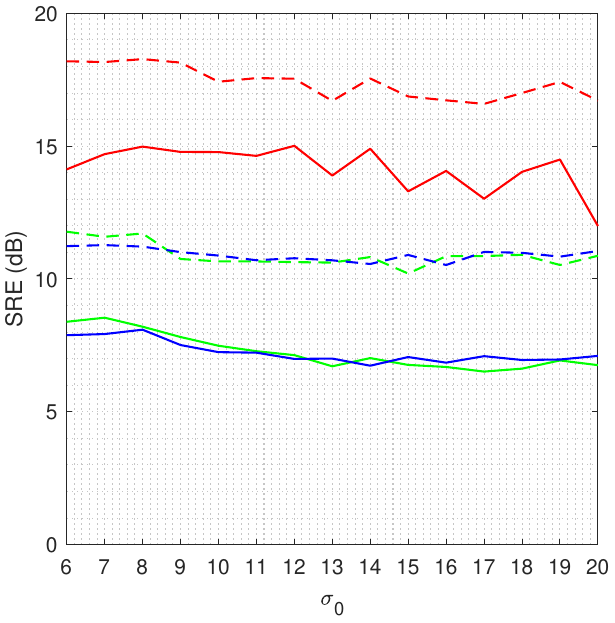}}

\subfigure{\label{fig:DCs_sensitivity_tau_outliers}\includegraphics[width=10em]{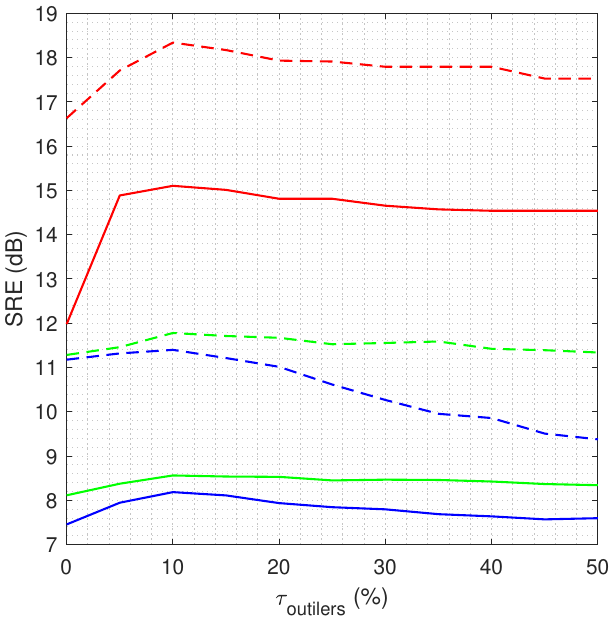}}
\subfigure{\label{fig:DCs_sensitivity_tau_homog}\includegraphics[width=10em]{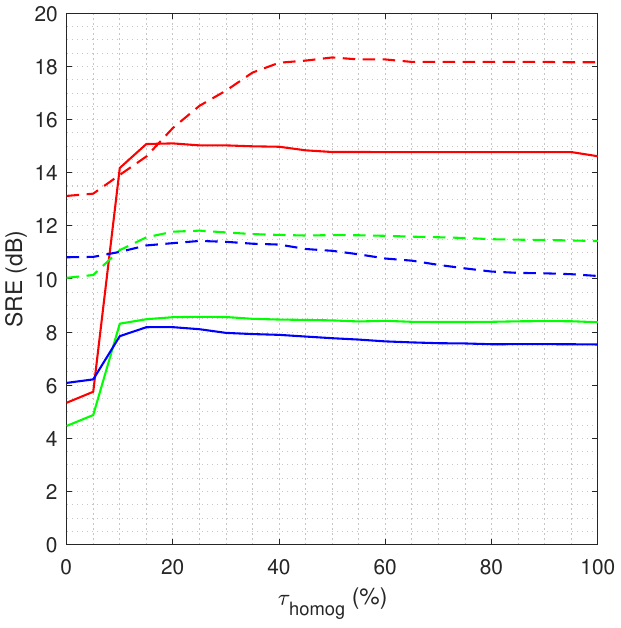}}
\caption{Variation of the SRE results due to changes in each parameter individually.}
\label{fig:DCs_sensitivity}
\end{figure}

\textit{\textbf{Sensitivity analysis}} -- Now, in the unmixing of the DC1, DC2 and DC3 data, under the 20~dB and 30~dB SNR scenarios, the sensitivity of the SRE result to the adjustment of the H$^2$BO input parameters is analysed.
The plots in Figure \ref{fig:DCs_sensitivity} illustrate the change of the method's SRE as a function of a wide range of values for each individual parameter with the other parameters set to their ideal values (Table \ref{tab:parameters_HMUA}).
Since they follow a nearly uniform trend and are constrained to the order $\sigma_1 > \sigma_2 > \sigma_3$, the optimal values of $\sigma_1$, $\sigma_2$ and $\sigma_3$ were left unchanged.
As can be seen, for a stable range of $\gamma$ and $\sigma_0$, the SRE value essentially remains steady. If $\gamma>0.1$, the reductions are the greatest. This results from the disproportionate preference for spatial regularity over spectral regularity during superpixel generation. H$^2$BO makes up for variations in $\sigma_0$ with additional rounds of oversegmentation using smaller superpixel sizes. The improved SRE produced by employing $\tau_{\text{outliers}}=10\%$ as opposed to $\tau_{\text{outliers}}=0\%$, with differences varied between 0.5 dB and 3 dB, shows how the outliers removal technique in the homogeneity assessment phase improves the quality of spectral unmixing.

\begin{figure}[!htb]
\centering
\subfigure{\label{fig:IMG1_30dB_statistical_hist}\includegraphics[width=11em]{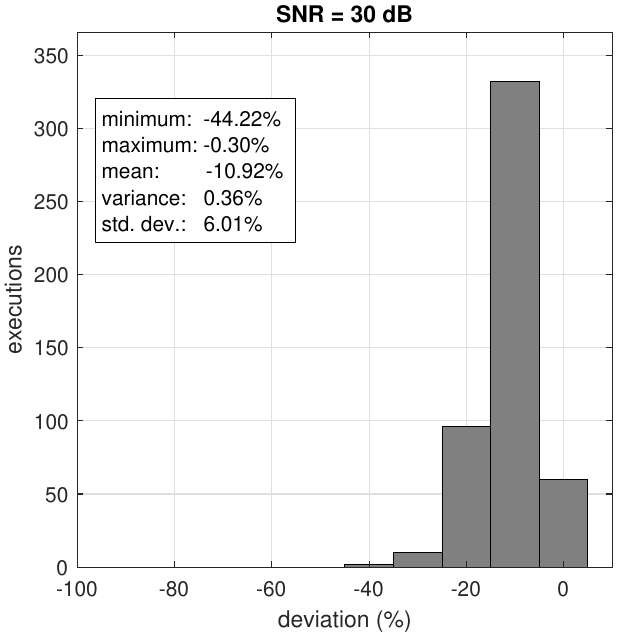}}
\subfigure{\label{fig:IMG2_30dB_statistical_hist}\includegraphics[width=11em]{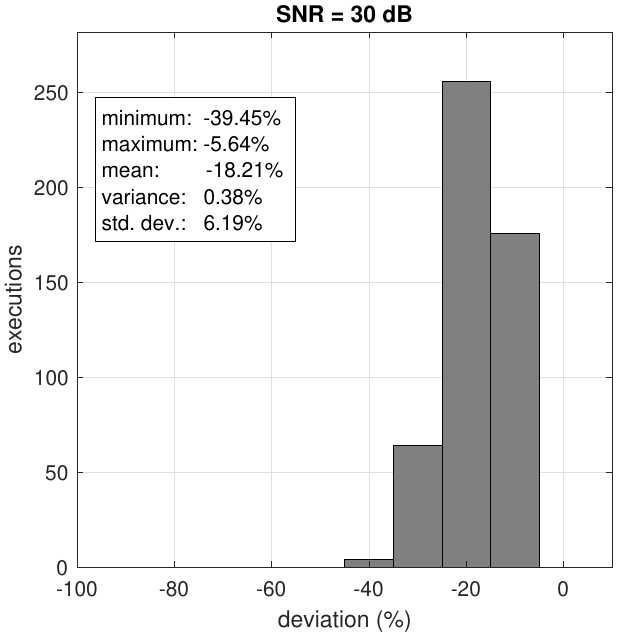}}
\subfigure{\label{fig:IMG3_30dB_statistical_hist}\includegraphics[width=11em]{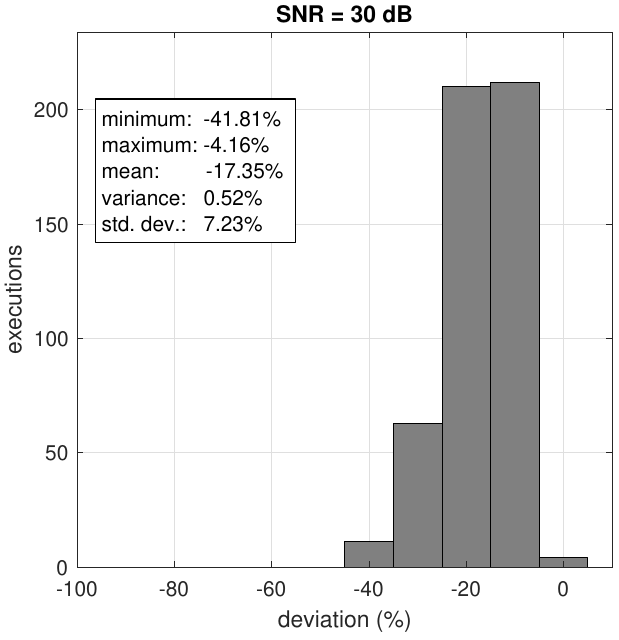}}

\subfigure{\label{fig:IMG1_20dB_statistical_hist}\includegraphics[width=11em]{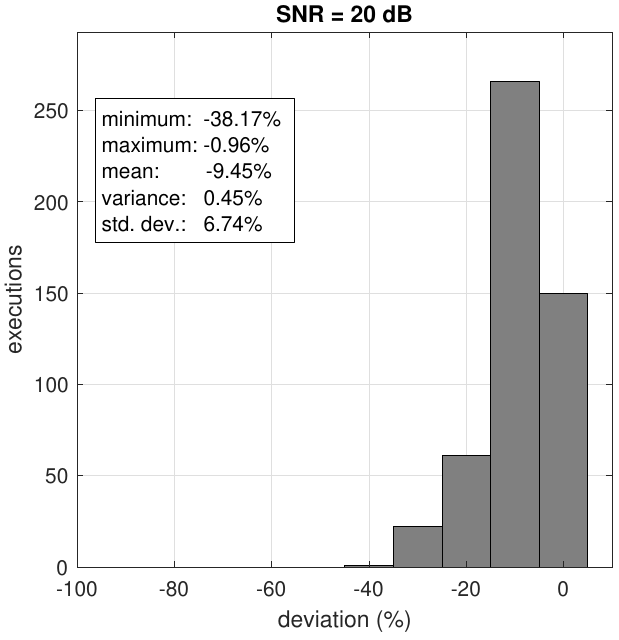}}
\subfigure{\label{fig:IMG2_20dB_statistical_hist}\includegraphics[width=11em]{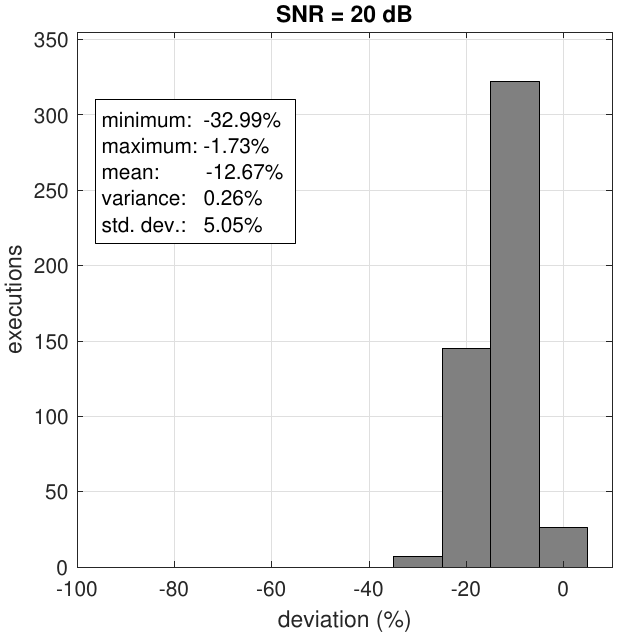}}
\subfigure{\label{fig:IMG3_20dB_statistical_hist}\includegraphics[width=11em]{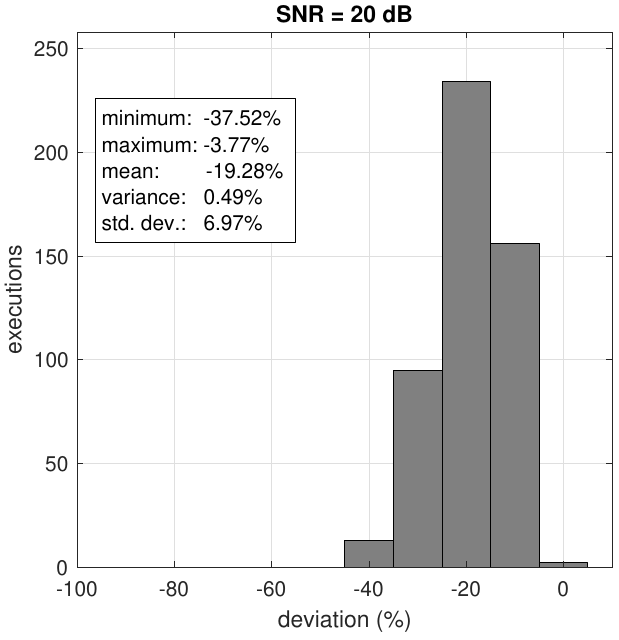}}

\caption{\cred{Deviation from the optimal SRE value of the proposed method for randomly chosen parameter values.}}
\label{fig:IMGs_statistical_hist}
\end{figure}

\begin{color}{black}
    \textit{\textbf{Statistical performance analysis}} -- As the $\text{MUA}_\text{H$^2$BO}$ has several parameters to be set, we evaluated its performance when the parameters are misspecified. Specifically, we evaluated how much the average performance of $\text{MUA}_\text{H$^2$BO}$ decreased when compared to its optimal value when the parameters were selected randomly within some intervals that were empirically selected.
    $\text{MUA}_\text{H$^2$BO}$ was executed 500 times for each image. Through a uniform distribution, in all executions a new random value for each parameter was chosen according to the following intervals: regulariser $\gamma \in [0,001,\ 0,02]$, superpixels size $\sigma_0 \in [5,\ 20]$, e thresholds $\tau_{outliers} \in [10\%,\ 20\%]$ and $\tau_{homog} \in [10\%,\ 50\%]$ -- regularisers $\lambda_C \in [0,001,\ 0,009]$, $\lambda \in [0,01,\ 0,9]$, $\beta \in [1,\ 50]$. For $\sigma_1$, $\sigma_2$ and $\sigma_3$ the following relationship was used: $\sigma_i \in [\frac{\sigma_{i-1}}{2},\ \sigma_{i-1}-1]$, rounded up to the upper integer and $\sigma_2$ and $\sigma_3$ limited to 3 and 2, respectively. 
After spectral unmixing, the deviation between the SRE value obtained in each execution and the optimal SRE value in Table \ref{tab:comparacao_algoritmos_SRE} was calculated as:

\begin{equation}
    \frac{\text{SRE}-\text{SRE}_{\text{optimal}}}{\text{SRE}_{\text{optimal}}} \times 100\% \,.
\end{equation}

Figure \ref{fig:IMGs_statistical_hist} shows the result of the variations. Through these histograms, it is possible to perceive an average decrease in performance between 10\% and 20\% and a low standard deviation, around 6\%. This shows that the probability of obtaining good spectral unmixing results is high for parameter values randomly chosen within a reasonable range.
\end{color}

\subsection{Experiments with real hyperspectral data}

In this section, we compare the performance of the $\text{MUA}_\text{H$^2$BO}$ technique to the $\text{MUA}_\text{SLIC}$\cred{, $\text{MUA}_\text{WPX}$} and S$^2$WSU algorithms using well-known real hyperspectral\footnote{Available online at http://lesun.weebly.com/hyperspectral-data-set.html.} data, specifically the Samson and Jasper Ridge HIs.
Two subimages were taken from these scenarios in to reduce the simulations' computation times and facilitate the evaluation of the results.
Samson's subimage is a 40 $\times$ 95 pixels region with 156 bands between 401 and 889 nm, and it is made up of three endmembers: soil, trees, and water. The region of interest for Jasper Ridge has 50 $\times$ 50 pixels, 198 bands between 380 and 2500 nm, and four key signatures: road, soil, water, and tree. The hyperspectral libraries utilised in this study are the same ones used by \cite{Borsoi2019}. These were created using an extraction method that extracts $\mathbf{A}$ directly from the HI \citep{Somers2012}, yielding $\mathbf{A} \in \mathbb{R}^{156 \times 105}$ for the Samson and $\mathbf{A} \in \mathbb{R}^{198 \times 529}$ for the Jasper Ridge subimages.
The additional parameters for each algorithm (included in Table~\ref{tab:parameters_real}) were determined based on a visual comparison between their estimated abundance maps and the RGB representation of the data. For H$^2$BO, $R=3$ was used.

\begin{table}[!htb] 
        \tbl{Parameters of the algorithms for the real HIs.}
	{\begin{tabular}{cccc}
	    \toprule
		Algorithm & Parameters & Samson & Jasper Ridge \\
		\midrule
		\multicolumn{4}{c}{$\text{MUA}_\text{H$^2$BO}$} \\
		\midrule
		\multirow{5}{*}{$\text{H$^2$BO}$} & $\gamma$ & 0.00125 & 0.00125 \\
		& $\sigma_0$ & 15 & 15 \\
		& $\sigma_1$ & 7 & 8 \\
		& $\tau_{\text{ outliers}}$ & 10\% & 10\% \\
		& $\tau_{\text{ homog}}$ & 120\% & 100\% \\
		\midrule
		\multirow{3}{*}{MUA} & $\lambda_{\mathcal{C}}$ & 0.1 & 0.003 \\
		& $\lambda$ & 0.01 & 0.03 \\
		& $\beta$ & 1 & 3 \\
		\midrule
		\multicolumn{4}{c}{$\text{MUA}_\text{SLIC}$} \\
		\midrule
		\multirow{2}{*}{SLIC} & $\gamma$ & 0.00125 & 0.00125 \\
		& $\sigma$ & 7 & 7 \\
	    \midrule
		\multirow{3}{*}{MUA} & $\lambda_{\mathcal{C}}$ & 0.1 & 0.003 \\
		& $\lambda$ & 0.01 & 0.03 \\
		& $\beta$ & 1 & 3 \\
		\midrule
            \multicolumn{4}{c}{$\text{\cred{MUA}}_{\cred{\text{WPX}}}$} \\
		\midrule
		\multirow{2}{*}{\cred{\text{WPX}}} & $\cred{k}$ & \cred{0.01} & \cred{0.01} \\
		& $\cred{\sigma}$ & \cred{8} & \cred{6} \\
	    \midrule
		\multirow{3}{*}{\cred{MUA}} & $\cred{\lambda_{\mathcal{C}}}$ & \cred{0.1} & \cred{0.003} \\
		& $\cred{\lambda}$ & \cred{0.01} & \cred{0.03} \\
		& $\cred{\beta}$ & \cred{1} & \cred{3} \\
		\midrule
		\multicolumn{4}{c}{S$^2$WSU} \\
		\midrule
		S$^2$WSU & $\lambda_{\text{swsp}}$ & 0.002 & 0.01 \\
		\bottomrule
	\end{tabular}}
\label{tab:parameters_real}
\end{table}

\begin{figure}[!htb]
	\centering
        \subfigure[]{\includegraphics[width=6em]{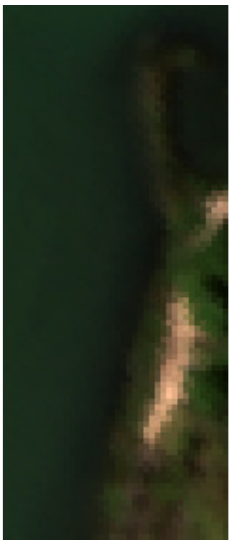}}
        \subfigure[]{\includegraphics[width=6em]{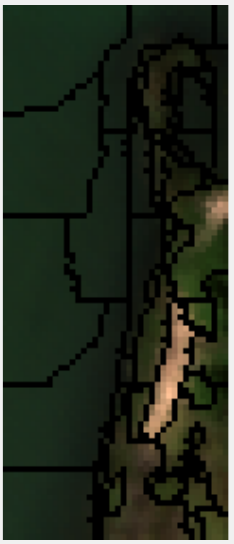}}
        \subfigure[]{\includegraphics[width=6em]{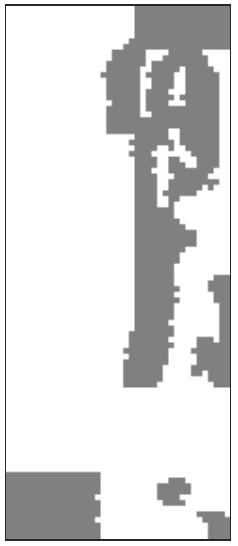}}
        \subfigure[]{\includegraphics[width=6em]{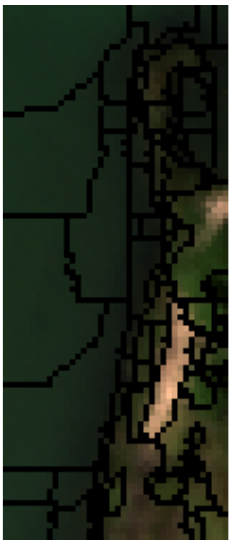}}	
	\caption{(a) Samson subimage RGB representation. (b) Initial oversegmentation, $\sigma_0 = 15$. (c) Initial homogeneity map (non-homogeneous regions in grey colour). (d) Result of the final oversegmentation from the subdivision of the non-homogeneous superpixels in (c), $\sigma_1 = 7$.}
	\label{fig:samson_HMUA}
\end{figure}

\begin{figure}[!htb]
	\centering
        \subfigure[]{\includegraphics[width=7.5em]{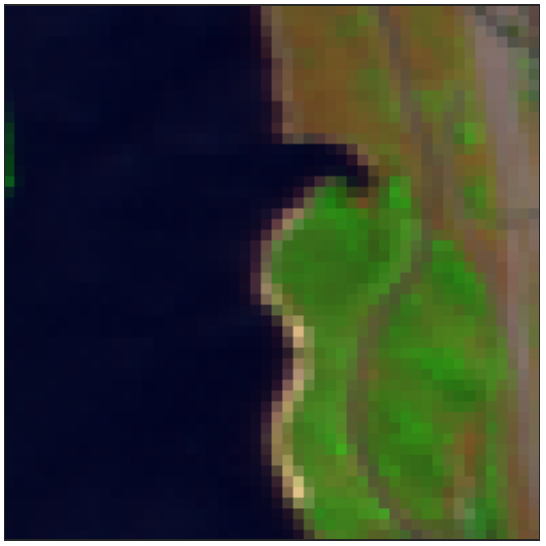}}
        \subfigure[]{\includegraphics[width=7.5em]{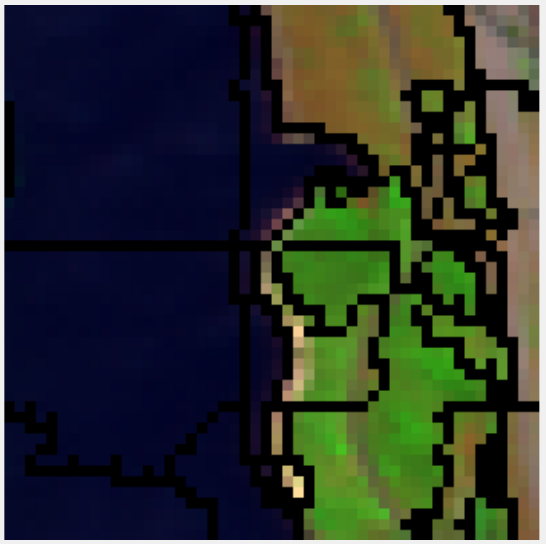}}
        \subfigure[]{\includegraphics[width=7.5em]{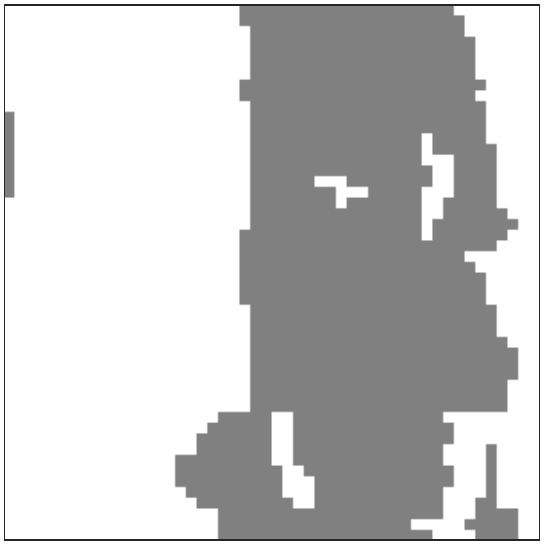}}
        \subfigure[]{\includegraphics[width=7.5em]{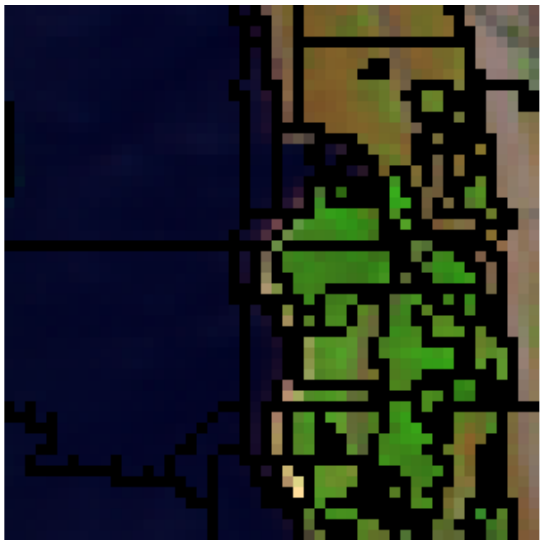}}
	\caption{(a) Jasper Ridge subimage RGB representation. (b) Initial oversegmentation, $\sigma_0 = 15$. (c) Initial homogeneity map (non-homogeneous regions in grey colour). (d) Result of the final oversegmentation from the subdivision of the non-homogeneous superpixels in (c), $\sigma_1 = 8$.}
	\label{fig:jasper_HMUA}
\end{figure}

The oversegmentation results of the Samson and Jasper Ridge are shown in Figures \ref{fig:samson_HMUA} and \ref{fig:jasper_HMUA}, respectively. We find large superpixels in more regular locations, like water, and smaller superpixels in more irregular places, demonstrating the effectiveness of the suggested strategy. As observed in Table~\ref{tab:comparare_real}, under these circumstances, just as with the synthetic data DC1, the proposed technique in H$^2$BO enables the depiction of the image with a markedly reduced number of superpixels as compared to the number employed by $\text{MUA}_\text{SLIC}$. \cred{The $\text{MUA}_\text{WPX}$ performed well for Tree endmember, but failed significantly for Soil and Water in the Samson image. In Jasper Ridge, however, the results were very consistent with those of other algorithms.} The execution time of the $\text{MUA}_\text{H$^2$BO}$ remained comparable to that of the $\text{MUA}_\text{SLIC}$ and far less than that of the S$^2$WSU even for simulations using real data. \cred{Still, the execution time of $\text{MUA}_\text{WPX}$ and $\text{MUA}_\text{SLIC}$ remain practically the same.}

\begin{table}[!htb]
      \tbl{Average execution time (in seconds) and number of superpixels for the real HIs.}
        {\begin{tabular}{cc|cc|cc|cc} 
            \toprule
            \multirow{2}{*}{Data} & S$^2$WSU & \multicolumn{2}{c|}{$\text{MUA}_\text{SLIC}$} & \multicolumn{2}{c|}{$\text{\cred{MUA}}_{\cred{\text{WPX}}}$} & \multicolumn{2}{c}{$\text{MUA}_\text{H$^2$BO}$} \\
            & ex. time & ex. time & superpixels & \cred{ex. time} & \cred{superpixels} & ex. time & superpixels \\
            \midrule
            Samson & 9 s & \textbf{5} s & 84 & \cred{5 s} & \cred{10} & 7 s & \textbf{51}  \\
            Jasper Ridge & 19 s & \textbf{7} s & 64 & \cred{7 s} & \cred{31} & 9 s & \textbf{42} \\
            \bottomrule
        \end{tabular}}
\label{tab:comparare_real}
\end{table}

\begin{figure}[!htb]
	\centerline{\includegraphics[width=21em]{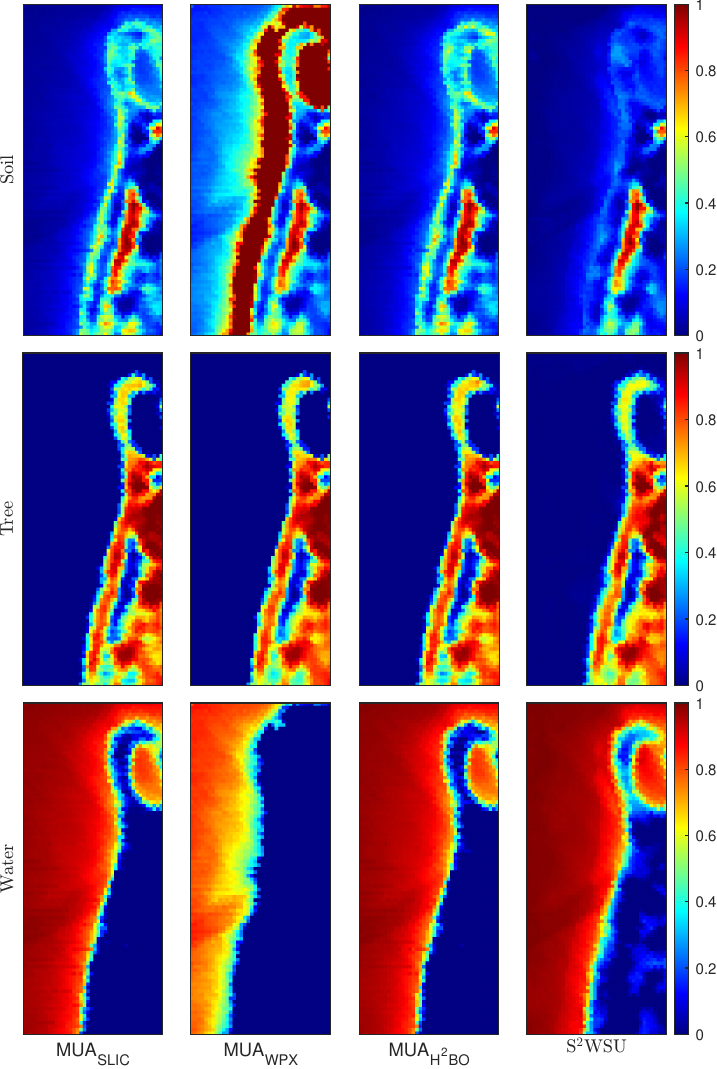}}
	\caption{\cred{Samson -- comparison between the estimated abundance maps for each endmember and algorithm.}}
	\label{fig:samson_compare}
\end{figure}

\begin{figure}[!htb]
	\centerline{\includegraphics[width=33em]{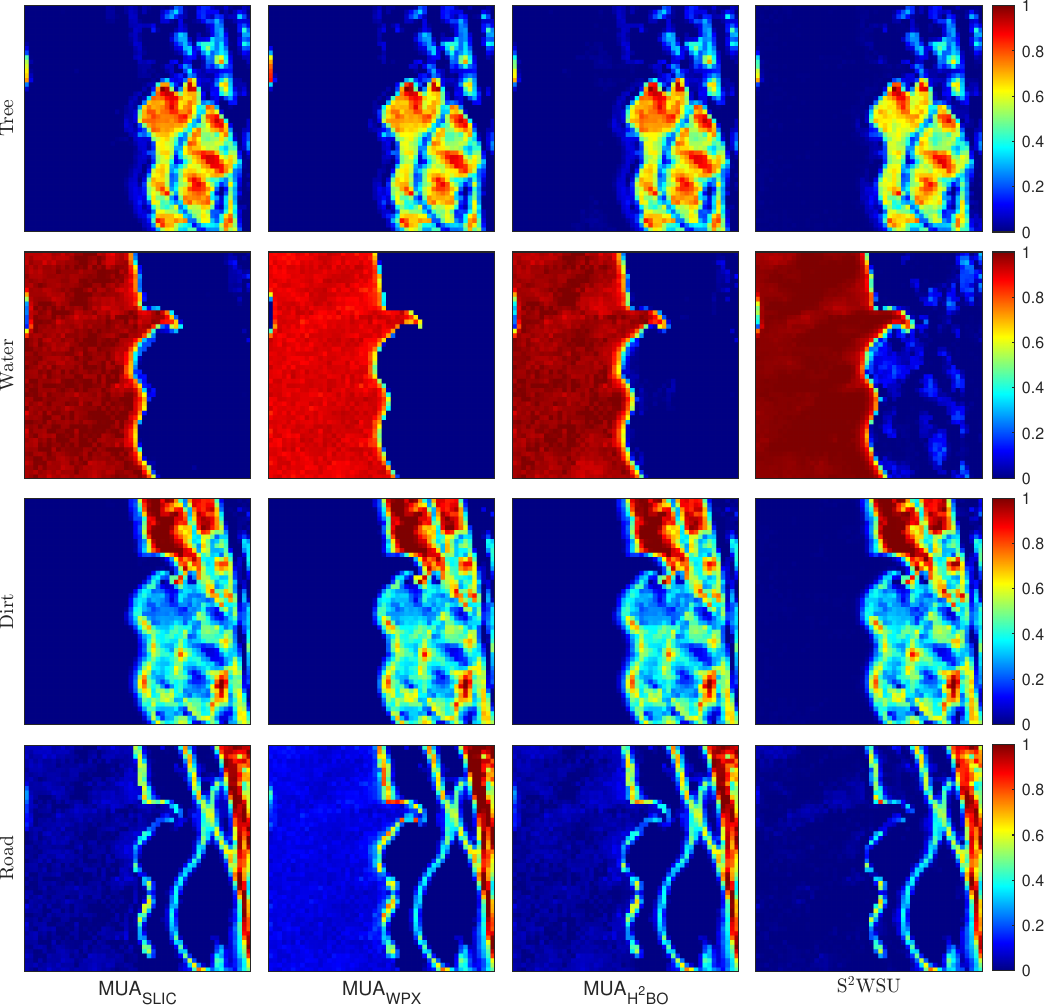}}
	\caption{\cred{Jasper Ridge -- comparison between the estimated abundance maps for each endmember and algorithm.}}
	\label{fig:jasper_compare}
\end{figure}

Due to the lack of ground-truth abundance maps for the widely used datasets made available online, the quantitative analysis of spectral unmixing methods based on tests with real hyperspectral data is limited.
As a result, only a visual examination of the abundance maps can be used to evaluate the results of abundance estimation with real data. Nonetheless, it is clear from the abundance maps calculated in Figures \ref{fig:samson_compare} and \ref{fig:jasper_compare} that the quality of the spectral unmixing is not diminished by this more straightforward representation. Although the outcomes of the algorithms are visually quite similar, if a ground-truth were available, tiny changes in the abundance maps produced by either $\text{MUA}_\text{H$^2$BO}$ and $\text{MUA}_\text{SLIC}$ may result in discrepancies in terms of quantitative metrics.

\section{Application to hyperspectral data classification} \label{sec:classification}

In this section, we illustrate the performance of the proposed H$^2$BO method when combined with hyperspectral image classification strategies that leverage spatial contextual information. More precisely, we consider the state-of-the-art semi-supervised graph convolutional neural network strategy proposed by~\cite{Liu2021}, which leverages the superpixel decomposition of the HI to construct a graph used in the neural network, which implicitly promotes pixels within a single superpixel to share the same class labels.

More precisely, \cite{Liu2021} proposed an HI classification method integrating a superpixel-based graph convolutional network (GCN) and a convolutional neural network (CNN) into a single network. The basic idea is to combine the potential of a GCN to model the various spatial structures of land covers in graphs with the ability of a CNN to learn local spectral-spatial features at the pixel level. The technique is known as CNN-enhanced GCN (CEGCN). The CEGCN algorithm performs semi-supervised classification: it learns a neural network to assign labels to each pixel of an HI based on a small set of labeled training samples using information contained in both the labeled and unlabeled data.

To assign each pixel a unique label, CEGCN is organised into the following stages: the superpixel-based graph constructor (SGC); the spectral transformation subnetwork (STsN); the superpixel-level graph subnetwork (SGsN); and the pixel-level convolutional subnetwork (PCsN).
The input HI is first transformed by an STsN, which reduces the irrelevant information of the spectra and improves the discrimination between different classes. Then, SGsN is built to model the large-scale spatial structure of the HI on graphs and produce superpixel-level features. In addition, a PCsN is employed to extract local spectral-spatial features at the pixel level. Lastly, the features extracted by SGsN and PCsN are merged to increase the classification performance.
The SGC is the step responsible for converting the structure of the hyperspectral data into a graph. The appropriate graph structure containing spectral-spatial information is built by applying the SLIC algorithm, where the similarity between each pair of superpixels is used to construct an undirected graph. Before image segmentation, a dimensionality reduction is also performed with linear discriminant analysis (LDA).
Although the use of superpixels decreases the number of nodes in a graph to be processed, at the superpixel scale, the pixels in a superpixel (node) are described by a single set of features. Hence, it is important for the superpixels to be homogeneous so that the features adequately represent the pixels contained therein. Thus, we shall compare the performance of CEGCN using \cred{three} segmentation steps, SLIC\cred{, waterpixels} and H$^2$BO, to evaluate the benefits the proposed method can bring in terms of classification performance.

\subsection{Experimental setup}

In this section, to validate the proposed method in the classification task, we propose to replace the LDA-SLIC step in the superpixel-based graph constructor (SGC) of the CEGCN algorithm with the proposed H$^2$BO homogeneity-based multiscale method, thus naming it CEGCN$_{\text{H$^2$BO}}$. We also refer to the original CEGCN algorithm using SLIC by CEGCN$_{\text{SLIC}}$\cred{, and CEGCN$_{\text{WPX}}$ by using waterpixels}. As performed for the unmixing task, simulations were conducted with synthetic and real datasets in order to show the potential of the proposed method to improve hyperspectral image analysis when used as preprocessing step. For the purpose of comparison between CEGCN$_{\text{H$^2$BO}}$, the original CEGCN$_{\text{SLIC}}$ \cred{ and CEGCN$_{\text{WPX}}$}, the same metrics used by \cite{Liu2021} were evaluated: overall accuracy (OA), average accuracy (AA), and kappa statistics (KPP) \citep{stehman1997selecting}.

\subsection{Simulation results with synthetic data}

To validate the presented method in the classification task, the superpixel generation step of the CEGCN algorithm was replaced by the proposed hierarchical segmentation strategy. For the purpose of a controlled evaluation, it was decided to analyse the application of the algorithm on a synthetic dataset, in which the true labels of all pixels are known in advance and can be used for a quantitative evaluation.

\begin{figure}[!htb]
	\centering
        \subfigure[False-colour]{\includegraphics[width=10em]{DC3_RGB.pdf}}
        \subfigure[Ground-truth]{\includegraphics[width=10em]{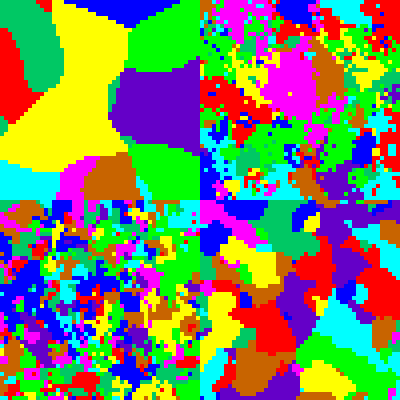}}

        \subfigure[CEGCN$_{\text{SLIC}}$]{\includegraphics[width=10em]{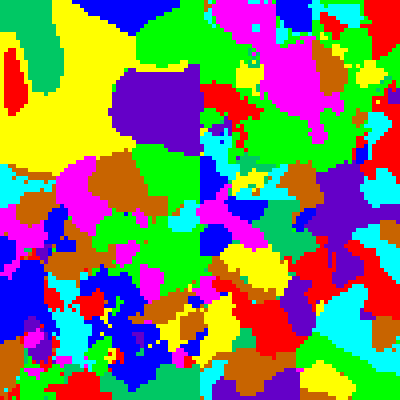}}
        \subfigure[CEGCN$_{\text{SLIC}}$, 30dB]{\includegraphics[width=10em]{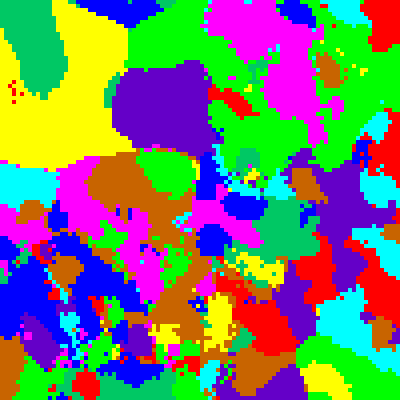}}
        \subfigure[CEGCN$_{\text{SLIC}}$, 20dB]{\includegraphics[width=10em]{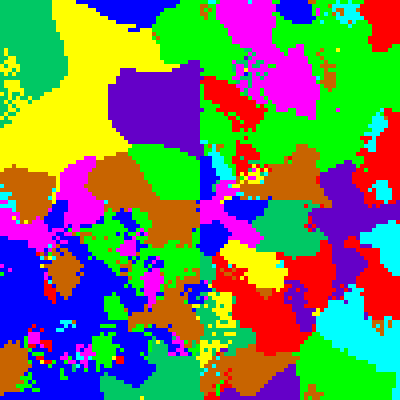}}

        \subfigure[\cred{CEGCN$_{\text{WPX}}$}]{\includegraphics[width=10em]{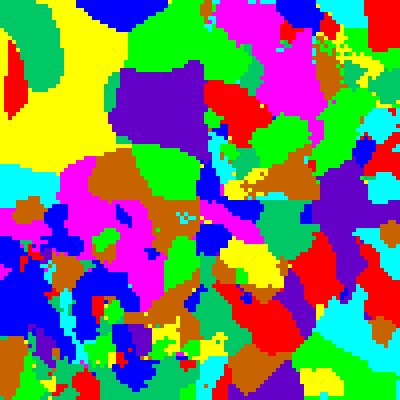}}
        \subfigure[\cred{CEGCN$_{\text{WPX}}$, 30dB}]{\includegraphics[width=10em]{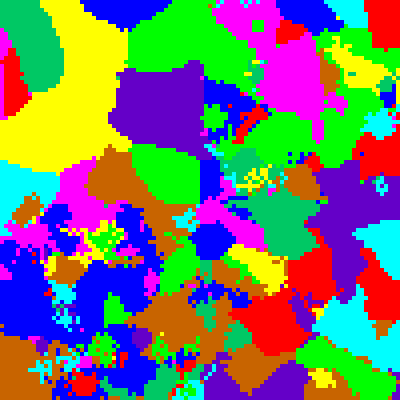}}
        \subfigure[\cred{CEGCN$_{\text{WPX}}$, 20dB}]{\includegraphics[width=10em]{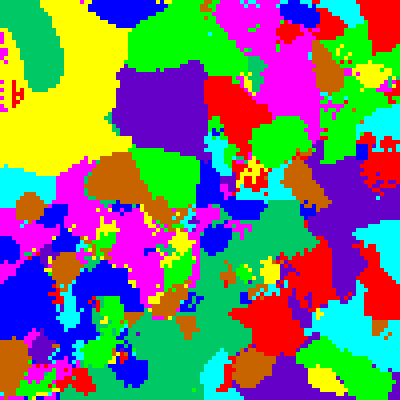}}

        \subfigure[CEGCN$_{\text{H$^2$BO}}$]{\includegraphics[width=10em]{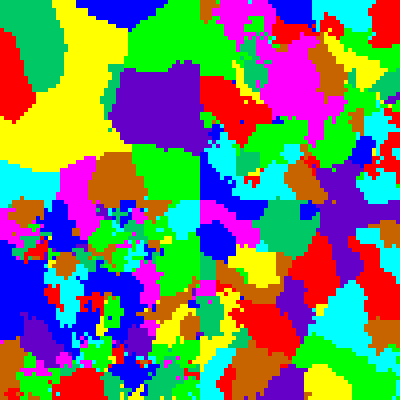}}
        \subfigure[CEGCN$_{\text{H$^2$BO}}$, 30dB]{\includegraphics[width=10em]{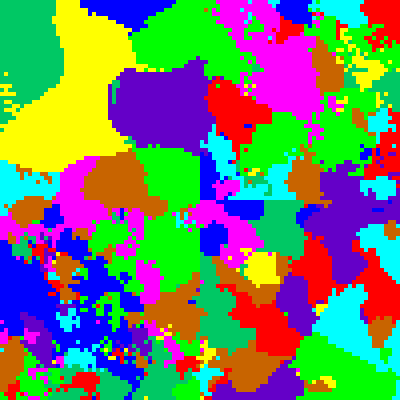}}
        \subfigure[CEGCN$_{\text{H$^2$BO}}$, 20dB]{\includegraphics[width=10em]{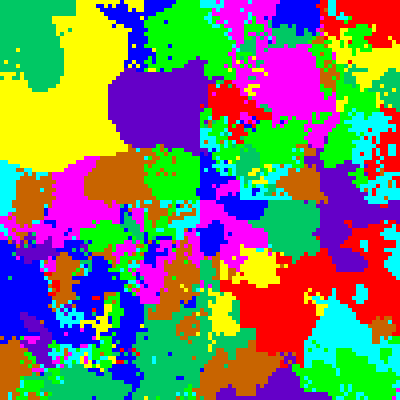}}

	\caption{DC3 - Classification maps}
	\label{fig:class_maps}
\end{figure}

The DC3 image from Section~\ref{sec:simuls_unmix_synth} (shown in Figures~\ref{fig:class_maps}a and~\ref{fig:class_maps}b) was chosen for the simulations with synthetic data, since it presents a greater variety of shapes in the spatial distribution of its content and allowed for a satisfactory evaluation of the unmixing task. We recall that this HI contained nine endmembers, which here will be considered as nine different classes, and their abundance proportions at each pixel are available.
To generate the ground-truth classification map of the DC3 image, for each pixel, the class of the endmember with the highest abundance contained in it was assigned. In addition to the noiseless condition, scenarios with an SNR of 30~dB and 20~dB were also considered. \cred{A grid search was performed to find the optimum values, which are shown in Table \ref{tab:parameters_classification_DC3}.} The parameters of CEGCN were kept with the same values as in the original work. From the total number of labeled pixels, \cred{10} samples per class are randomly selected for training, \cred{1\% for validation, and the remaining for testing.}

\begin{table}[!htb]
  \tbl{\cred{Parameters of the classification CEGCN algorithm for DC3.}}
  {\color{black}  %
  \begin{tabular}{ccccc}
    \toprule
      \multicolumn{5}{c}{CEGCN} \\
      \multirow{2}{*}{Algorithm} & \multirow{2}{*}{Parameters} & \multicolumn{3}{c}{DC3} \\
        & & noiseless & 30 dB & 20 dB \\
      \midrule
      \multirow{7}{*}{$\text{H$^2$BO}$} & $\gamma$ & 0.00125 & 0.00125 & 0.00125\\
        & $\sigma_0$ & 8 & 8 & 8 \\
        & $\sigma_1$ & 7 & 7 & 7\\
        & $\sigma_2$ & 4 & 4 & 4 \\
        & $\sigma_3$ & 3 & 3 & 3\\
        & $\tau_{\text{ outliers}}$ & 10\% & 10\% & 10\% \\
        & $\tau_{\text{ homog}}$ & 20\% & 30\% & 20\% \\
      \midrule
      \multirow{2}{*}{SLIC} & $\gamma$ & 0.01 & 0.1 & 0.1 \\
        & $\sigma$ & 60 & 60 & 80 \\
        \midrule
        \multirow{2}{*}{Waterpixels} & $k$ & 0.001 & 0.001 & 0.01 \\
        & $\sigma$ & 8 & 14 & 11 \\
      \bottomrule
    \end{tabular}}
  \label{tab:parameters_classification_DC3}
  {\color{black}} %
\end{table}

Figure \ref{fig:class_maps} shows the resulting classification maps generated by the \cred{three} methods. It is possible to notice that the use of the proposed H$^2$BO oversegmentation method leads to a smaller amount of misclassified pixels in CEGCN$_{\text{H$^2$BO}}$ when compared to the original method CEGCN$_{\text{SLIC}}$ \cred{and CEGCN$_{\text{WPX}}$}. This difference becomes more evident when in the noisier condition of 20~dB SNR and in the lower left quarter of the image, where the distribution of classes is more irregular. Observing Figures \ref{fig:class_maps}e and \ref{fig:class_maps}h, even in the upper left region, where the arrangements are larger and uniform, the proposed method presents better classification performance, with emphasis on neighbouring pixels at the boundaries between different classes. This same result can be seen from the overall and per class quantitative results shown in Table~\ref{tab:class_results}, as the proposed CEGCN$_{\text{H$^2$BO}}$ method led to improvements of up to about \cred{$2\%$} with respect to the CEGCN$_{\text{SLIC}}$ \cred{and $22\%$ over CEGCN$_{\text{WPX}}$ in the cases of 20~dB SNR}.

\begin{table}[!htb]
\addtolength{\tabcolsep}{-3pt}
\scriptsize
	\tbl{Classification results of CEGCN$_{\text{SLIC}}$\cred{, $\text{CEGCN}_{\text{WPX}}$ algorithms} and the proposed CEGCN$_{\text{H$^2$BO}}$ on the DC3 data set \cred{-- Training samples: $10$}.}{
\begin{tabular}{cccc|ccc|ccc}
	\toprule
\multicolumn{10}{c}{DC3 dataset} \\
\midrule
\multirow{2}{*}{Metric} & \multicolumn{3}{c|}{CEGCN$_{\text{SLIC}}$} & \multicolumn{3}{c|}{\cred{$\text{CEGCN}_{\text{WPX}}$}} & \multicolumn{3}{c}{CEGCN$_{\text{H$^2$BO}}$} \\
& Noiseless & 30 dB & 20 dB & \cred{Noiseless} & \cred{30 dB} & \cred{20 dB} & Noiseless & 30 dB & 20 dB\\
\midrule
OA(\%) & {69.5}\textpm{1.6} & {71.8}\textpm{1.6} & {65.4}\textpm{1.6} & {74.8}\textpm{1.6} & {66.5}\textpm{2.0} & {54.8}\textpm{2.8} & {79.4}\textpm{0.3} & {77.4}\textpm{1.7} & {66.7}\textpm{3.7} \\
AA(\%)& {69.3}\textpm{1.8} & {71.7}\textpm{1.2} & {64.9}\textpm{2.0} & {74.6}\textpm{1.8} & {65.8}\textpm{1.7} & {54.5}\textpm{2.6} & {79.2}\textpm{0.4} & {77.2}\textpm{1.9} & {66.3}\textpm{3.3} \\
KPP(x100) & {65.5}\textpm{1.8} & {68.1}\textpm{1.7} & {60.9}\textpm{1.8} & {71.5}\textpm{1.8} & {62.1}\textpm{2.2} & {49.0}\textpm{3.1} & {76.7}\textpm{0.4} & {74.4}\textpm{2.0} & {62.4}\textpm{4.2}\\
\bottomrule
\end{tabular}}
\label{tab:class_results}
\end{table}

\begin{figure}[!htb]
\centering
\subfigure[]{\label{fig:DC3_homog_20dB_sens_sigma0}\includegraphics[width=11em]{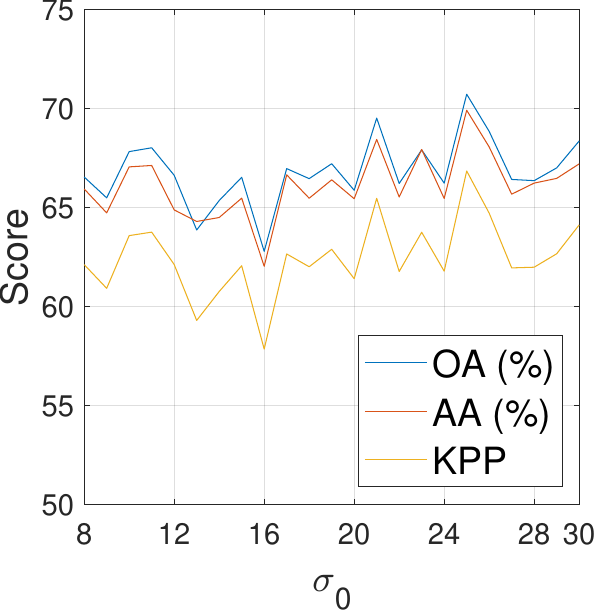}}
\subfigure[]{\label{fig:DC3_homog_20dB_sens_tau_outliers}\includegraphics[width=11em]{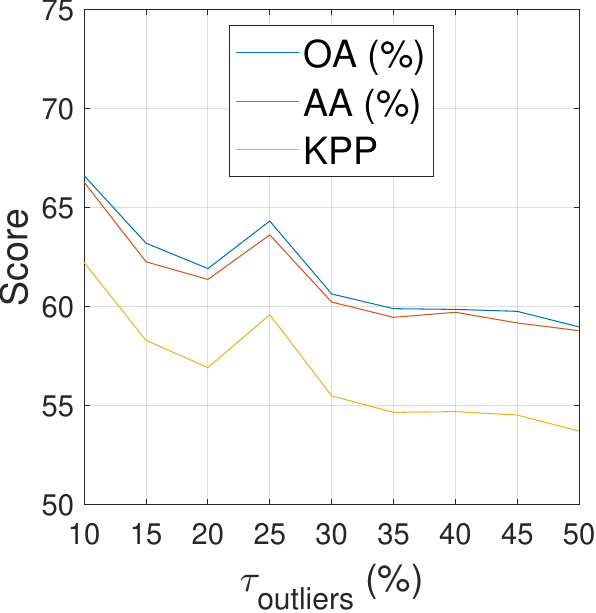}}
\subfigure[]{\label{fig:DC3_homog_20dB_sens_tau_homog}\includegraphics[width=11em]{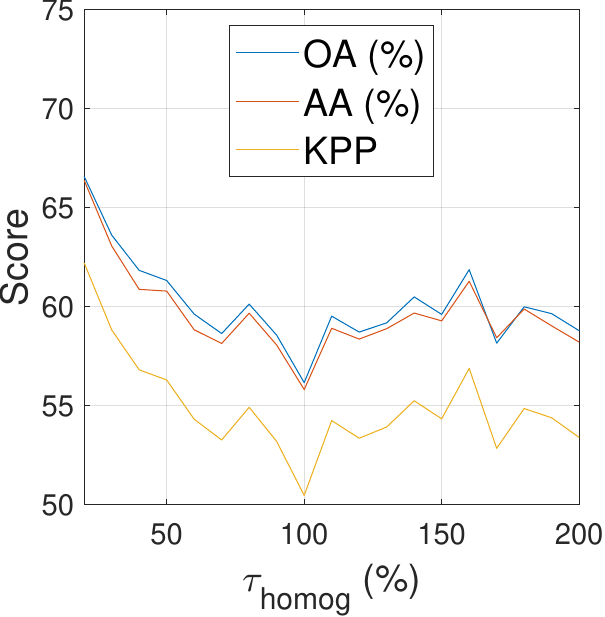}}

\subfigure[\credd{Deviation from reference overall accuracy}]{\label{fig:DC3_homog_20dB_statistical_hist}\includegraphics[width=18em]{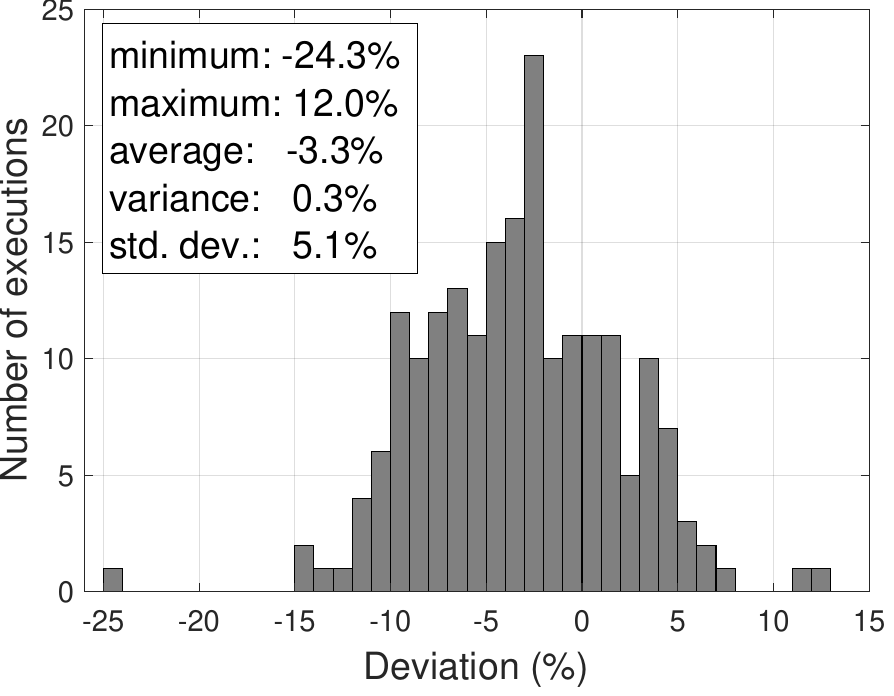}}

\caption{\credd{Parameter sensitivity and statistical performance analysis of the classification task in DC3 using CEGCN$_{\text{H$^2$BO}}$, 20 dB SNR. The top row shows the score ($y$-axis) of the OA(\%), AA(\%) and KPP metrics for the variation of the parameters ($x$-axis) of (a) initial superpixel size, and (b) outliers and (c) homogeneity thresholds. The (d) histogram in the bottom row shows the percentage deviation ($x$-axis) from the reference value OA(\%) $= 66.7\%$ considering 200 executions ($y$-axis) of the algorithm.}}
\label{fig:DC3_analysis_class}
\end{figure}

\begin{color}{black}
\textit{\textbf{Sensitivity and statistical performance analysis}} -- Similar to the tests conducted for the unmixing task, the sensitivity of the OA, AA and KPP classification metrics to the parameter values of the proposed \text{H$^2$BO} algorithm is analysed. The DC3 synthetic data and the 20~dB scenario were chosen for these experiments. In the statistical performance analysis for randomly chosen parameters, in 200 executions, a new random value for each parameter was sampled according uniform distributions within the following intervals: regulariser $\gamma \in [0,001,\ 0,02]$, superpixels size $\sigma_0 \in [8,\ 30]$, thresholds $\tau_{outliers} \in [10\%,\ 50\%]$ and $\tau_{homog} \in [20\%,\ 200\%]$.
For $\sigma_1$, $\sigma_2$ and $\sigma_3$ the following relationship was used: $\sigma_i \in [\frac{\sigma_{i-1}}{2},\ \sigma_{i-1}-1]$, rounded up to the upper integer and $\sigma_2$ and $\sigma_3$ limited to 3 and 2, respectively.
We also varied the values of parameters $\sigma_0$, $\tau_{outliers}$ and $\tau_{homog}$ while keeping the remaining parameters at their optimal values, and evaluated the effect on the classification performance.
The results are shown in Figure~\ref{fig:DC3_analysis_class}. 

\credd{In Figure~\ref{fig:DC3_analysis_class}a, for larger values of the initial superpixels size $\sigma_0$ the score results tends to improve, since it is possible that the large superpixels will be re-segmented at a later scale. As the values of $\tau_{outliers}$ increase (Figure~\ref{fig:DC3_analysis_class}b), performance worsens, as relevant information is removed in addition to unwanted outliers. As expected, the $\tau_{homog}$ values in Figure~\ref{fig:DC3_analysis_class}c have the most influence on the results, since they directly determine the degree of homogeneity for subdividing the superpixels and providing the finest representation of HI.}

\credd{The histogram in Figure \ref{fig:DC3_analysis_class}d shows the percentage deviation from the reference value of OA $= 66.7\%$ presented in Table \ref{tab:class_results} considering the 200 executions of CEGCN$_{\text{H$^2$BO}}$ algorithm, 20 dB SNR scenario, with randomly selected parameters. It is possible to perceive an average variation of approximately 3\% and a low standard deviation, around 5\%. This shows that the probability of obtaining good results in the classification task is high for parameter values randomly chosen in a reasonable range according to the parameter selection methodology presented in Section~\ref{sec:parameter_selection}.}

\end{color}

\subsection{Experiments with real hyperspectral data}

For experimental analysis on real datasets, the CEGCN$_{\text{SLIC}}$\cred{, CEGCN$_{\text{WPX}}$} and CEGCN$_{\text{H$^2$BO}}$ algorithms were applied to the classification of the popular Indian Pines, Pavia University, and Salinas scenes in the noiseless, 20~dB and 30~dB SNR scenarios. \cred{For the Indian Pines, Salinas, and Pavia University datasets, 15, 5, and 10 training samples per class were used, respectively, 1\% for validation and the remaining for testing.} The multiscale representation parameters for \cred{all algorithms were determined by grid search and are shown in Table \ref{tab:parameters_classification_real}.}

\begin{table}[!htb]
\addtolength{\tabcolsep}{-3.2pt}
	\scriptsize
	\tbl{\cred{Parameters of the classification CEGCN algorithm for each HI.}}
	{\color{black} %
	\begin{tabular}{ccccc|ccc|ccc}
		\toprule
	\multicolumn{11}{c}{CEGCN} \\
		\multirow{2}{*}{Algorithm} & \multirow{2}{*}{Parameters} & \multicolumn{3}{c}{Indian Pines} & \multicolumn{3}{c}{Salinas} & \multicolumn{3}{c}{Pavia University}\\
 & & noiseless & 30 dB & 20 dB & noiseless & 30 dB & 20 dB & noiseless & 30 dB & 20 dB \\
		\midrule
		\multirow{7}{*}{$\text{H$^2$BO}$} & $\gamma$ & 0.00125 & 0.00125 & 0.001 & 0.00125 & 0.00125 & 0.00125 & 0.00125 & 0.00125 & 0.00125\\
		& $\sigma_0$ & 40 & 50 & 20 & 60 & 70 & 20 & 60 & 70 & 40 \\
		& $\sigma_1$ & 15 & 15 & 20 & 15 & 15 & 15 & 15 & 15 & 15 \\
		& $\sigma_2$ & 10 & 10 & 10 & 10 & 10 & 10 & 10 & 10 & 10 \\
		& $\sigma_3$ & 8 & 8 & 8 & 8 & 8 & 8 & 8 & 8 & 8 \\
		& $\tau_{\text{ outliers}}$ & 10\% & 10\% & 10\% & 10\% & 10\% & 10\% & 10\% & 10\% & 10\%\\
		& $\tau_{\text{ homog}}$ & 50\% & 20\% & 175\% & 150\% & 150\% & 200\% & 100\% & 100\% & 100\%\\
		\midrule
		\multirow{2}{*}{SLIC} & $\gamma$ & 0.001 & 0.1 & 0.1 & 0.001 & 0.01 & 0.1 & 0.001 & 0.01 & 0.01\\
		& $\sigma$ & 60 & 60 & 80 & 30 & 80 & 80 & 60 & 70 & 60\\
 \midrule
 \multirow{2}{*}{Waterpixels} & $k$ & 0.01 & 0.01 & 0.001 & 0.1 & 0.001 & 0.001 & 0.001 & 0.01 & 0.01\\
 & $\sigma$ & 20 & 20 & 11 & 15 & 14 & 10 & 12 & 12 & 10\\
		\bottomrule
	\end{tabular}}
	\label{tab:parameters_classification_real}
    \end{table}

\begin{figure}[!htb]
	\centering
        \subfigure[False-colour]{\includegraphics[width=10em]{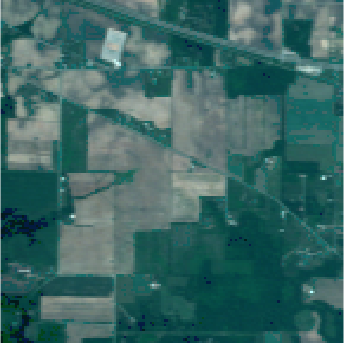}}
        \subfigure[Ground-truth]{\includegraphics[width=10em]{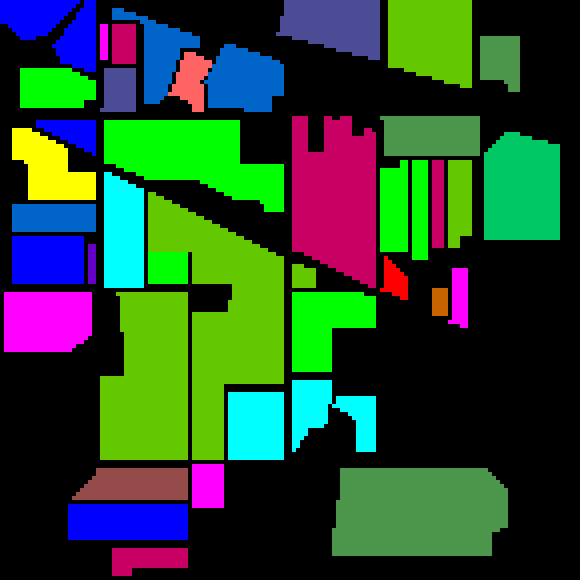}}
 
        \subfigure[CEGCN$_{\text{SLIC}}$]{\includegraphics[width=10em]{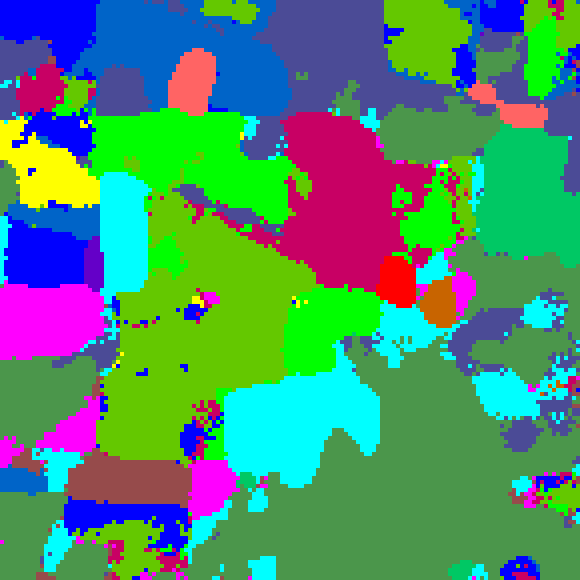}}
        \subfigure[CEGCN$_{\text{SLIC}}$, 30dB]{\includegraphics[width=10em]{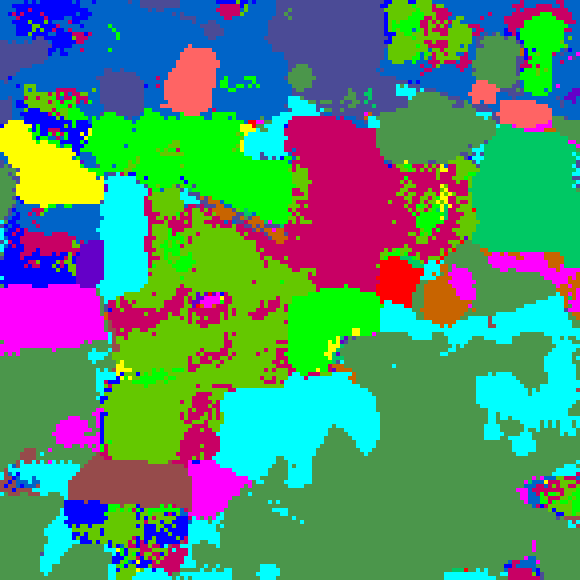}}
        \subfigure[CEGCN$_{\text{SLIC}}$, 20dB]{\includegraphics[width=10em]{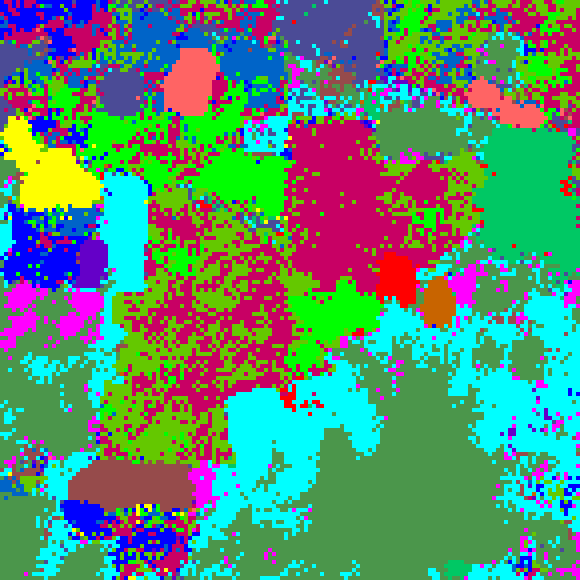}}

        \subfigure[\cred{CEGCN$_{\text{WPX}}$}]{\includegraphics[width=10em]{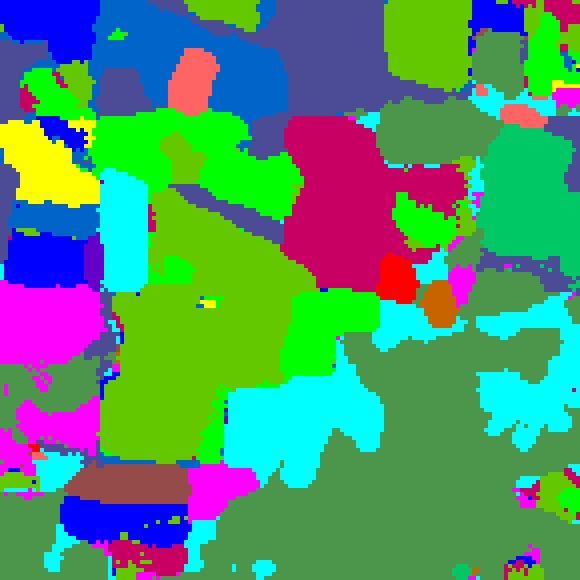}}
        \subfigure[\cred{CEGCN$_{\text{WPX}}$, 30dB}]{\includegraphics[width=10em]{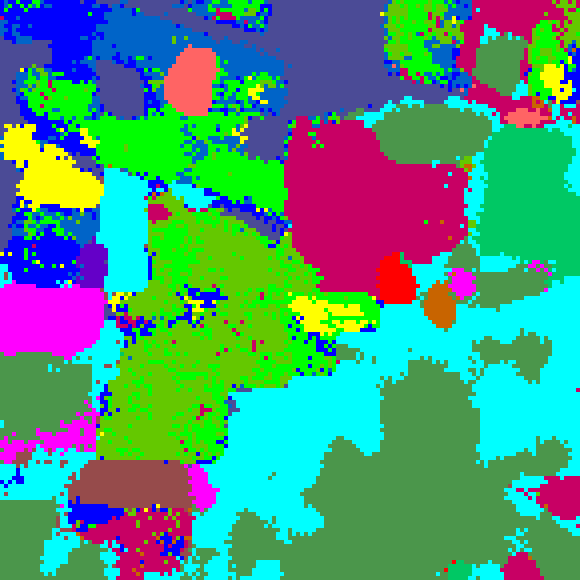}}
        \subfigure[\cred{CEGCN$_{\text{WPX}}$, 20dB}]{\includegraphics[width=10em]{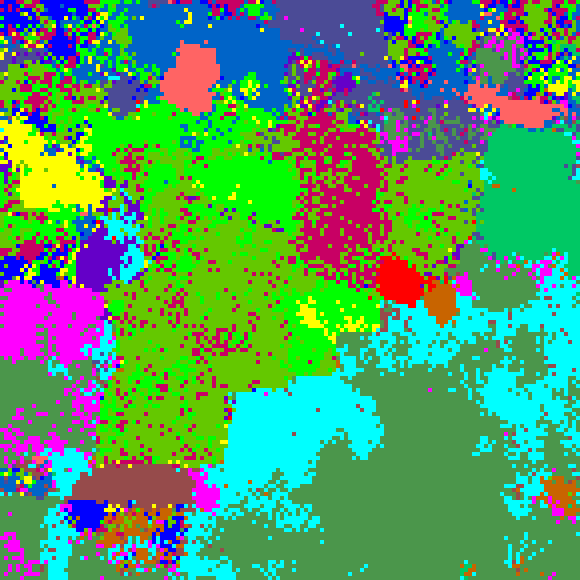}}

        \subfigure[CEGCN$_{\text{H$^2$BO}}$]{\includegraphics[width=10em]{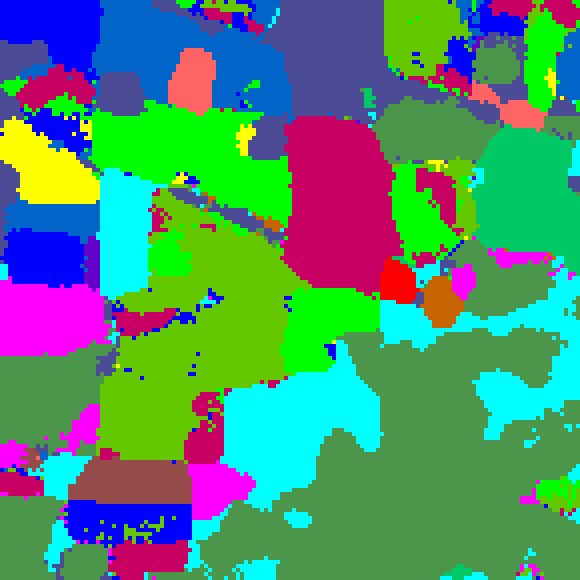}}
        \subfigure[CEGCN$_{\text{H$^2$BO}}$, 30dB]{\includegraphics[width=10em]{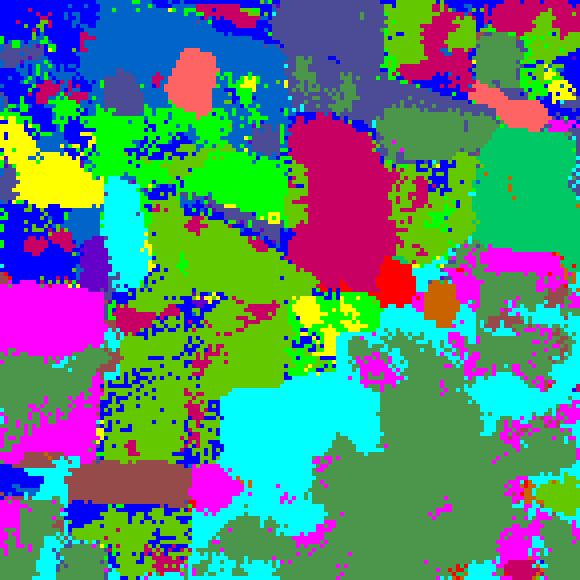}}
        \subfigure[CEGCN$_{\text{H$^2$BO}}$, 20dB]{\includegraphics[width=10em]{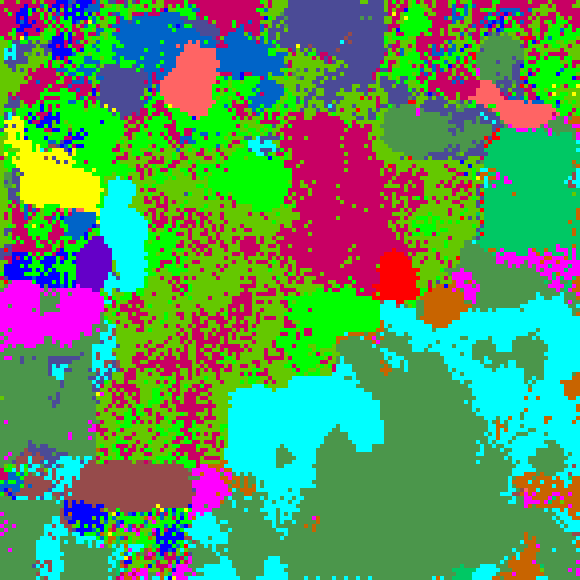}}

	\caption{Indian Pines -- false-colour representation, ground-truth and classification maps.}
	\label{fig:class_maps_indian}
\end{figure}

The Indian Pines dataset is composed mostly of cropland and forest from Indiana, USA, and has a spatial size of 145$\times$145 pixels and 200 spectral bands (after removal of noisy and water absorption bands) with wavelengths between 400 and 2500~nm observed by the AVIRIS sensor. A false-colour representation and the ground-truth of the scene with its 16 different classes are presented in Figure~\ref{fig:class_maps_indian}a and~\ref{fig:class_maps_indian}b, respectively. The resulting classification maps are shown in Figure~\ref{fig:class_maps_indian}c-h.

\begin{figure}[!htb]
	\centering
        \subfigure[False-colour]{\includegraphics[width=4.5em]{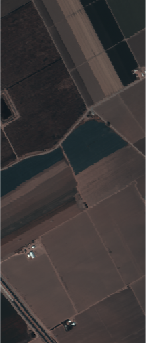}}
        \subfigure[Ground-truth]{\includegraphics[width=4.5em]{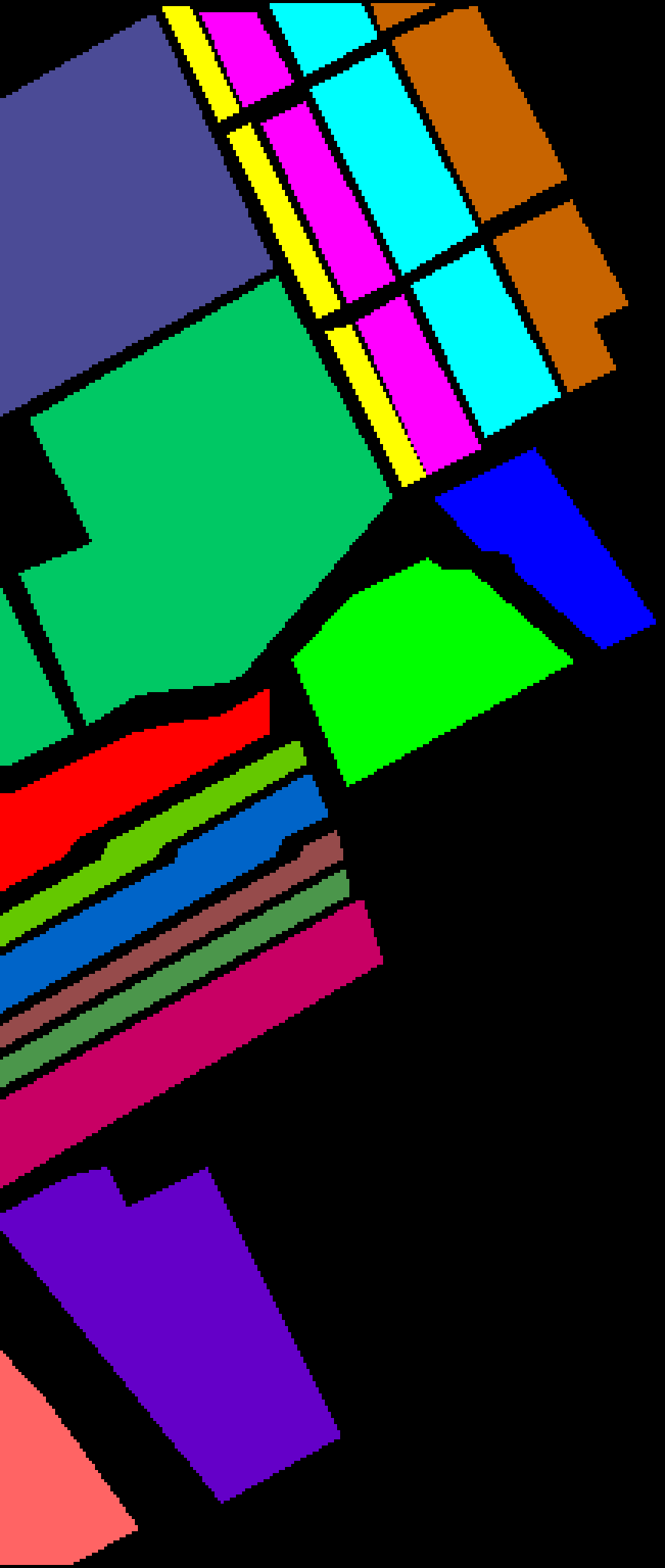}}
 
        \subfigure[CEGCN$_{\text{SLIC}}$]{\includegraphics[width=4.5em]{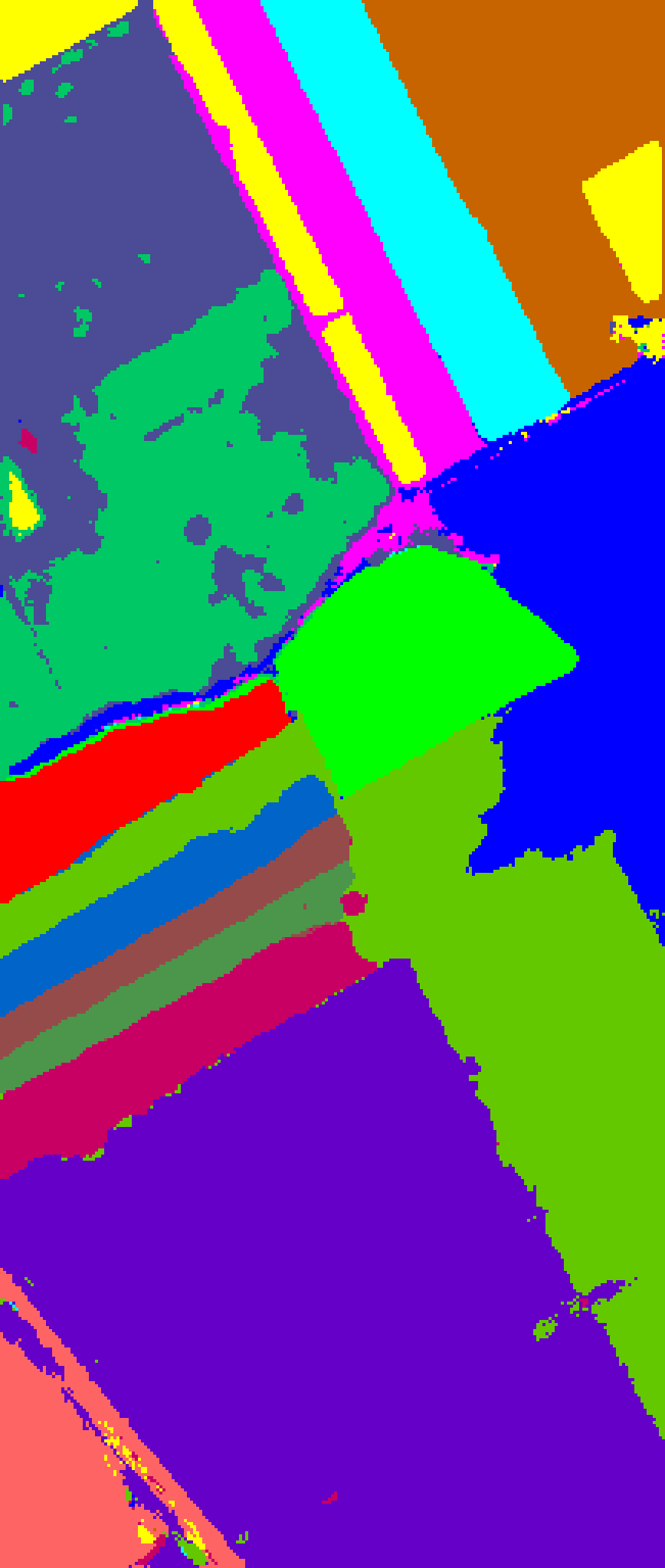}}
        \subfigure[CEGCN$_{\text{SLIC}}$, 30dB]{\includegraphics[width=4.5em]{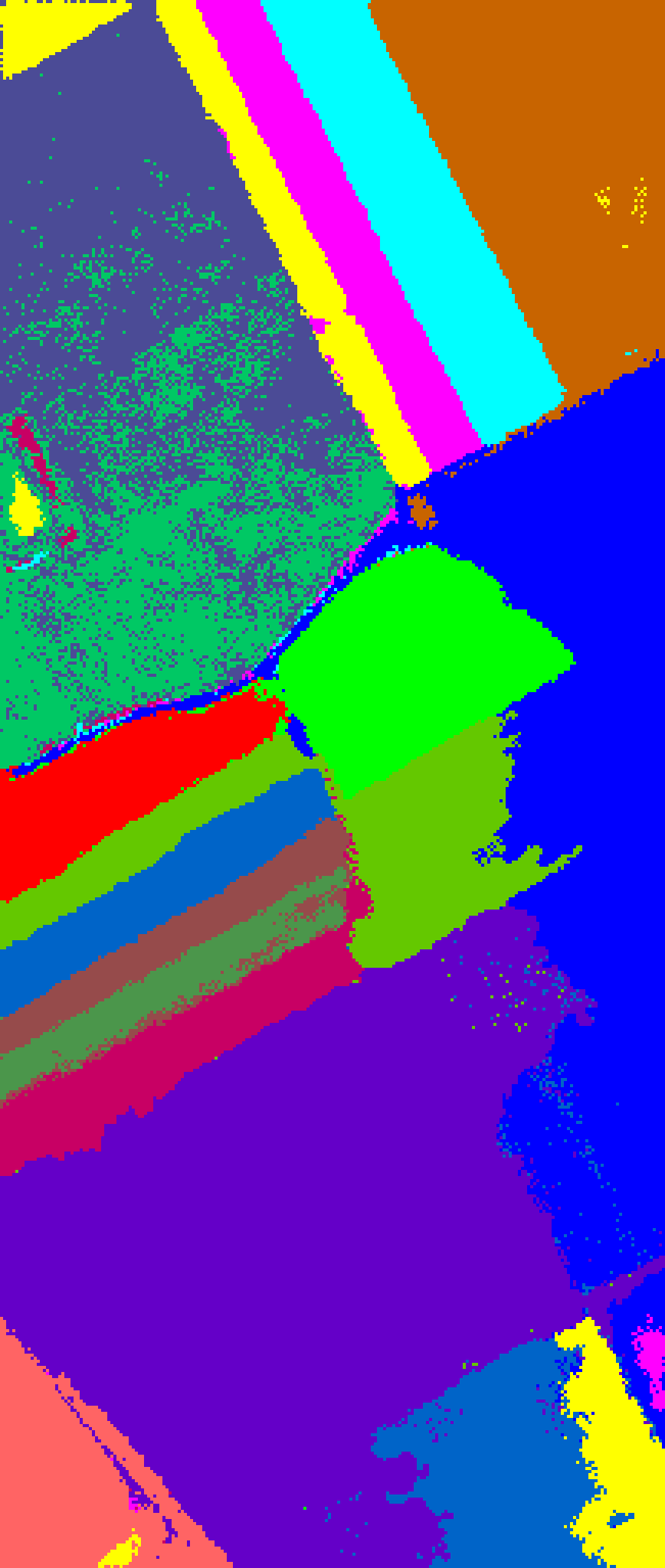}}
        \subfigure[CEGCN$_{\text{SLIC}}$, 20dB]{\includegraphics[width=4.5em]{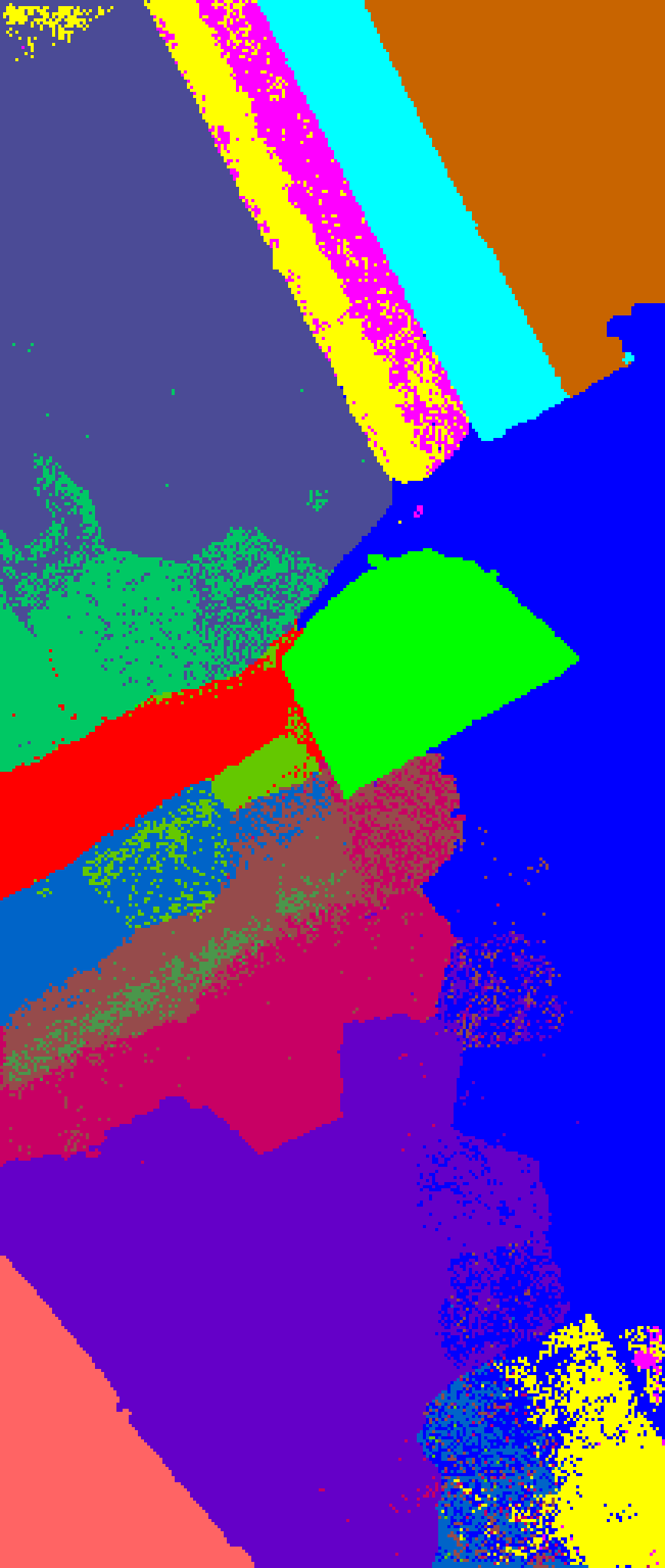}}

        \subfigure[\cred{CEGCN$_{\text{WPX}}$}]{\includegraphics[width=4.5em]{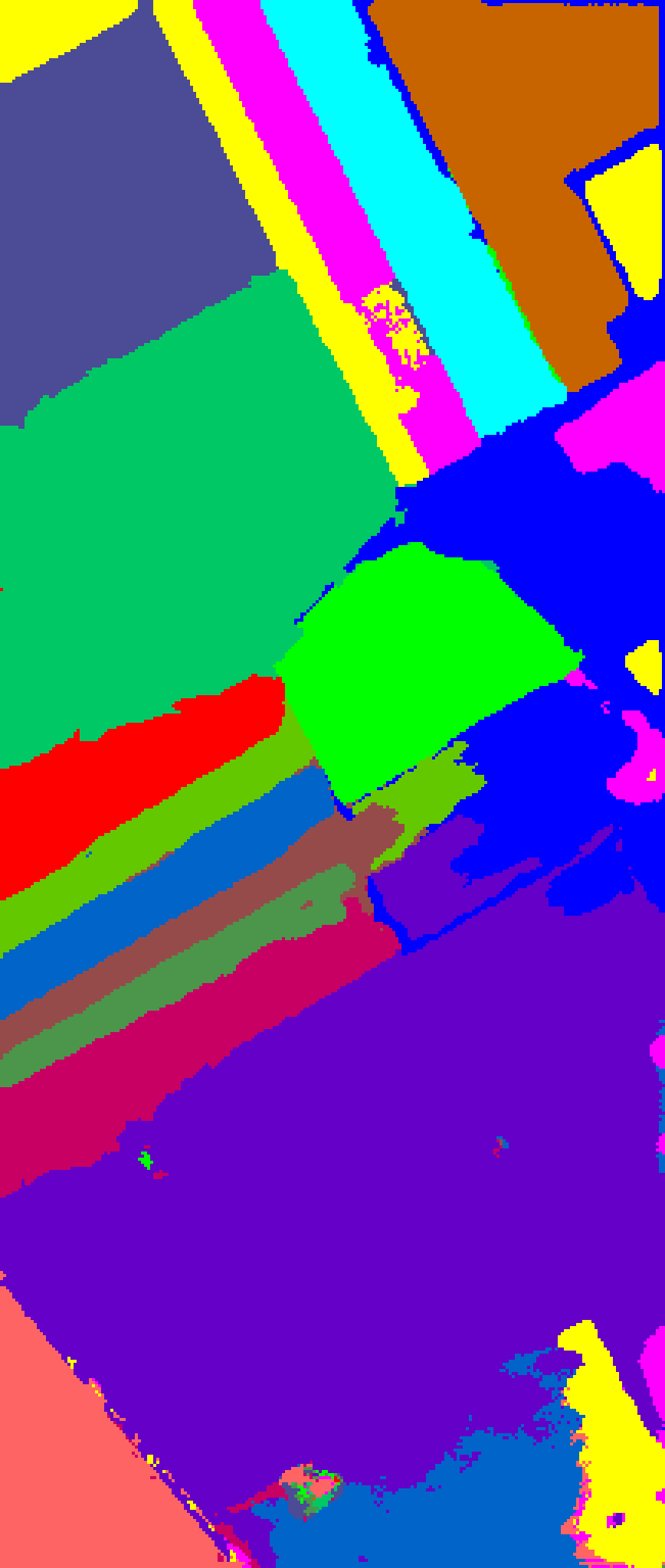}}
        \subfigure[\cred{CEGCN$_{\text{WPX}}$, 30dB}]{\includegraphics[width=4.5em]{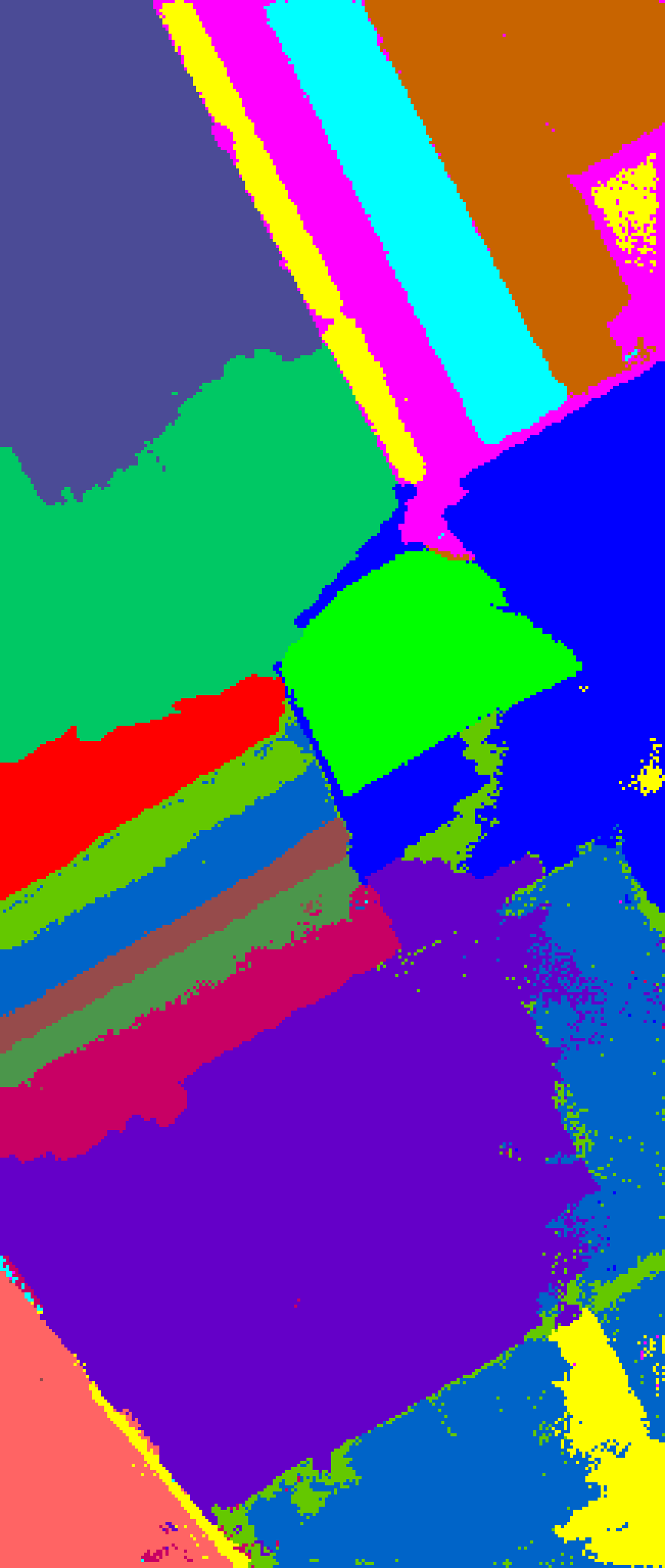}}
        \subfigure[\cred{CEGCN$_{\text{WPX}}$, 20dB}]{\includegraphics[width=4.5em]{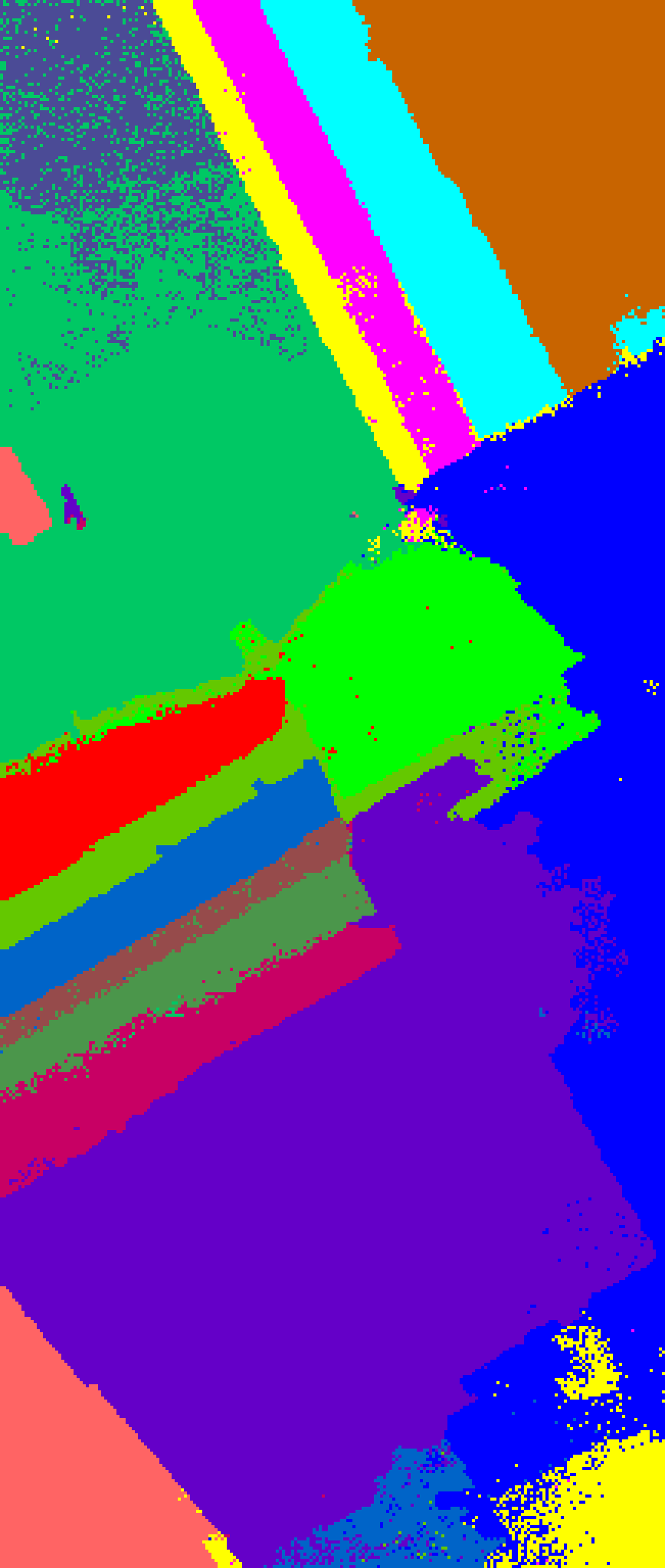}}

        \subfigure[CEGCN$_{\text{H$^2$BO}}$]{\includegraphics[width=4.5em]{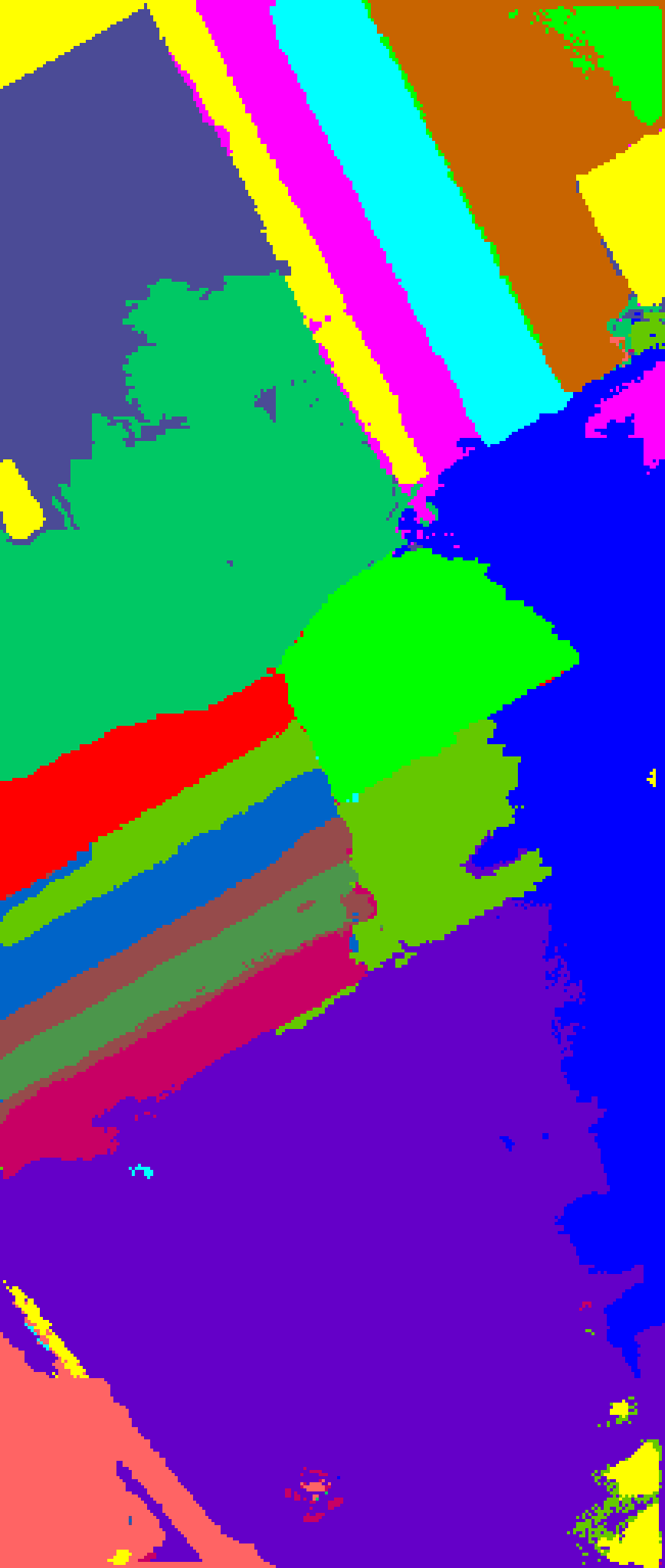}}
        \subfigure[CEGCN$_{\text{H$^2$BO}}$, 30dB]{\includegraphics[width=4.5em]{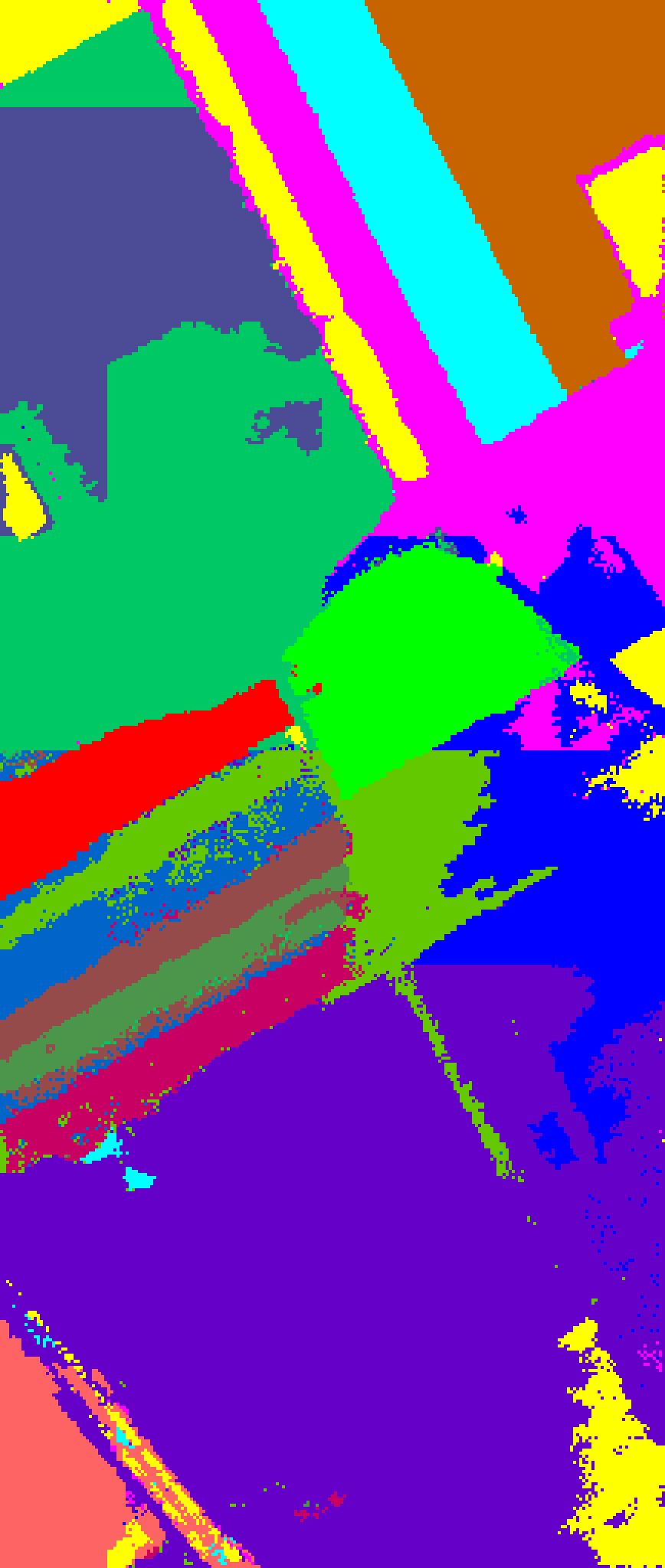}}
        \subfigure[CEGCN$_{\text{H$^2$BO}}$, 20dB]{\includegraphics[width=4.5em]{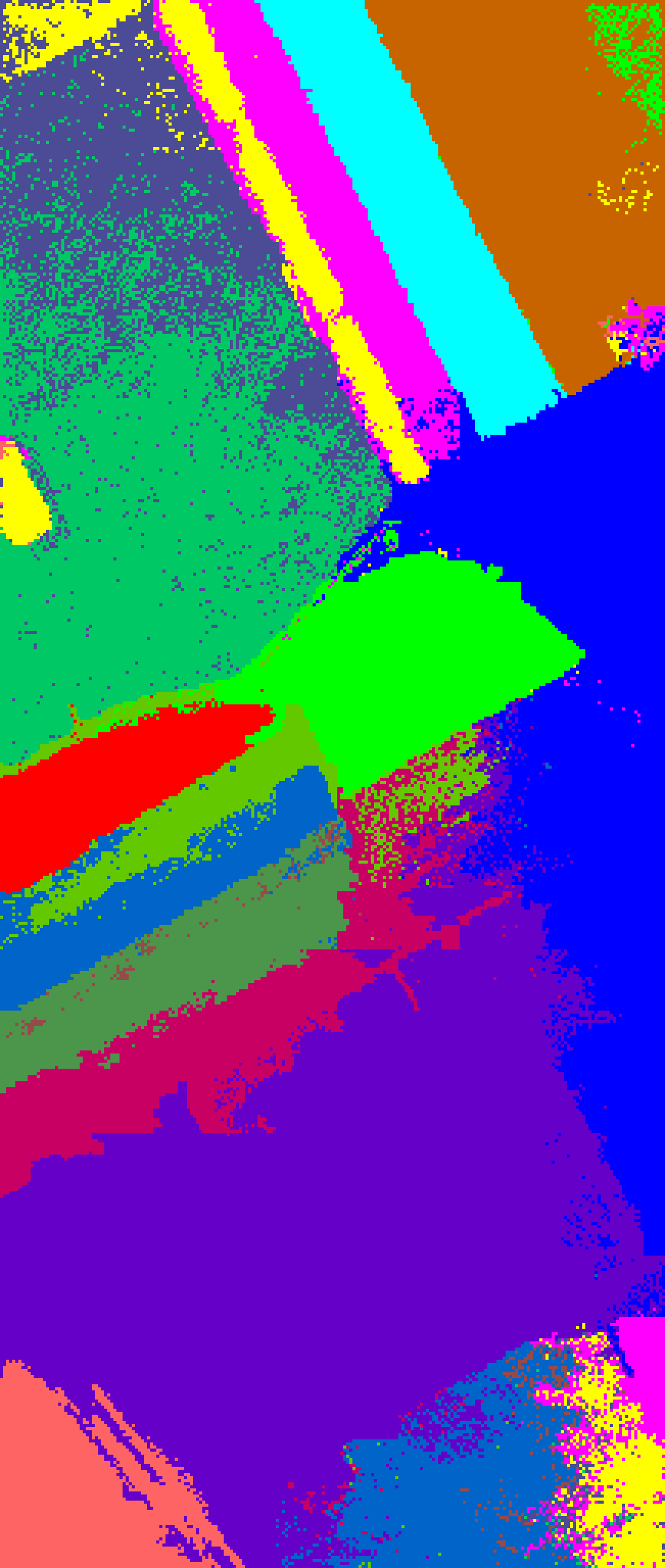}}

	\caption{Salinas -- false-colour representation, ground-truth and classification maps.}
	\label{fig:class_maps_salinas}
\end{figure}

Salinas is a scene also collected by the AVIRIS sensor, from Salinas Valley, California, USA, with size 512$\times$217 pixels and 16 classes annotated in its ground-truth, including fields and fruit and vegetable plantations. The false-colour image, the ground-truth and the classification maps generated by the algorithms are shown in Figure \ref{fig:class_maps_salinas}.  

\begin{figure}[!htb]
	\centering
        \centering
        \subfigure[False-colour]{\includegraphics[width=6em]{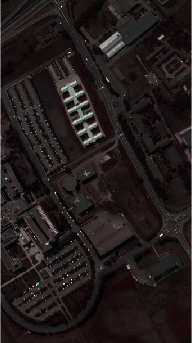}}
        \subfigure[Ground-truth]{\includegraphics[width=6em]{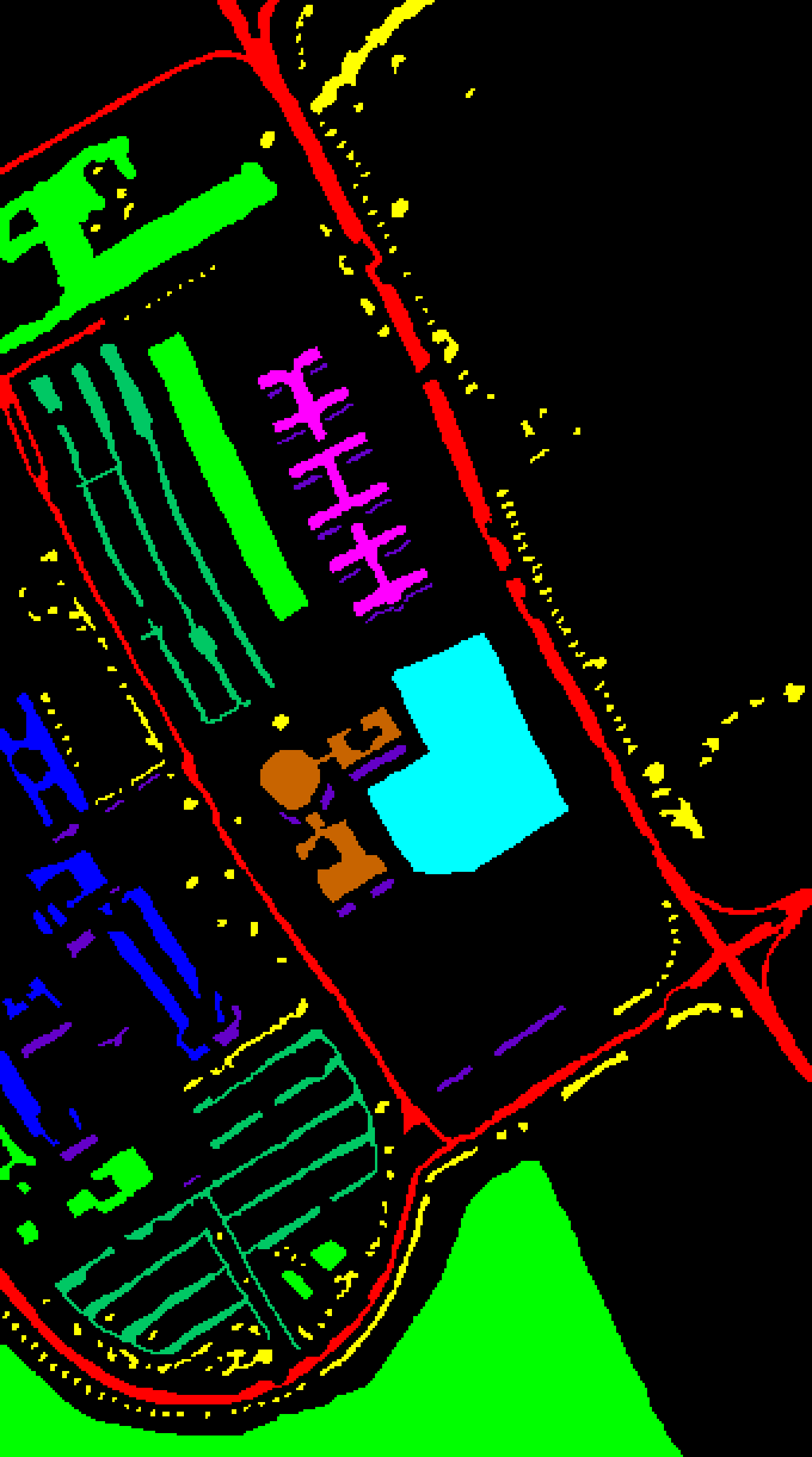}}
 
        \subfigure[CEGCN$_{\text{SLIC}}$]{\includegraphics[width=6em]{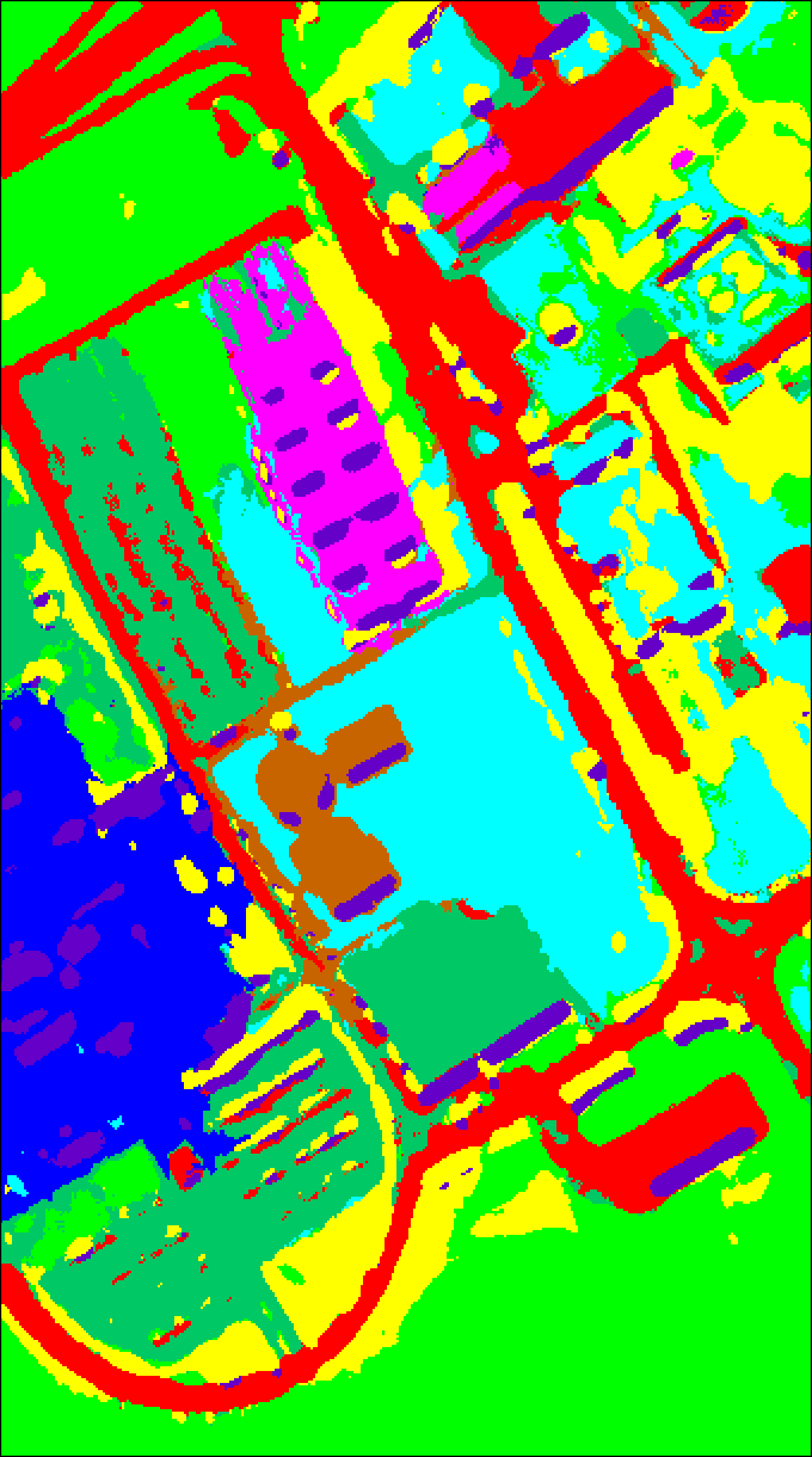}}
        \subfigure[CEGCN$_{\text{SLIC}}$, 30dB]{\includegraphics[width=6em]{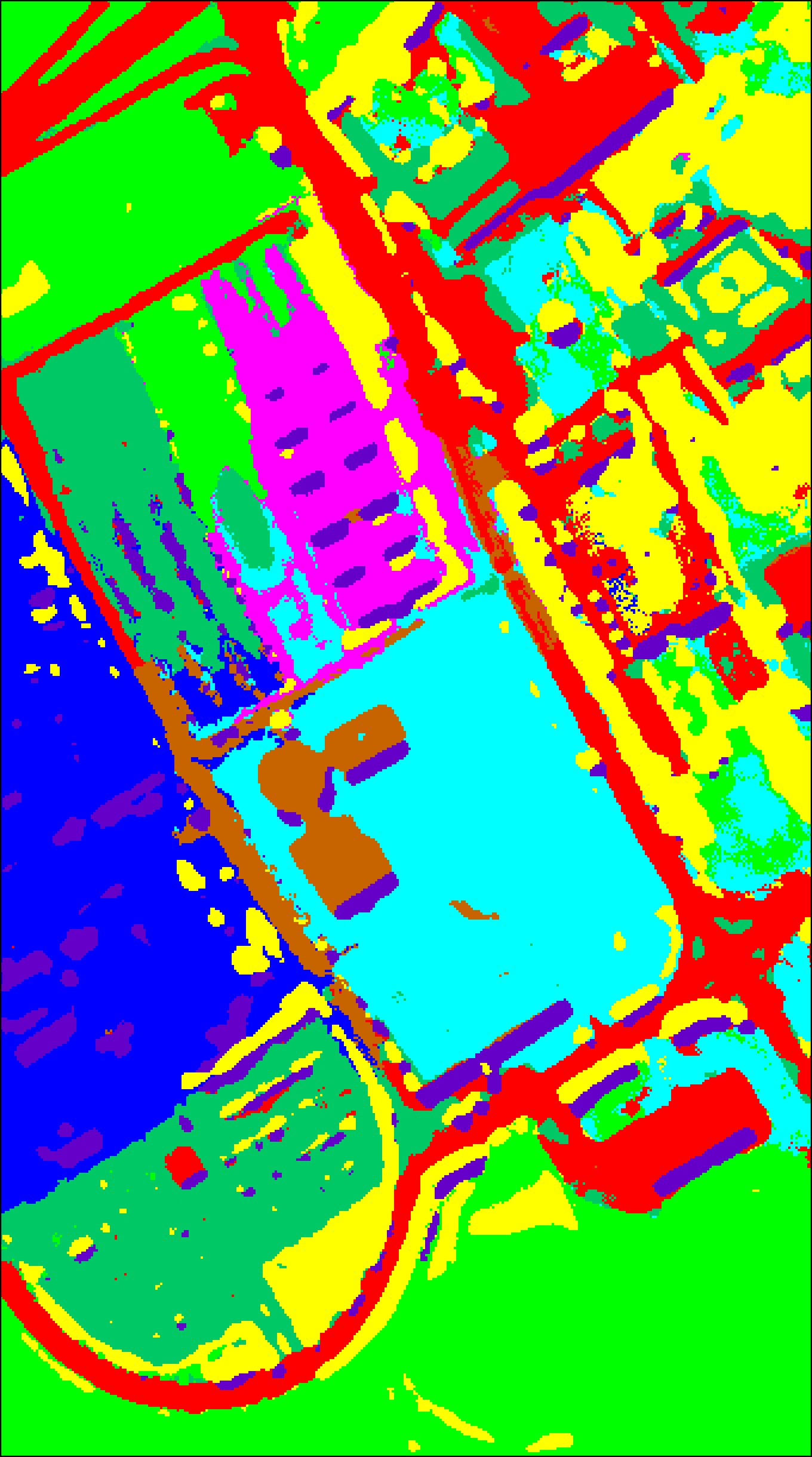}}
        \subfigure[CEGCN$_{\text{SLIC}}$, 20dB]{\includegraphics[width=6em]{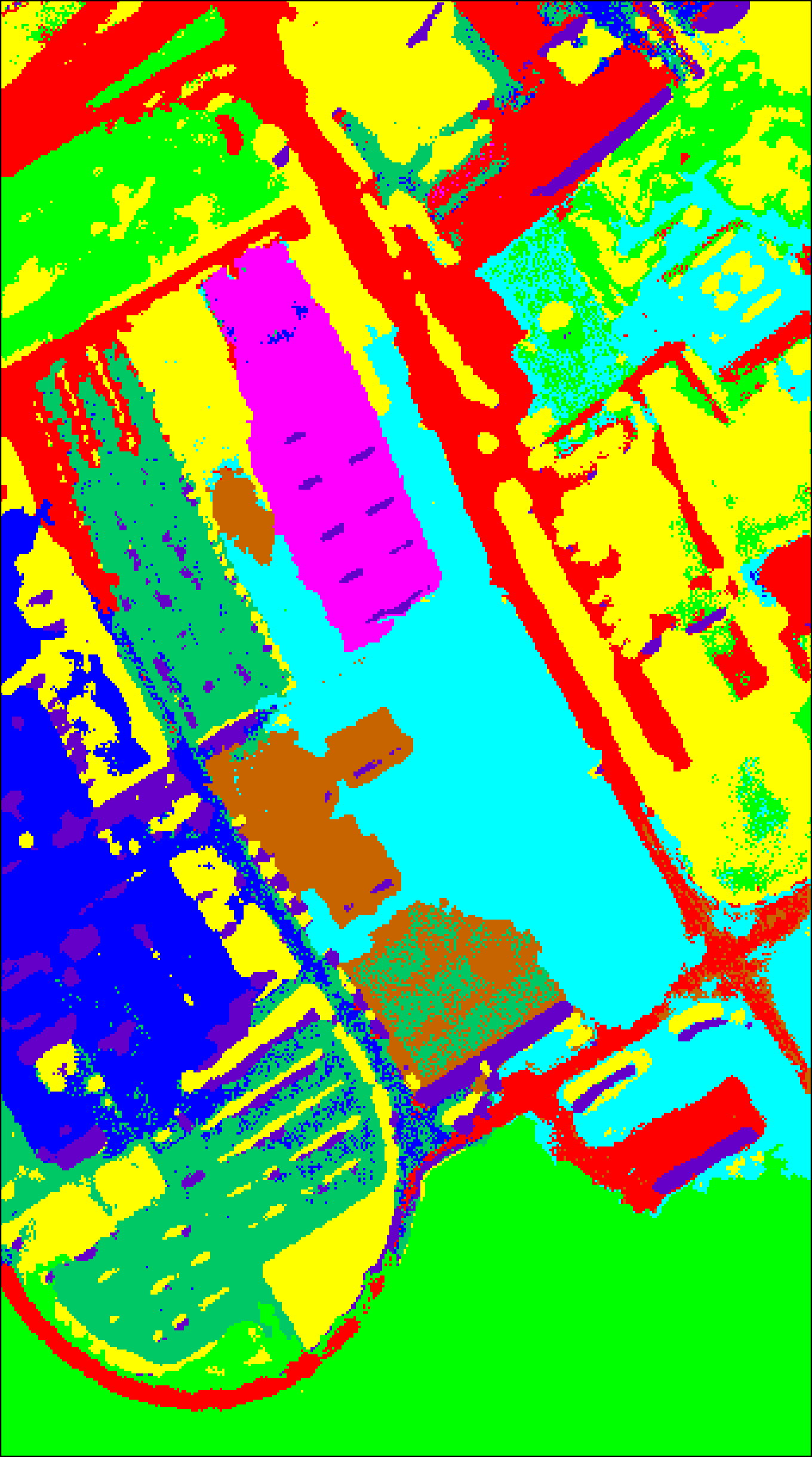}}

        \subfigure[\cred{CEGCN$_{\text{WPX}}$}]{\includegraphics[width=6em]{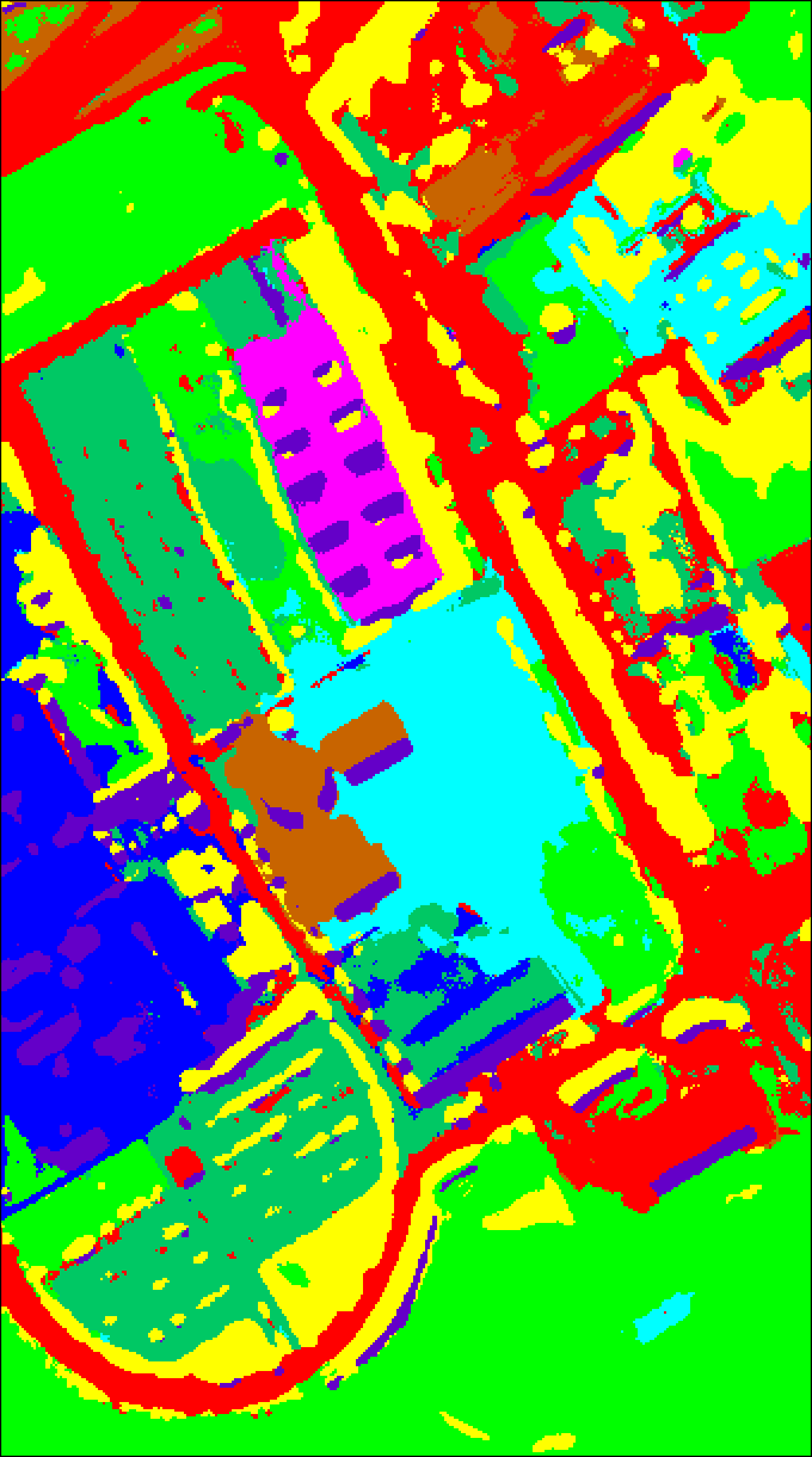}}
        \subfigure[\cred{CEGCN$_{\text{WPX}}$, 30dB}]{\includegraphics[width=6em]{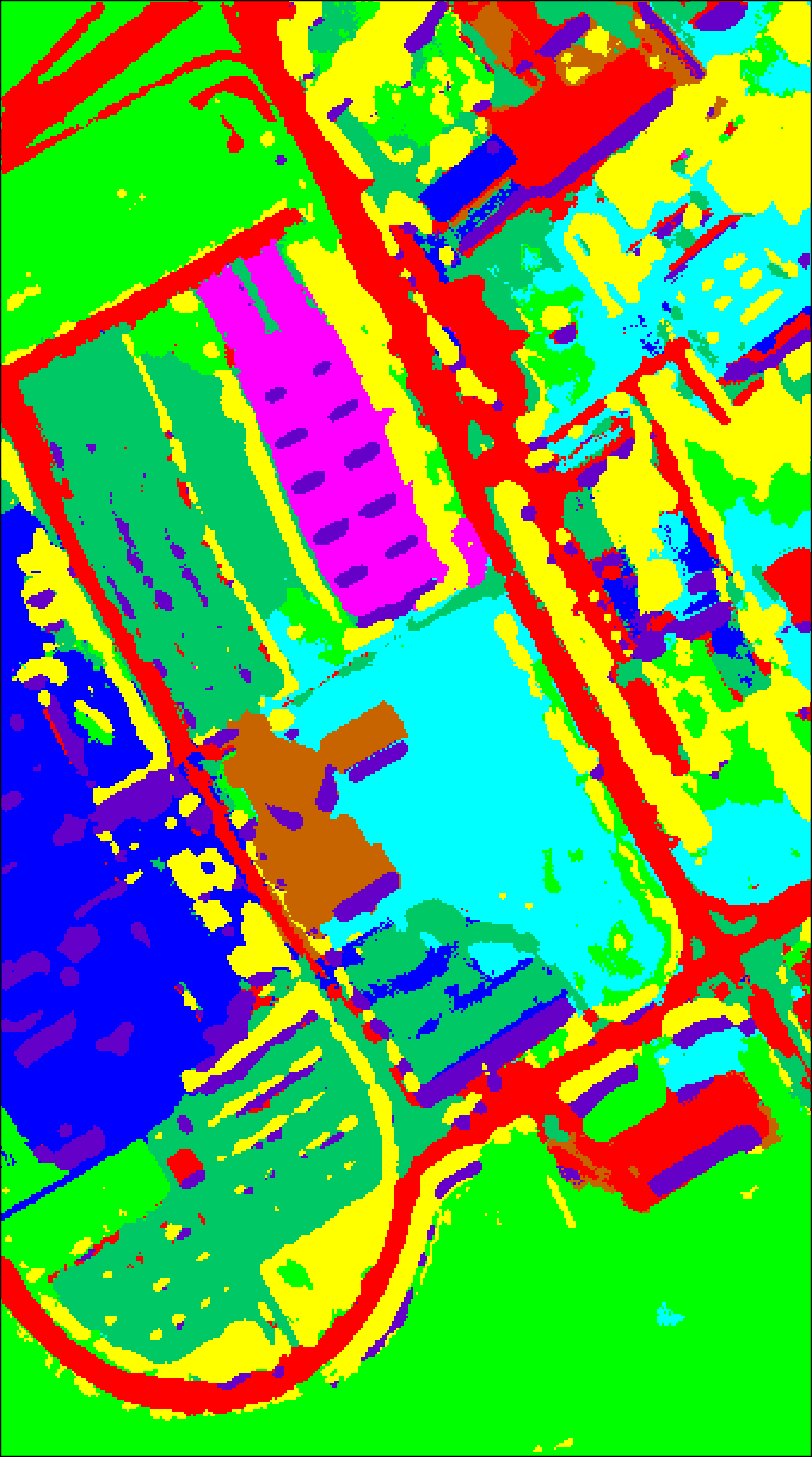}}
        \subfigure[\cred{CEGCN$_{\text{WPX}}$, 20dB}]{\includegraphics[width=6em]{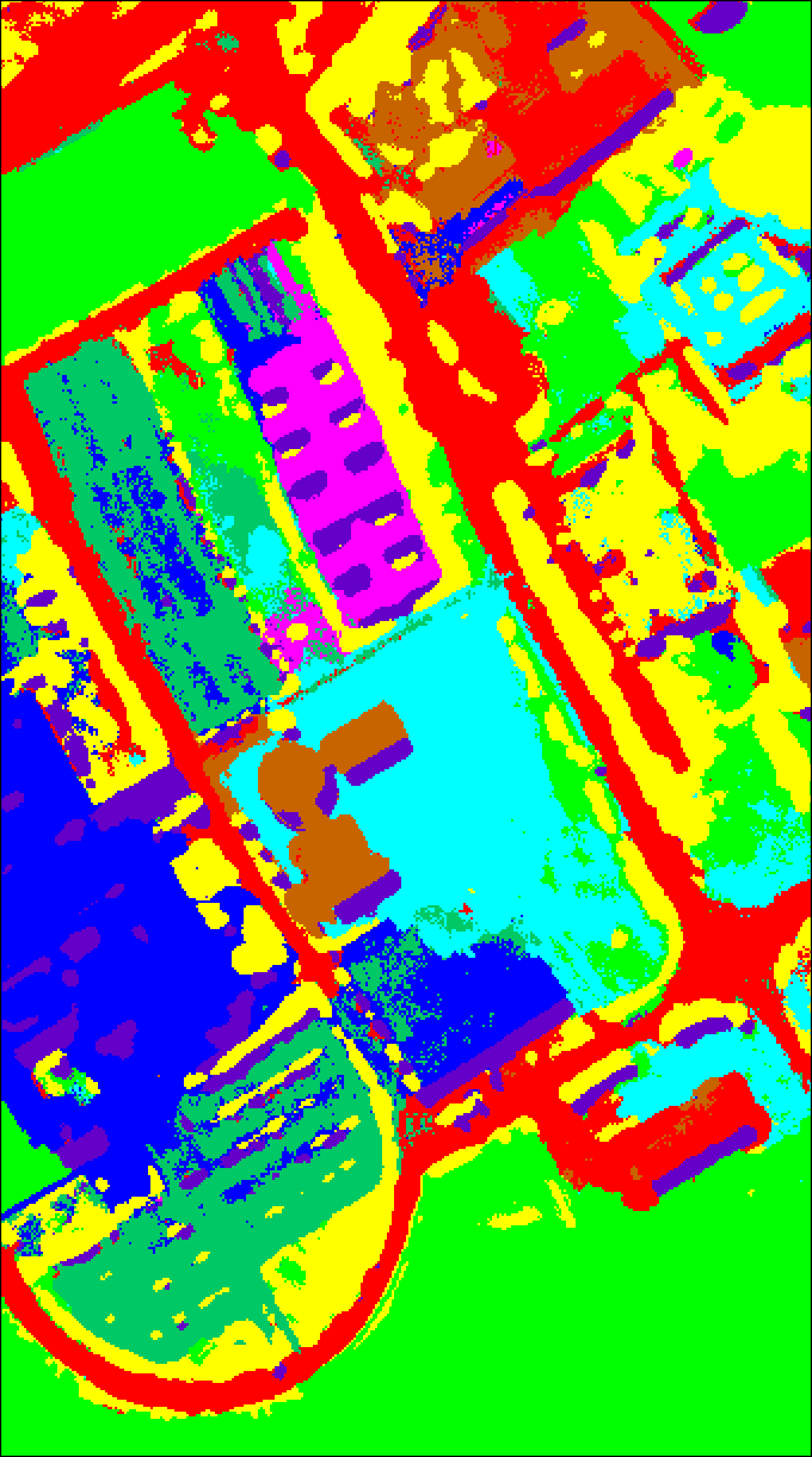}}

        \subfigure[CEGCN$_{\text{H$^2$BO}}$]{\includegraphics[width=6em]{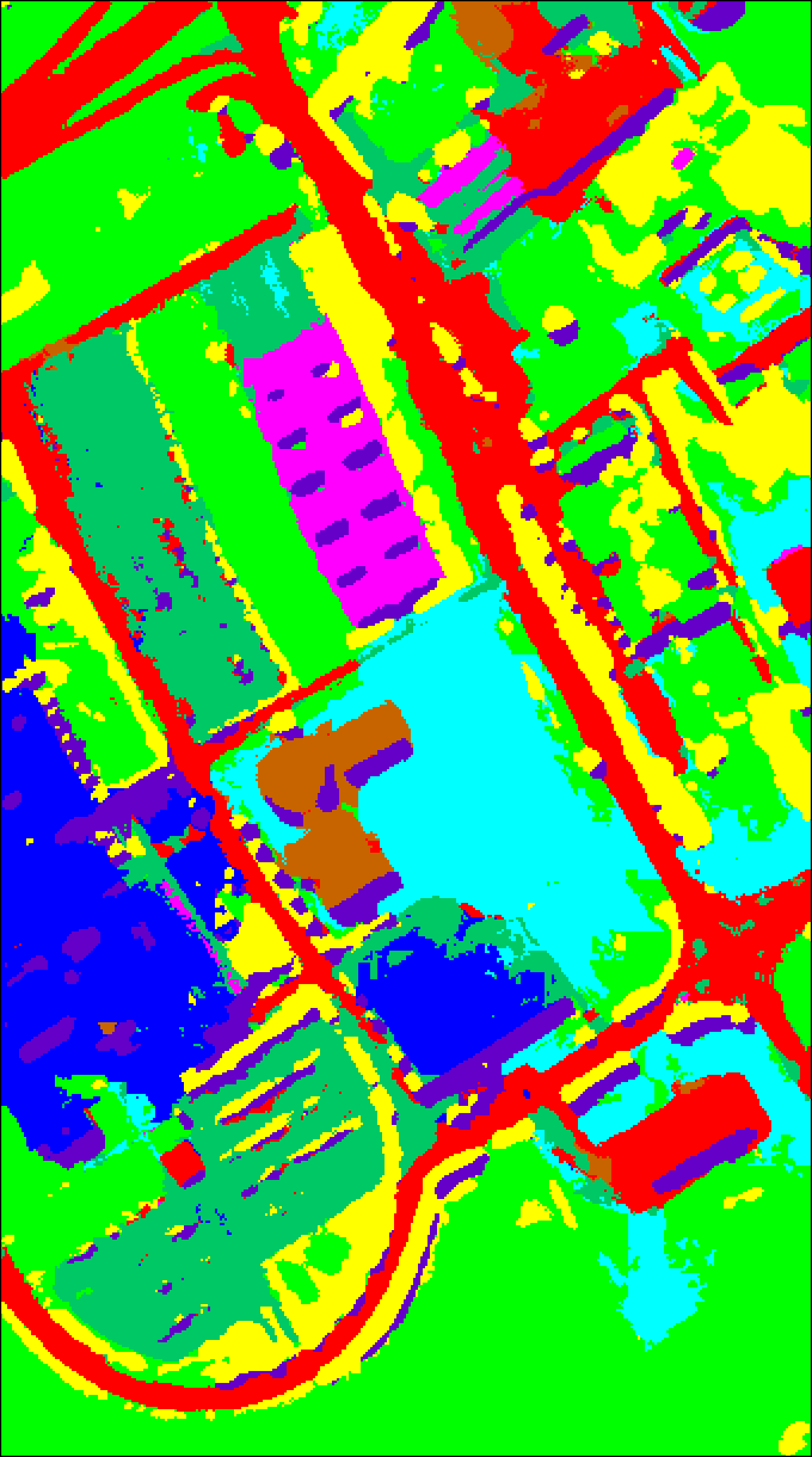}}
        \subfigure[CEGCN$_{\text{H$^2$BO}}$, 30dB]{\includegraphics[width=6em]{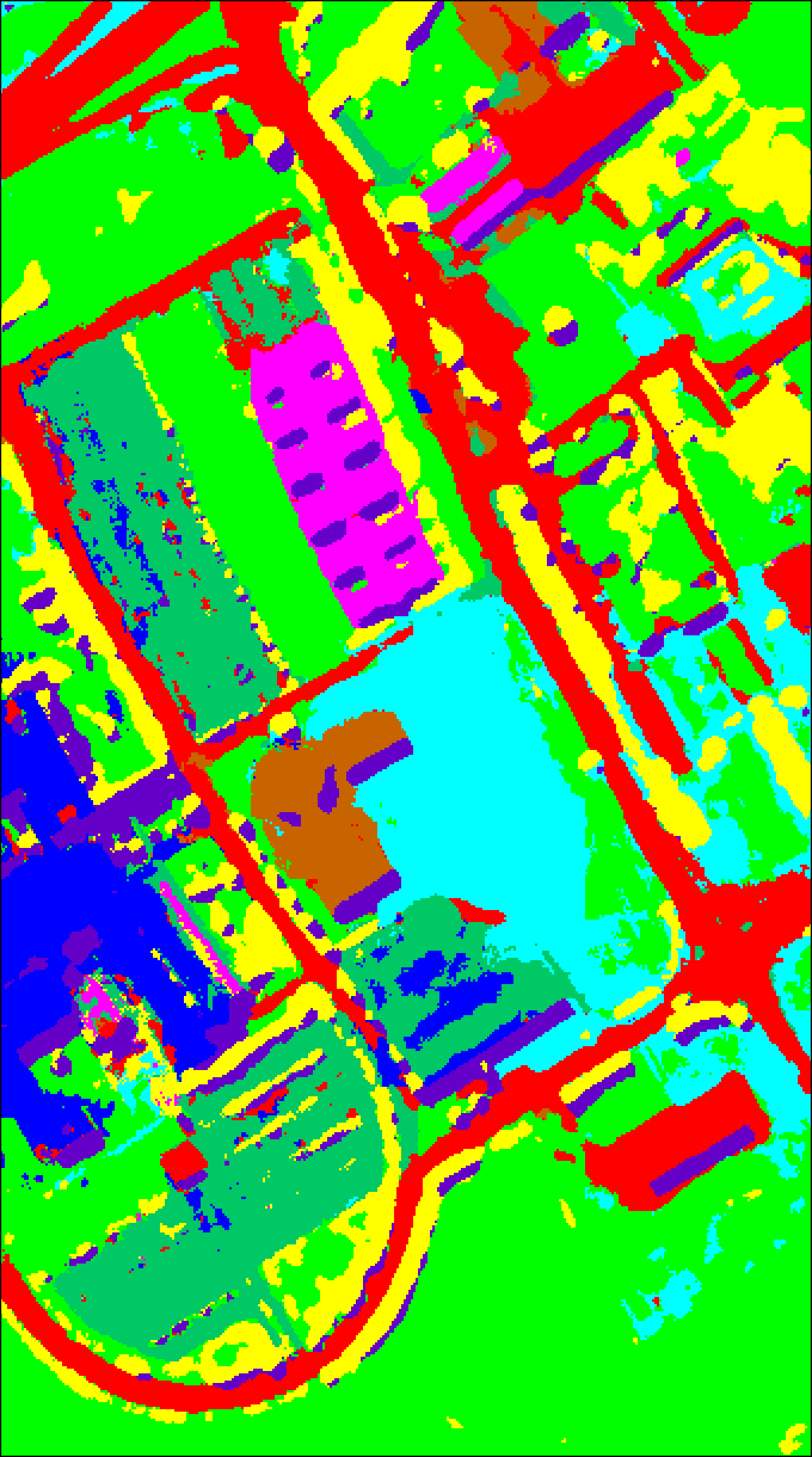}}
        \subfigure[CEGCN$_{\text{H$^2$BO}}$, 20dB]{\includegraphics[width=6em]{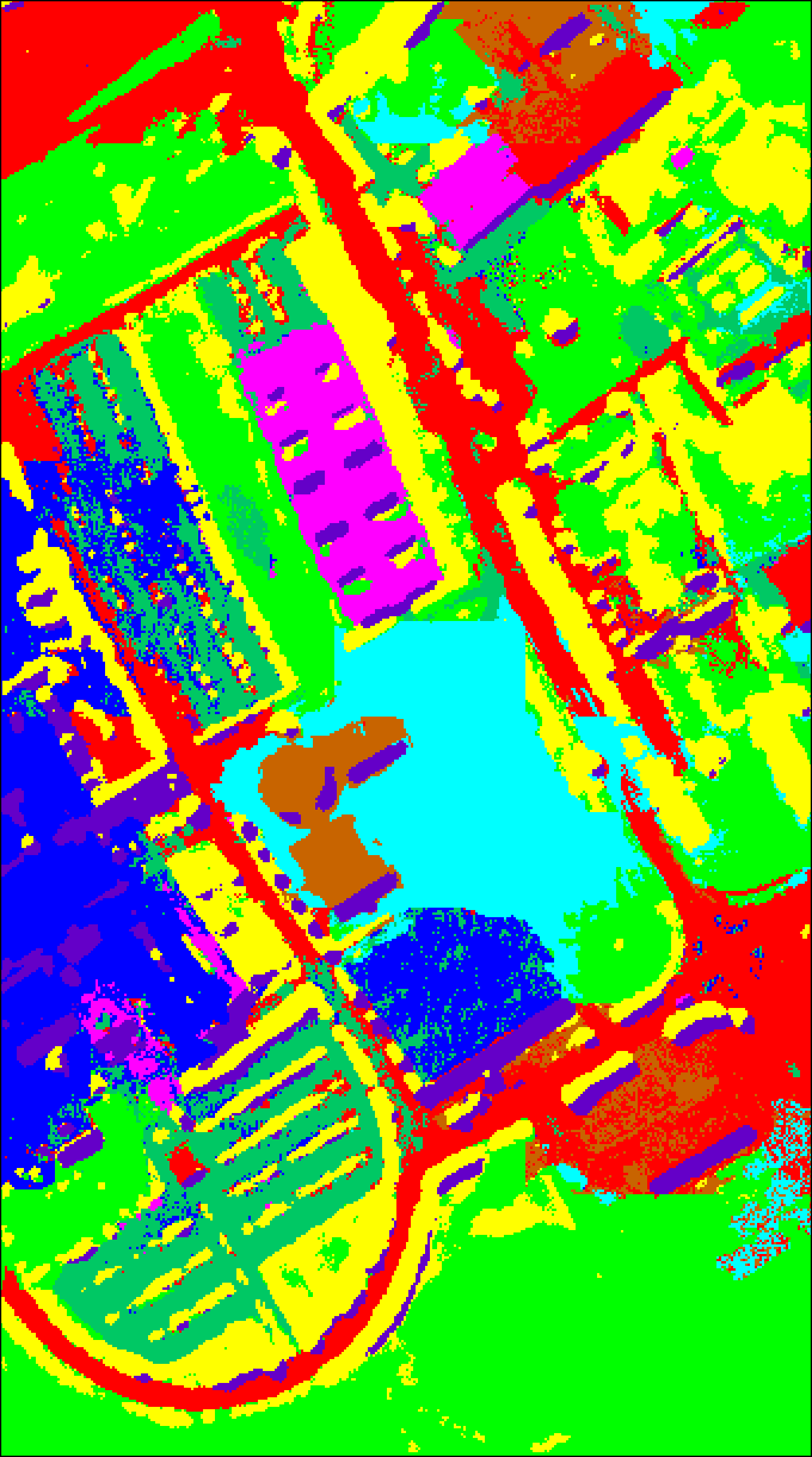}}

	\caption{Pavia University -- false-colour representation, ground-truth and classification maps.}
	\label{fig:class_maps_paviaU}
\end{figure}

The last dataset, Pavia University, is an urban scene of Pavia, Italy, with a size of 610$\times$340 pixels obtained by the ROSIS sensor, with 103 bands (after removal of noisy and water absorption bands) with wavelengths between 430 and 860~nm. The false-colour image and ground-truth of the scene with the 9 annotated classes are shown in Figures \ref{fig:class_maps_paviaU}a and \ref{fig:class_maps_paviaU}b, respectively. The resulting classification maps are given in Figures \ref{fig:class_maps_paviaU}c-h.

Tables \ref{tab:class_results_indian}, \ref{tab:class_results_salinas} and \ref{tab:class_results_paviaU} show the results of the OA, AA and KPP metrics obtained by applying the classification algorithms to the Indian Pines, Salinas and Pavia University datasets, respectively. 
Analysing these quantitative results, the classification maps and their differences for the ground-truth, it can be seen that the accuracies of both CEGCN$_{\text{SLIC}}$\cred{, CEGCN$_{\text{WPX}}$} and CEGCN$_{\text{H$^2$BO}}$ \cred{were similar between the different algorithms and SNR scenarios.}
\cred{The CEGCN$_{\text{WPX}}$ obtained better results for the Salinas dataset, given that the image has large regions with a very regular shape, which could benefit from a less elaborate approach to segmentation than \cblue{the one used in} \text{H$^2$BO}. However, for datasets with more complex and irregular shapes, the CEGCN$_{\text{H$^2$BO}}$ approach may be more suitable.}
\cred{When observing the classification maps of Figures \ref{fig:class_maps_indian}, \ref{fig:class_maps_salinas} and \ref{fig:class_maps_paviaU}, it is difficult to evaluate the influence of different superpixel segmentation methods on the performance of the CEGCN algorithm, particularly due to the non-contiguous nature of the available ground truth labels.}
\cred{Since the annotated classes are not spatially contiguous, they do not allow for an evaluation in the challenging condition of classifying the pixels located at the boundary between classes, where the use of the proposed hierarchical homogeneity-based strategy could provide more effective and robust results in CEGCN$_{\text{H$^2$BO}}$. This highlights the importance of simulations with synthetic data.}

\begin{table}[!b]
\addtolength{\tabcolsep}{-3pt}
\scriptsize
	\tbl{Classification results of CEGCN$_{\text{SLIC}}$\cred{, $\text{CEGCN}_{\text{WPX}}$ algorithms} and the proposed CEGCN$_{\text{H$^2$BO}}$ on the Indian Pines data set \cred{-- Training samples: $15$}.}{
\begin{tabular}{cccc|ccc|ccc}
\toprule
\multicolumn{10}{c}{Indian Pines dataset} \\
\midrule
\multirow{2}{*}{Metric} & \multicolumn{3}{c|}{CEGCN$_{\text{SLIC}}$} & \multicolumn{3}{c|}{\cred{$\text{CEGCN}_{\text{WPX}}$}} & \multicolumn{3}{c}{CEGCN$_{\text{H$^2$BO}}$} \\
& Noiseless & 30 dB & 20 dB & \cred{Noiseless} & \cred{30 dB} & \cred{20 dB} & Noiseless & 30 dB & 20 dB\\
\midrule
OA(\%)& {88.6}\textpm{2.0} & {79.1}\textpm{2.5} & {71.2}\textpm{2.3} & {89.5}\textpm{1.2} & {80.8}\textpm{1.6} & {71.4}\textpm{2.4} & {88.7}\textpm{1.5} & {77.3}\textpm{1.8} & {71.6}\textpm{3.9} \\
AA(\%)& {94.0}\textpm{0.5} & {88.0}\textpm{1.1} & {83.2}\textpm{1.2} & {94.7}\textpm{0.4} & {89.1}\textpm{1.2} & {82.5}\textpm{1.1} & {93.6}\textpm{0.8} & {86.4}\textpm{1.6} & {81.6}\textpm{2.3} \\
KPP(x100) & {87.0}\textpm{2.2} & {76.2}\textpm{2.8} & {67.5}\textpm{2.3} & {88.1}\textpm{1.3} & {78.2}\textpm{1.7} & {67.5}\textpm{2.7} & {87.2}\textpm{1.7} & {74.1}\textpm{2.0} & {67.5}\textpm{4.3}\\
\bottomrule
\end{tabular}}
\label{tab:class_results_indian}
\end{table}

\begin{table}[!htb]
\addtolength{\tabcolsep}{-3pt}
\scriptsize
	\tbl{Classification results of CEGCN$_{\text{SLIC}}$\cred{, $\text{CEGCN}_{\text{WPX}}$ algorithms} and the proposed CEGCN$_{\text{H$^2$BO}}$ on the Salinas data set \cred{-- Training samples: $5$}.}{
\begin{tabular}{cccc|ccc|ccc}
\toprule
\multicolumn{10}{c}{Salinas dataset} \\
\midrule
\multirow{2}{*}{Metric} & \multicolumn{3}{c|}{CEGCN$_{\text{SLIC}}$} & \multicolumn{3}{c|}{\cred{$\text{CEGCN}_{\text{WPX}}$}} & \multicolumn{3}{c}{CEGCN$_{\text{H$^2$BO}}$} \\
& Noiseless & 30 dB & 20 dB & \cred{Noiseless} & \cred{30 dB} & \cred{20 dB} & Noiseless & 30 dB & 20 dB\\
\midrule
OA(\%)& {96.4}\textpm{1.0} & {90.9}\textpm{4.0} & {84.6}\textpm{3.5} & {99.1}\textpm{0.5} & {91.3}\textpm{1.5} & {90.0}\textpm{5.0} & {95.9}\textpm{1.4} & {84.6}\textpm{3.4} & {85.5}\textpm{3.4} \\
AA(\%)  & {98.0}\textpm{0.7} & {94.2}\textpm{1.8} & {83.6}\textpm{1.4} & {98.8}\textpm{0.5} & {94.0}\textpm{1.3} & {92.2}\textpm{3.7} & {97.5}\textpm{0.7} & {85.5}\textpm{2.0} & {88.0}\textpm{2.0} \\
KPP(x100) & {96.0}\textpm{1.1} & {89.9}\textpm{4.3} & {83.0}\textpm{3.8} & {99.0}\textpm{0.5} & {90.3}\textpm{1.6} & {88.8}\textpm{5.6} & {95.5}\textpm{1.6} & {90.7}\textpm{2.1} & {83.9}\textpm{3.7}\\
\bottomrule
\end{tabular}}
\label{tab:class_results_salinas}
\end{table}

\begin{table}[!htb]
\addtolength{\tabcolsep}{-3pt}
\scriptsize
	\tbl{Classification results of CEGCN$_{\text{SLIC}}$\cred{, $\text{CEGCN}_{\text{WPX}}$ algorithms} and the proposed CEGCN$_{\text{H$^2$BO}}$ on the Pavia University data set \cred{-- Training samples: $10$}.}{
\begin{tabular}{cccc|ccc|ccc}
\toprule
\multicolumn{10}{c}{Pavia University dataset} \\
\midrule
\multirow{2}{*}{Metric} & \multicolumn{3}{c|}{CEGCN$_{\text{SLIC}}$} & \multicolumn{3}{c|}{\cred{$\text{CEGCN}_{\text{WPX}}$}} & \multicolumn{3}{c}{CEGCN$_{\text{H$^2$BO}}$} \\
& Noiseless & 30 dB & 20 dB & \cred{Noiseless} & \cred{30 dB} & \cred{20 dB} & Noiseless & 30 dB & 20 dB\\
\midrule
OA(\%)& {94.3}\textpm{1.6} & {93.6}\textpm{2.1} & {91.0}\textpm{3.0} & {95.6}\textpm{2.0} & {96.3}\textpm{1.1} & {92.9}\textpm{1.7} & {95.5}\textpm{1.3} & {95.9}\textpm{2.1} & {93.2}\textpm{2.6} \\
AA(\%)& {96.9}\textpm{0.6} & {95.9}\textpm{0.4} & {94.8}\textpm{1.6} & {97.2}\textpm{0.8} & {97.7}\textpm{0.3} & {95.9}\textpm{0.9} & {97.6}\textpm{0.5} & {97.2}\textpm{1.0} & {95.0}\textpm{2.0} \\
KPP(x100) & {92.6}\textpm{2.0} & {91.6}\textpm{2.6} & {88.4}\textpm{3.8} & {94.3}\textpm{2.5} & {95.2}\textpm{1.4} & {90.8}\textpm{2.1} & {94.1}\textpm{1.7} & {94.6}\textpm{2.7} & {91.1}\textpm{3.4}\\
\bottomrule
\end{tabular}}
\label{tab:class_results_paviaU}
\end{table}

\begin{table}[!htb]
\tbl{\credd{Average training time (in seconds) of the classification algorithms and their superpixel segmentation step.}}{
\begin{tabular}{ccc|cc|cc}
\toprule
\multirow{2}{*}{Dataset} & \multicolumn{2}{c|}{CEGCN$_{\text{SLIC}}$} & \multicolumn{2}{c|}{$\text{CEGCN}_{\text{WPX}}$} & \multicolumn{2}{c}{CEGCN$_{\text{H$^2$BO}}$}  \\ [0.2cm]
& Total & SLIC & Total & WPX & Total & \text{H$^2$BO} \\
\midrule
DC3              &  57.04 & 0.04 &  54.25 & 0.25 &  73.25 &  1.25 \\
Indian Pines     &  56.09 & 0.09 &  55.51 & 0.51 &  57.23 &  6.23 \\
Salinas          & 277.57 & 0.50 & 328.73 & 2.73 & 169.21 & 28.21 \\
Pavia University & 575.52 & 0.52 & 565.52 & 4.52 & 500.07 & 95.07 \\     
\bottomrule
\end{tabular}}
\label{tab:class_exec_time}
\end{table}

\credd{The classification methods were also compared in terms of average training time and the time of the superpixel generation step. From the data presented in Table~\ref{tab:class_exec_time}, it is noticeable that SLIC is the fastest in the segmentation stage, but this does not necessarily result in reduced training times.
On the other hand, although the proposed H$^2$BO presents the longest time in the superpixel generation process, its ability to represent scales with different superpixel sizes can lead to faster convergence and a reduction in total training time. This effect becomes clearer when the dimensions of the images are larger, as in the case of the Salinas and Pavia University datasets.
Although the extra rounds of segmentation and homogeneity assessment can impact the training time for small images, it still remains comparable to those of the other algorithms. 
As for waterpixels, their performance is in the middle ground between SLIC and H$^2$BO, still proving to be an efficient algorithm in general.
}

\cred{The obtained results show that the proposed method leads to improved classification performance when the regions to be classified have a more homogeneous composition, and when areas corresponding to different classes tend to have significantly different sizes.  This is the expected performance, as the proposed algorithm was designed with the aim of identifying homogeneous regions of the image, and to allow significantly different numbers of pixels for different superpixels.  Of course, not all hyperspectral images will have these characteristics.  Another interesting characteristic of the proposed algorithm was that it has led to a smaller performance degradation as the image SNR is reduced when compared with the competing algorithms.}
\cred{These characteristics were illustrated more clearly in the simulation results using synthetic images, where we were able to test different relative sizes of class regions and different degrees of homogeneity within each class.  For the real images, the results are more difficult to interpret since the pixels belonging to single class are not necessarily spectrally homogeneous (e.g., it is possible that a large region assigned to a given class is composed of pixels with large variability). In such cases, the use of superpixels decomposition algorithms based on the spatial-spectral homogeneity property, such as SLIC and the proposed H$^2$BO, might be inadequate to aid in a classification task. 
}

\cred{Finally, the performance of the proposed algorithm was still on par when compared with the competing algorithms even when the conditions for which it was designed were not satisfied. This illustrates the important robustness of the algorithm, making it useful for hyperspectral imaging tasks.}

\section{Conclusions} \label{sec:conclusions}

In this paper, we presented a novel hierarchical superpixels segmentation technique for hyperspectral image analysis. We progressively split an HI into irregular superpixels with increased spectral homogeneity, which better depict the spatial information of the materials in a scene, using the SLIC oversegmentation algorithm and an unique robust homogeneity testing approach. A multiscale methodology is then applied to the final superpixel decomposition to incorporate spatial information into the sparse unmixing and convolutional graph neural networks-based classification challenges. The results of experiments using hyperspectral data with different spatial compositions demonstrated that, in noisy situations, homogeneity-based approaches outperform state-of-the-art algorithms while preserving computational complexity.

\bibliographystyle{tfcad}
\bibliography{library, libref}

\begin{thebibliography}{86}
\newcommand{\enquote}[1]{``#1''}
\providecommand{\natexlab}[1]{#1}
\providecommand{\url}[1]{\normalfont{#1}}
\providecommand{\urlprefix}{}

\bibitem[Acci{\'o}n, Arg{\"u}ello, and Heras(2020)]{accion2020dual}
Acci{\'o}n, {\'A}lvaro, Francisco Arg{\"u}ello, and Dora~B Heras. 2020.
  ``Dual-window superpixel data augmentation for hyperspectral image
  classification.'' \emph{Applied Sciences} 10 (24): 8833.

\bibitem[Achanta et~al.(2012)]{achanta2012slic}
Achanta, Radhakrishna, A.~Shaji, K.~Smith, A.~Lucchi, P.~Fua, and Sabine
  S{\"{u}}sstrunk. 2012. ``{SLIC Superpixels Compared to State-of-the-Art
  Superpixel Methods}.'' \emph{IEEE Transactions on Pattern Analysis and
  Machine Intelligence} 34 (11): 2274--2282.
  https://doi.org/{10.1109/TPAMI.2012.120}.

\bibitem[Arg{\"u}ello et~al.(2021)]{arguello2021watershed}
Arg{\"u}ello, Francisco, Dora~B Heras, Alberto~S Garea, and Pablo
  Quesada-Barriuso. 2021. ``Watershed monitoring in galicia from UAV
  multispectral imagery using advanced texture methods.'' \emph{Remote Sensing}
  13 (14): 2687.

\bibitem[Audebert, Le~Saux, and Lef{\`e}vre(2019)]{audebert2019}
Audebert, Nicolas, Bertrand Le~Saux, and S{\'e}bastien Lef{\`e}vre. 2019.
  ``Deep learning for classification of hyperspectral data: A comparative
  review.'' \emph{IEEE geoscience and remote sensing magazine} 7 (2): 159--173.

\bibitem[Ayres et~al.(2021)]{ayres2021}
Ayres, L.~C., Sergio~J.M. {De Almeida}, Jose~C.M. Bermudez, and Ricardo~A.
  Borsoi. 2021. ``{A Homogeneity-based Multiscale Hyperspectral Image
  Representation for Sparse Spectral Unmixing}.'' In \emph{ICASSP, IEEE
  International Conference on Acoustics, Speech and Signal Processing -
  Proceedings}, Vol. 2021-June, 1460--1464. Institute of Electrical and
  Electronics Engineers Inc.

\bibitem[Ayres et~al.(2024)]{ayres2024generalizedMUA}
Ayres, Luciano~C, Ricardo~A Borsoi, Jos{\'e}~CM Bermudez, and S{\'e}rgio~JM
  De~Almeida. 2024. ``A Generalized Multiscale Bundle-Based Hyperspectral
  Sparse Unmixing Algorithm.'' \emph{IEEE Geoscience and Remote Sensing
  Letters} 21.

\bibitem[Beucher and Meyer(1993)]{Beucher1993}
Beucher, S, and F~Meyer. 1993. ``{The Morphological Approach to Segmentation:
  The Watershed Transformation}.'' In \emph{Mathematical Morphology in Image
  Processing},  edited by E.R Dougherty, 433--481. CRC Press.

\bibitem[Bioucas-dias et~al.(2013)]{Bioucas-dias2013}
Bioucas-dias, Jos{\'{e}}~M, Antonio Plaza, Gustavo Camps-valls, Paul
  Scheunders, Nasser~M Nasrabadi, and Jocelyn Chanussot. 2013. ``{Hyperspectral
  Remote Sensing Data Analysis and Future Challenges}.'' \emph{IEEE Geoscience
  and Remote Sensing Magazine} 1 (June): 6--36.
  https://doi.org/{10.1109/MGRS.2013.2244672}.

\bibitem[Bioucas-Dias et~al.(2012)]{Bioucas-Dias2012}
Bioucas-Dias, Jos{\'{e}}~M., Antonio Plaza, Nicolas Dobigeon, Mario Parente,
  Qian Du, Paul Gader, and Jocelyn Chanussot. 2012. ``{Hyperspectral unmixing
  overview: Geometrical, statistical, and sparse regression-based
  approaches}.'' \emph{IEEE Journal of Selected Topics in Applied Earth
  Observations and Remote Sensing} 5 (2): 354--379.
  https://doi.org/{10.1109/JSTARS.2012.2194696}.

\bibitem[Bischof and Leonardis(1998)]{bischof1998}
Bischof, Horst, and Ales Leonardis. 1998. ``Finding optimal neural networks for
  land use classification.'' \emph{IEEE transactions on Geoscience and Remote
  Sensing} 36 (1): 337--341.

\bibitem[Borsoi, Imbiriba, and Bermudez(2020)]{Borsoi2020Data}
Borsoi, Ricardo~Augusto, Tales Imbiriba, and Jose Carlos~Moreira Bermudez.
  2020. ``{A Data Dependent Multiscale Model for Hyperspectral Unmixing with
  Spectral Variability}.'' \emph{IEEE Transactions on Image Processing} 29:
  3638--3651. https://doi.org/{10.1109/TIP.2020.2963959}.

\bibitem[Borsoi et~al.(2019)]{Borsoi2019}
Borsoi, Ricardo~Augusto, Tales Imbiriba, Jos{\'{e}} Carlos~Moreira Bermudez,
  and Cedric Richard. 2019. ``{A Fast Multiscale Spatial Regularization for
  Sparse Hyperspectral Unmixing}.'' \emph{IEEE Geoscience and Remote Sensing
  Letters} 16 (4): 598--602. https://doi.org/{10.1109/LGRS.2018.2878394}.

\bibitem[Borsoi et~al.(2020)]{Borsoi2019a}
Borsoi, Ricardo~Augusto, Tales Imbiriba, Jos{\'{e}} Carlos~Moreira Bermudez,
  and C{\'{e}}dric Richard. 2020. ``{A Blind Multiscale Spatial Regularization
  Framework for Kernel-based Spectral Unmixing}.'' \emph{IEEE Transactions on
  Image Processing} 29: 4965--4979.

\bibitem[Borsoi et~al.(2021)]{borsoi2021spectralVariabilityReview}
Borsoi, Ricardo~Augusto, Tales Imbiriba, Jos{\'e} Carlos~Moreira Bermudez,
  C{\'e}dric Richard, Jocelyn Chanussot, Lucas Drumetz, Jean-Yves Tourneret,
  Alina Zare, and Christian Jutten. 2021. ``Spectral variability in
  hyperspectral data unmixing: A comprehensive review.'' \emph{IEEE geoscience
  and remote sensing magazine} 9 (4): 223--270.

\bibitem[Cao
  et~al.(2023{\natexlab{a}})]{cao2023semiSupervisedDisjointClassification}
Cao, Xianghai, Chenguang Li, Jie Feng, and Licheng Jiao. 2023{\natexlab{a}}.
  ``Semi-supervised feature learning for disjoint hyperspectral imagery
  classification.'' \emph{Neurocomputing} 526: 9--18.

\bibitem[Cao
  et~al.(2023{\natexlab{b}})]{cao2023transformerContrastiveLossClassification}
Cao, Xianghai, Haifeng Lin, Shuaixu Guo, Tao Xiong, and Licheng Jiao.
  2023{\natexlab{b}}. ``Transformer-based masked autoencoder with contrastive
  loss for hyperspectral image classification.'' \emph{IEEE Transactions on
  Geoscience and Remote Sensing} .

\bibitem[Chen et~al.(2022)]{chen2022superpixel}
Chen, Tao, Yang Liu, Yuxiang Zhang, Bo~Du, and Antonio Plaza. 2022.
  ``Superpixel-Based Collaborative and Low-Rank Regularization for Sparse
  Hyperspectral Unmixing.'' \emph{IEEE Transactions on Geoscience and Remote
  Sensing} .

\bibitem[Cihan, Ceylan, and Ornek(2022)]{cihan2022spectral}
Cihan, Mucahit, Murat Ceylan, and Ahmet~Haydar Ornek. 2022. ``Spectral-spatial
  classification for non-invasive health status detection of neonates using
  hyperspectral imaging and deep convolutional neural networks.''
  \emph{Spectroscopy Letters} 55 (5): 336--349.

\bibitem[Civco(1993)]{civco1993}
Civco, Daniel~L. 1993. ``Artificial neural networks for land-cover
  classification and mapping.'' \emph{International journal of geographical
  information science} 7 (2): 173--186.

\bibitem[Di et~al.(2021)]{liao2021}
Di, Shuanhu, Miao Liao, Yuqian Zhao, Yang Li, and Yezhan Zeng. 2021. ``Image
  superpixel segmentation based on hierarchical multi-level LI-SLIC.''
  \emph{Optics and Laser Technology} 135: 106703.
  https://doi.org/{https://doi.org/10.1016/j.optlastec.2020.106703}.

\bibitem[Dieste, Arg{\"u}ello, and Heras(2023)]{dieste2023resbagan}
Dieste, {\'A}lvaro~G, Francisco Arg{\"u}ello, and Dora~B Heras. 2023.
  ``ResBaGAN: A Residual Balancing GAN with Data Augmentation for Forest
  Mapping.'' \emph{IEEE Journal of Selected Topics in Applied Earth
  Observations and Remote Sensing} .

\bibitem[Ding et~al.(2022)]{DING2022246}
Ding, Yao, Zhili Zhang, Xiaofeng Zhao, Danfeng Hong, Wei Cai, Chengguo Yu,
  Nengjun Yang, and Weiwei Cai. 2022. ``Multi-feature fusion: Graph neural
  network and CNN combining for hyperspectral image classification.''
  \emph{Neurocomputing} 501: 246--257.
  https://doi.org/{https://doi.org/10.1016/j.neucom.2022.06.031}.

\bibitem[Dobigeon et~al.(2014)]{Dobigeon2014}
Dobigeon, Nicolas, Jean-Yves Tourneret, Cedric Richard, Jos{\'{e}}
  Carlos~Moreira Bermudez, Stephen McLaughlin, and Alfred~O. Hero. 2014.
  ``{Nonlinear Unmixing of Hyperspectral Images: Models and Algorithms}.''
  \emph{IEEE Signal Processing Magazine} 31 (1): 82--94.
  https://doi.org/{10.1109/MSP.2013.2279274}.

\bibitem[Felzenszwalb and Huttenlocher(2004)]{Felzenszwalb2004}
Felzenszwalb, Pedro~F, and Daniel~P Huttenlocher. 2004. ``{Efficient
  Graph-Based Image Segmentation}.'' \emph{International Journal of Computer
  Vision} 59 (2): 167--181.
  https://doi.org/{10.1023/B:VISI.0000022288.19776.77}.

\bibitem[Halicek et~al.(2019)]{halicek2019cancer}
Halicek, Martin, Himar Fabelo, Samuel Ortega, James~V Little, Xu~Wang, Amy~Y
  Chen, Gustavo~Marrero Callico, Larry~L Myers, Baran~D Sumer, and Baowei Fei.
  2019. ``Cancer detection using hyperspectral imaging and evaluation of the
  superficial tumor margin variance with depth.'' In \emph{Medical Imaging
  2019: Image-Guided Procedures, Robotic Interventions, and Modeling}, Vol.
  10951, 109511A. International Society for Optics and Photonics.

\bibitem[He et~al.(2018)]{He2018}
He, Lin, Jun Li, Chenying Liu, and Shutao Li. 2018. ``{Recent Advances on
  Spectral–Spatial Hyperspectral Image Classification: An Overview and New
  Guidelines}.'' \emph{IEEE Transactions on Geoscience and Remote Sensing} 56
  (3): 1579--1597. https://doi.org/{10.1109/TGRS.2017.2765364}.

\bibitem[Ince(2020)]{Ince2020}
Ince, Taner. 2020. ``{Superpixel-Based Graph Laplacian Regularization for
  Sparse Hyperspectral Unmixing}.'' \emph{IEEE Geoscience and Remote Sensing
  Letters} 1--5. https://doi.org/{10.1109/lgrs.2020.3027055}.

\bibitem[Ince(2021)]{Ince2021}
Ince, Taner. 2021. ``{Double Spatial Graph Laplacian Regularization for Sparse
  Unmixing}.'' \emph{IEEE Geoscience and Remote Sensing Letters} 1--5.
  https://doi.org/{10.1109/LGRS.2021.3065989}.

\bibitem[Iordache, Bioucas-Dias, and Plaza(2011)]{Iordache2011}
Iordache, Marian-Daniel, J~M Bioucas-Dias, and Antonio Plaza. 2011. ``{Sparse
  Unmixing of Hyperspectral Data}.'' \emph{IEEE Trans. Geosc. Rem. Sens.} 49
  (6): 2014--2039. https://doi.org/{10.1109/TGRS.2010.2098413}.

\bibitem[Iordache, Bioucas-Dias, and Plaza(2012)]{Iordache2012}
Iordache, Marian-Daniel, Jos{\'{e}}~M. Bioucas-Dias, and Antonio Plaza. 2012.
  ``{Total variation spatial regularization for sparse hyperspectral
  unmixing}.'' \emph{IEEE Transactions on Geoscience and Remote Sensing} 50 (11
  PART1): 4484--4502. https://doi.org/{10.1109/TGRS.2012.2191590}.

\bibitem[Iordache, Bioucas-Dias, and Plaza(2014)]{Iordache2014}
Iordache, Marian-Daniel, Jose~M. Bioucas-Dias, and Antonio Plaza. 2014.
  ``{Collaborative Sparse Regression for Hyperspectral Unmixing}.'' \emph{IEEE
  Transactions on Geoscience and Remote Sensing} 52 (1): 341--354.
  https://doi.org/{10.1109/TGRS.2013.2240001}.

\bibitem[Iordache, Plaza, and Bioucas-Dias(2010)]{Iordache2010}
Iordache, Marian-Daniel, Antonio Plaza, and Jos{\'{e}} Bioucas-Dias. 2010.
  ``{On the use of spectral libraries to perform sparse unmixing of
  hyperspectral data}.'' \emph{2nd Workshop on Hyperspectral Image and Signal
  Processing: Evolution in Remote Sensing, WHISPERS 2010 - Workshop Program}
  1--4. https://doi.org/{10.1109/WHISPERS.2010.5594888}.

\bibitem[Jampani et~al.(2018)]{jampani2018superpixelNetworks}
Jampani, Varun, Deqing Sun, Ming-Yu Liu, Ming-Hsuan Yang, and Jan Kautz. 2018.
  ``Superpixel sampling networks.'' In \emph{Proceedings of the European
  Conference on Computer Vision (ECCV)}, 352--368.

\bibitem[Keshava and Mustard(2002)]{Keshava2002}
Keshava, Nirmal, and J.F. Mustard. 2002. ``{Spectral unmixing}.'' \emph{IEEE
  Signal Processing Magazine} 19 (1): 44--57.
  https://doi.org/{10.1109/79.974727}.

\bibitem[Kotzagiannidis and Schönlieb(2022)]{Kotzagiannidis2021a}
Kotzagiannidis, Madeleine~S., and Carola-Bibiane Schönlieb. 2022.
  ``Semi-Supervised Superpixel-Based Multi-Feature Graph Learning for
  Hyperspectral Image Data.'' \emph{IEEE Transactions on Geoscience and Remote
  Sensing} 60: 1--12. https://doi.org/{10.1109/TGRS.2021.3112298}.

\bibitem[Kumar, Singh, and Dua(2022)]{Kumar2022}
Kumar, Vinod, Ravi~Shankar Singh, and Yaman Dua. 2022. ``Morphologically
  dilated convolutional neural network for hyperspectral image
  classification.'' \emph{Signal Processing: Image Communication} 101: 116549.
  https://doi.org/{https://doi.org/10.1016/j.image.2021.116549}.

\bibitem[Landgrebe(2002)]{Landgrebe2002}
Landgrebe, David. 2002. ``{Hyperspectral image data analysis}.'' \emph{IEEE
  Signal Processing Magazine} 19 (1): 17--28.
  https://doi.org/{10.1109/79.974718}.

\bibitem[Levinshtein et~al.(2009)]{Levinshtein2009}
Levinshtein, Alex, Adrian Stere, Kiriakos~N. Kutulakos, David~J. Fleet, Sven~J.
  Dickinson, and Kaleem Siddiqi. 2009. ``{TurboPixels: Fast superpixels using
  geometric flows}.'' \emph{IEEE Transactions on Pattern Analysis and Machine
  Intelligence} 31 (12): 2290--2297. https://doi.org/{10.1109/TPAMI.2009.96}.

\bibitem[Li et~al.(2023{\natexlab{a}})]{li2023hyperspectral}
Li, Xueying, Pingping Fan, Zongmin Li, Huimin Qiu, Guangli Hou, Guangyuan Chen,
  and Guoxing Ren. 2023{\natexlab{a}}. ``Hyperspectral images classification of
  small sample based on the strategy of sample enlargement by superpixel pair
  method.'' \emph{International Journal of Remote Sensing} 44 (20): 6259--6279.

\bibitem[Li et~al.(2023{\natexlab{b}})]{li2023spectral}
Li, Zhi, Ruyi Feng, Lizhe Wang, and Tieyong Zeng. 2023{\natexlab{b}}.
  ``Spectral-spatial-sparse unmixing with superpixel-oriented graph
  Laplacian.'' \emph{International Journal of Remote Sensing} 44 (8):
  2573--2589.

\bibitem[Liu et~al.(2021)]{Liu2021}
Liu, Qichao, Liang Xiao, Jingxiang Yang, and Zhihui Wei. 2021. ``{CNN-Enhanced
  Graph Convolutional Network with Pixel- And Superpixel-Level Feature Fusion
  for Hyperspectral Image Classification}.'' \emph{IEEE Transactions on
  Geoscience and Remote Sensing} 59 (10): 8657--8671.
  https://doi.org/{10.1109/TGRS.2020.3037361}.

\bibitem[Liu, Pu, and Sun(2017)]{liu2017hyperspectral}
Liu, Yuwei, Hongbin Pu, and Da-Wen Sun. 2017. ``Hyperspectral imaging technique
  for evaluating food quality and safety during various processes: A review of
  recent applications.'' \emph{Trends in food science \& technology} 69:
  25--35.

\bibitem[Ma et~al.(2013)]{ma2013signal}
Ma, Wing-Kin, Jos{\'e}~M Bioucas-Dias, Tsung-Han Chan, Nicolas Gillis, Paul
  Gader, Antonio~J Plaza, ArulMurugan Ambikapathi, and Chong-Yung Chi. 2013.
  ``A signal processing perspective on hyperspectral unmixing: Insights from
  remote sensing.'' \emph{IEEE Signal Processing Magazine} 31 (1): 67--81.

\bibitem[Machairas et~al.(2015)]{machairas2015waterpixels}
Machairas, Va{\"\i}a, Matthieu Faessel, David C{\'a}rdenas-Pe{\~n}a,
  Th{\'e}odore Chabardes, Thomas Walter, and Etienne Decenciere. 2015.
  ``Waterpixels.'' \emph{IEEE Transactions on Image Processing} 24 (11):
  3707--3716.

\bibitem[MacQueen(1967)]{MacQueenJamesandothers1967}
MacQueen, James. 1967. ``{Some methods for classification and analysis of
  multivariate observations}.'' \emph{Proceedings of the fifth Berkeley
  symposium on mathematical statistics and probability} 1 (14): 281--297.

\bibitem[Mei et~al.(2018)]{Mei2018}
Mei, Xiaoguang, Yong Ma, Chang Li, Fan Fan, Jun Huang, and Jiayi Ma. 2018.
  ``{Robust GBM hyperspectral image unmixing with superpixel segmentation based
  low rank and sparse representation}.'' \emph{Neurocomputing} 275: 2783--2797.
  https://doi.org/{10.1016/j.neucom.2017.11.052}.

\bibitem[Melgani and Bruzzone(2004)]{melgani2004}
Melgani, Farid, and Lorenzo Bruzzone. 2004. ``Classification of hyperspectral
  remote sensing images with support vector machines.'' \emph{IEEE Transactions
  on geoscience and remote sensing} 42 (8): 1778--1790.

\bibitem[Mookambiga and Gomathi(2021)]{Mookambiga2021}
Mookambiga, A., and V.~Gomathi. 2021. ``Kernel eigenmaps based multiscale
  sparse model for hyperspectral image classification.'' \emph{Signal
  Processing: Image Communication} 99: 116416.
  https://doi.org/{https://doi.org/10.1016/j.image.2021.116416}.

\bibitem[Naik et~al.(2023)]{naik2023spatio}
Naik, Nitesh, Kandasamy Chandrasekaran, Venkatesan Meenakshi~Sundaram, and
  Prabhavathy Panneer. 2023. ``Spatio-temporal analysis of land use/land cover
  change detection in small regions using self-supervised lightweight deep
  learning.'' \emph{Stochastic Environmental Research and Risk Assessment} 37
  (12): 5029--5049.

\bibitem[Noyel, Angulo, and Jeulin(2020)]{noyel2020morphological}
Noyel, Guillaume, Jesus Angulo, and Dominique Jeulin. 2020. ``Morphological
  segmentation of hyperspectral images.'' \emph{arXiv preprint
  arXiv:2010.00853} .

\bibitem[{P. Ghamisi} et~al.(2017)]{P.Ghamisi2017}
{P. Ghamisi}, {J. Plaza}, {Y. Chen}, {J. Li}, and {A. J. Plaza}. 2017.
  ``{Advanced Spectral Classifiers for Hyperspectral Images: A review}.''
  \emph{IEEE Geoscience and Remote Sensing Magazine} 5 (1): 8--32.

\bibitem[Pu, Wei, and Sun(2023)]{pu2023recent}
Pu, Hongbin, Qingyi Wei, and Da-Wen Sun. 2023. ``Recent advances in muscle food
  safety evaluation: Hyperspectral imaging analyses and applications.''
  \emph{Critical Reviews in Food Science and Nutrition} 63 (10): 1297--1313.

\bibitem[Qi et~al.(2019)]{QI201997}
Qi, Lin, Jie Li, Xinbo Gao, Ying Wang, Chongyue Zhao, and Yu~Zheng. 2019. ``A
  novel joint dictionary framework for sparse hyperspectral unmixing
  incorporating spectral library.'' \emph{Neurocomputing} 356: 97--106.
  https://doi.org/{https://doi.org/10.1016/j.neucom.2019.04.053}.

\bibitem[Qin et~al.(2018)]{qin2018}
Qin, Anyong, Zhaowei Shang, Jinyu Tian, Yulong Wang, Taiping Zhang, and
  Yuan~Yan Tang. 2018. ``Spectral--spatial graph convolutional networks for
  semisupervised hyperspectral image classification.'' \emph{IEEE Geoscience
  and Remote Sensing Letters} 16 (2): 241--245.

\bibitem[Saranathan and Parente(2016)]{Saranathan2016}
Saranathan, Arun~M., and Mario Parente. 2016. ``{Uniformity-based superpixel
  segmentation of hyperspectral images}.'' \emph{IEEE Transactions on
  Geoscience and Remote Sensing} 54 (3): 1419--1430.
  https://doi.org/{10.1109/TGRS.2015.2480863}.

\bibitem[Schowengerdt(2006)]{schowengerdt2006remote}
Schowengerdt, R.A. 2006. \emph{Remote Sensing: Models and Methods for Image
  Processing}. 3rd ed. Elsevier Science.

\bibitem[Sellars, Aviles-Rivero, and Schönlieb(2020)]{Sellars2020}
Sellars, Philip, Angelica~I. Aviles-Rivero, and Carola-Bibiane Schönlieb.
  2020. ``Superpixel Contracted Graph-Based Learning for Hyperspectral Image
  Classification.'' \emph{IEEE Transactions on Geoscience and Remote Sensing}
  58 (6): 4180--4193. https://doi.org/{10.1109/TGRS.2019.2961599}.

\bibitem[Shi and Malik(2000)]{Shi2000}
Shi, Jianbo, and Jitendra Malik. 2000. ``{Normalized cuts and image
  segmentation}.'' \emph{IEEE Transactions on Pattern Analysis and Machine
  Intelligence} 22 (8): 888--905. https://doi.org/{10.1109/34.868688}.

\bibitem[Shimoni, Haelterman, and Perneel(2019)]{Shimoni2019}
Shimoni, Michal, Rob Haelterman, and Christiaan Perneel. 2019. ``{Hyperspectral
  imaging for military and security applications: Combining Myriad processing
  and sensing techniques}.'' \emph{IEEE Geoscience and Remote Sensing Magazine}
  7 (2): 101--117. https://doi.org/{10.1109/MGRS.2019.2902525}.

\bibitem[Somers et~al.(2012)]{Somers2012}
Somers, Ben, MacIel Zortea, Antonio Plaza, and {Gregory P.} Asner. 2012.
  ``Automated extraction of image-based endmember bundles for improved spectral
  unmixing.'' \emph{IEEE Journal of Selected Topics in Applied Earth
  Observations and Remote Sensing} 5 (2): 396--408.
  https://doi.org/{10.1109/JSTARS.2011.2181340}.

\bibitem[Stehman(1997)]{stehman1997selecting}
Stehman, Stephen~V. 1997. ``Selecting and interpreting measures of thematic
  classification accuracy.'' \emph{Remote Sensing of Environment} 62 (1):
  77--89.

\bibitem[Stutz, Hermans, and Leibe(2018)]{stutz2018}
Stutz, David, Alexander Hermans, and Bastian Leibe. 2018. ``Superpixels: An
  evaluation of the state-of-the-art.'' \emph{Computer Vision and Image
  Understanding} 166: 1--27.
  https://doi.org/{https://doi.org/10.1016/j.cviu.2017.03.007}.

\bibitem[Subudhi et~al.(2021)]{Subudhi2021}
Subudhi, Subhashree, Ram~Narayan Patro, Pradyut~Kumar Biswal, and Fabio
  Dell'acqua. 2021. ``{A Survey on Superpixel Segmentation as a Preprocessing
  Step in Hyperspectral Image Analysis}.'' \emph{IEEE Journal of Selected
  Topics in Applied Earth Observations and Remote Sensing} 14: 5015--5035.
  https://doi.org/{10.1109/JSTARS.2021.3076005}.

\bibitem[Tao et~al.(2015)]{tao2015unsupervised}
Tao, Chao, Hongbo Pan, Yansheng Li, and Zhengrou Zou. 2015. ``Unsupervised
  spectral--spatial feature learning with stacked sparse autoencoder for
  hyperspectral imagery classification.'' \emph{IEEE Geoscience and remote
  sensing letters} 12 (12): 2438--2442.

\bibitem[Tu et~al.(2018)]{tu2018knn_superpixels_classification}
Tu, Bing, Jinping Wang, Xudong Kang, Guoyun Zhang, Xianfeng Ou, and Longyuan
  Guo. 2018. ``{KNN}-based representation of superpixels for hyperspectral
  image classification.'' \emph{IEEE Journal of Selected Topics in Applied
  Earth Observations and Remote Sensing} 11 (11): 4032--4047.

\bibitem[Van~den Bergh et~al.(2012)]{van2012seeds}
Van~den Bergh, Michael, Xavier Boix, Gemma Roig, Benjamin De~Capitani, and Luc
  Van~Gool. 2012. ``Seeds: Superpixels extracted via energy-driven sampling.''
  In \emph{Computer Vision--ECCV 2012: 12th European Conference on Computer
  Vision, Florence, Italy, October 7-13, 2012, Proceedings, Part VII 12},
  13--26. Springer.

\bibitem[Vasquez and Scharcanski(2018)]{Vasquez2018}
Vasquez, Dionicio, and Jacob Scharcanski. 2018. ``{An iterative approach for
  obtaining multi-scale superpixels based on stochastic graph contraction
  operations}.'' \emph{Expert Systems with Applications} 102: 57--69.
  https://doi.org/{10.1016/j.eswa.2018.02.027}.

\bibitem[Vedaldi and Soatto(2008)]{Vedaldi2008}
Vedaldi, Andrea, and Stefano Soatto. 2008. ``{Quick Shift and Kernel Methods
  for Mode Seeking}.'' \emph{Computer Vision – ECCV 2008} 705--718.

\bibitem[Veganzones et~al.(2014)]{Veganzones2014}
Veganzones, Miguel~A., Guillaume Tochon, Mauro Dalla-Mura, Antonio~J. Plaza,
  and Jocelyn Chanussot. 2014. ``{Hyperspectral image segmentation using a new
  spectral unmixing-based binary partition tree representation}.'' \emph{IEEE
  Transactions on Image Processing} 23 (8): 3574--3589.
  https://doi.org/{10.1109/TIP.2014.2329767}.

\bibitem[Wan et~al.(2019)]{wan2019multiscale}
Wan, Sheng, Chen Gong, Ping Zhong, Bo~Du, Lefei Zhang, and Jian Yang. 2019.
  ``Multiscale dynamic graph convolutional network for hyperspectral image
  classification.'' \emph{IEEE Transactions on Geoscience and Remote Sensing}
  58 (5): 3162--3177.

\bibitem[Wang et~al.(2017{\natexlab{a}})]{Wang20177}
Wang, Murong, Xiabi Liu, Yixuan Gao, Xiao Ma, and Nouman~Q. Soomro.
  2017{\natexlab{a}}. ``Superpixel segmentation: A benchmark.'' \emph{Signal
  Processing: Image Communication} 56: 28--39.
  https://doi.org/{https://doi.org/10.1016/j.image.2017.04.007}.

\bibitem[Wang et~al.(2016)]{Wang2016}
Wang, Rui, Heng-Chao Li, Wenzhi Liao, and Aleksandra Pizurica. 2016. ``{Double
  reweighted sparse regression for hyperspectral unmixing}.'' In \emph{2016
  IEEE International Geoscience and Remote Sensing Symposium (IGARSS)}, jul,
  6986--6989. IEEE.

\bibitem[Wang et~al.(2017{\natexlab{b}})]{Wang2017a}
Wang, Rui, Heng-Chao Li, Aleksandra Pizurica, Jun Li, Antonio Plaza, and
  William~J. Emery. 2017{\natexlab{b}}. ``{Hyperspectral Unmixing Using Double
  Reweighted Sparse Regression and Total Variation}.'' \emph{IEEE Geoscience
  and Remote Sensing Letters} 14 (7): 1146--1150.
  https://doi.org/{10.1109/LGRS.2017.2700542}.

\bibitem[Wang et~al.(2017{\natexlab{c}})]{Wang2017spatial}
Wang, Xinyu, Yanfei Zhong, Liangpei Zhang, and Yanyan Xu. 2017{\natexlab{c}}.
  ``{Spatial Group Sparsity Regularized Nonnegative Matrix Factorization for
  Hyperspectral Unmixing}.'' \emph{IEEE Transactions on Geoscience and Remote
  Sensing} 55 (11): 6287--6304. https://doi.org/{10.1109/TGRS.2017.2724944}.

\bibitem[Wei et~al.(2018)]{wei2018}
Wei, Xing, Qingxiong Yang, Yihong Gong, Narendra Ahuja, and Ming-Hsuan Yang.
  2018. ``Superpixel Hierarchy.'' \emph{IEEE Transactions on Image Processing}
  27 (10): 4838--4849. https://doi.org/{10.1109/TIP.2018.2836300}.

\bibitem[Yang et~al.(2018)]{yang2018superpixel}
Yang, Chen, Lorenzo Bruzzone, Haishi Zhao, Yulei Tan, and Renchu Guan. 2018.
  ``Superpixel-based unsupervised band selection for classification of
  hyperspectral images.'' \emph{IEEE Transactions on Geoscience and Remote
  Sensing} 56 (12): 7230--7245.

\bibitem[Yao et~al.(2015)]{yao2015real}
Yao, Jian, Marko Boben, Sanja Fidler, and Raquel Urtasun. 2015. ``Real-time
  coarse-to-fine topologically preserving segmentation.'' In \emph{Proceedings
  of the IEEE conference on computer vision and pattern recognition},
  2947--2955.

\bibitem[Ye et~al.(2022)]{ye2022combining}
Ye, Chuanlong, Shanwei Liu, Mingming Xu, and Zhiru Yang. 2022. ``Combining
  low-rank constraint for similar superpixels and total variation sparse
  unmixing for hyperspectral image.'' \emph{International Journal of Remote
  Sensing} 43 (12): 4331--4351.

\bibitem[{Yi} and {Velez-Reyes}(2018)]{Yi2018}
{Yi}, Jiarui, and Miguel {Velez-Reyes}. 2018. ``{Low-dimensional enhanced
  superpixel representation with homogeneity testing for unmixing of
  hyperspectral imagery}.'' In \emph{Algorithms and Technologies for
  Multispectral, Hyperspectral, and Ultraspectral Imagery XXIV}, Vol. 10644 of
  \emph{Society of Photo-Optical Instrumentation Engineers (SPIE) Conference
  Series}, May, 1064422.

\bibitem[Yuan et~al.(2021)]{yuan2021contentAdaptiveSuperpixels}
Yuan, Ye, Wei Zhang, Hai Yu, and Zhiliang Zhu. 2021. ``Superpixels With
  Content-Adaptive Criteria.'' \emph{IEEE Transactions on Image Processing} 30:
  7702--7716.

\bibitem[Zhang et~al.(2018)]{Zhang2018}
Zhang, Shaoquan, Jun Li, Heng-Chao Li, Chengzhi Deng, and Antonio Plaza. 2018.
  ``{Spectral–Spatial Weighted Sparse Regression for Hyperspectral Image
  Unmixing}.'' \emph{IEEE Transactions on Geoscience and Remote Sensing} 56
  (6): 3265--3276. https://doi.org/{10.1109/TGRS.2018.2797200}.

\bibitem[Zhang et~al.(2017)]{Zhang2017a}
Zhang, Shaoquan, Jun Li, Javier Plaza, Heng-Chao Li, and Antonio Plaza. 2017.
  ``{Spatial weighted sparse regression for hyperspectral image unmixing}.'' In
  \emph{2017 IEEE International Geoscience and Remote Sensing Symposium
  (IGARSS)}, jul, 225--228. IEEE.

\bibitem[Zhang, Yuan, and Li(2022)]{zhang2022reweighted}
Zhang, Xinxin, Yuan Yuan, and Xuelong Li. 2022. ``Reweighted Low-Rank and
  Joint-Sparse Unmixing With Library Pruning.'' \emph{IEEE Transactions on
  Geoscience and Remote Sensing} 60: 1--16.

\bibitem[Zhao, Yuan, and Wang(2019)]{zhao2019}
Zhao, Yang, Yuan Yuan, and Qi~Wang. 2019. ``Fast Spectral Clustering for
  Unsupervised Hyperspectral Image Classification.'' \emph{Remote Sensing} 11
  (4). https://doi.org/{10.3390/rs11040399}.

\bibitem[Zheng et~al.(2016)]{Zheng2016}
Zheng, Cheng~Yong, Hong Li, Qiong Wang, and C.L. {Philip Chen}. 2016.
  ``{Reweighted Sparse Regression for Hyperspectral Unmixing}.'' \emph{IEEE
  Transactions on Geoscience and Remote Sensing} 54 (1): 479--488.
  https://doi.org/{10.1109/TGRS.2015.2459763}.

\bibitem[Zou and Lan(2019)]{Zou2019}
Zou, Jinlin, and Jinhui Lan. 2019. ``{A multiscale hierarchical model for
  sparse Hyperspectral unmixing}.'' \emph{Remote Sensing} 11 (5).
  https://doi.org/{10.3390/rs11050500}.

\end{thebibliography}
\end{document}